\documentclass[fleqn,usenatbib]{mnras}

\usepackage{newtxtext,newtxmath}

\usepackage[T1]{fontenc}

\DeclareRobustCommand{\VAN}[3]{#2}
\let\VANthebibliography\thebibliography
\def\thebibliography{\DeclareRobustCommand{\VAN}[3]{##3}\VANthebibliography}

\usepackage{graphicx}	
\usepackage{amsmath}	

\def\Omm{{\Omega_m}}
\def\Omdm{{\Omega_{DM}}}
\def\Omb{{\Omega_b}}
\def\Oml{{\Omega_{\Lambda}}}
\def\beq{\begin{equation}}
\def\eeq{\end{equation}}

\newcommand{\Msun}{\ensuremath{M_{\odot}}}

\title[Plane of satellites for Milky Way-{ mass} galaxies in  cosmological simulations]{Analysis of the plane of satellites around Milky Way-{ mass} galaxies in the IllustrisTNG simulation}

\author[Xinghai Zhao et al.]{
Xinghai Zhao,$^{1}$\thanks{E-mail: xzhao@iona.edu}
Guobao Tang,$^{2}$
Paola Gonzalez,$^{3}$
Grant J. Mathews$^{2}$
and Lara Arielle Phillips$^{2}$
\\
$^{1}$Mathematics \& Physics Department, Iona University, New Rochelle, New York 10801, USA\\
$^{2}$Center for Astrophysics, Department of Physics and Astronomy, University of Notre Dame, Notre Dame, Indiana 46556, USA\\
$^{3}$Department of Physical Sciences, Dalton State College, Dalton, Georgia 30720, USA\\
}

\date{Accepted XXX. Received YYY; in original form ZZZ}

\pubyear{\the\year{}}

\begin{document}
\label{firstpage}
\pagerange{\pageref{firstpage}--\pageref{lastpage}}
\maketitle

\begin{abstract}
It has been suggested that the Plane of Satellites (PoS) phenomenon may imply a tension with current $\Lambda$CDM cosmology since a Milky Way (MW)-like PoS is very rare in simulations. In this study, we analyze a large sample of satellite systems of MW-{ mass}  galaxies identified within the IllustrisTNG simulations. We analyze their spatial aspect ratio, orbital pole dispersion, Gini coefficient, radial distribution, and bulk satellite velocity relative to the host galaxy. These are compared to properties of the observed Milky~Way PoS. We identified galaxy samples in two virial mass ranges ($0.1 - 0.8 \times 10^{12} $ M$_\odot$ and $0.8 - 3.0 \times 10^{12}$ M$_\odot$). { We find for both mass ranges that only $\sim$ 1 percent of MW-like galaxies contain a PoS similar to that of the MW. Nevertheless, these outliers occur naturally in $\Lambda$CDM cosmology. We analyze the formation, environment, and evolution of the PoS for nine systems that are most MW-like in the TNG50 simulation. By using a new analysis of the local large scale structure surrounding the satellite systems, we suggest that a PoS can form from one or more of at least five different processes. Moreover, consistent with the SAGA DR3 results, we find that about 1/3 of the identified systems contain a recent infall of a LMC and/or a SMC like system. We find a tendency for about half of the satellites to have recently arrived at $z < 0.2$, indicating that a MW-like PoS is a recent and transient phenomenon.}

\end{abstract}




\section{Introduction}
The fact that a number of satellites of the Milky Way (MW) form a thin plane with a nearly polar orientation has been noted for decades \citep{Lynden-Bell:1976,Kunkel:1976,Kroupa:2005}, but has not been well understood.  This phenomenon has been referred to \citep{Kang:2005,Kroupa:2005,Metz:2007,Metz:2009} as the "plane of satellites" (henceforth PoS in this paper), the ``disk of satellites''\citep{Maji17, Zhao23},  or the ``vast polar structure'' (VPoS), [e.g. \cite{Pawlowski:2012, Taibi24}].  The question as to whether the existence of the  PoS  poses a challenge to the current $\Lambda$CDM theory of structure formation  has been actively debated in the literature [e.g.~\cite{Kroupa:2005,Libeskind:2005,Zentner:2005, Metz:2007,Kroupa:2010,Kroupa:2012,Pawlowski:2012a,Pawlowski:2020,Sawala:2022a,Pham:2023,Xu:2023,Zhao23}]. 

In addition to the alignment of the 11 "classical" luminous satellites of the MW, similar anisotropic spatial distributions have also been found for the faint satellites of the MW \citep{Metz:2009,Kroupa:2010}, the globular clusters, streams of stars and gas of the MW \citep{Keller:2012, Pawlowski:2012}, the satellites of the Andromeda galaxy (M31) \citep{Grebel:1999,Hartwick:2000,Koch:2006,McConnachie:2006,Metz:2007,Metz:2009, Ibata:2013}, the satellites of the Centaurus A galaxy \citep{Tully:2015,Muller:2018}, possibly the dwarf galaxies in the M101 group \citep{Muller:2017} and the satellites of even more distant galaxies \citep{Heesters:2021}.

It has also been suggested that at least some of the satellites in the PoS have a coherent and/or rotational motion in phase space \citep{Lynden-Bell:1995,Kroupa:2005,Metz:2007,Metz:2008, Ibata:2013, Muller:2018}. Moreover, with the available proper motion studies \citep{McConnachie20, Li21,Pace22,Battaglia22} of the satellites from the Gaia mission \citep{Gaia18,Gaia21}, it has become possible to understand the accretion history and motion of the satellites in more detail \citep{Helmi20, Hammer21, Hammer23, Taibi24}.

{ Of particular interest for the present study is the recent SAGA DR3 census \citep{Mao:2024} of 101 satellite systems around Milky-Way mass galaxies within a distance of 25-40.75 Mpc.   The SAGA satellite radial distributions were found to be less concentrated than the Milky~Way PoS and they do not appear to be co-rotating as has been suggested in the MW and M31 systems.  It was concluded that the best predictor of the abundance of satellite systems is the existence of an LMC-mass system. We will investigate the impact of an LMC like satellite on the radial profile of the satellite systems below in this manuscript.

The purpose of this paper is to make a new independent analysis of the formation of MW-like PoS systems within the context of the IllustrisTNG cosmological simulations.  We identify the most MW-like PoS systems.  These are compared with previous studies and the results of observations, particularly the SAGA DR-3 results. We examine various properties of PoS systems and the likelihood of finding MW-like PoS systems.  We also examine the formation of and cosmic environment of the most MW-like simulated PoS systems.}

\subsection{Background}
There is a long history of attempts to explain the PoS phenomenon. The first such studies \citep{Holmberg:1969,Zaritsky:1997} suggested that the observed alignment of the satellites of the MW is simply due to the fact that it is hard to observe the satellites near the equatorial region of the host galaxy. However, the discoveries of PoS systems outside of the MW make this explanation unlikely.

Numerous efforts have been made to study the PoS phenomenon using numerical simulations. In particular, one of the most puzzling questions is why the satellites  closely align in a certain direction while the dark matter is  more broadly distributed. Early studies using dark-matter only simulations \citep{Kang:2005} or semianalytic models \citep{Libeskind:2005,Zentner:2005} in a $\Lambda$CDM cosmology found that some satellites formed in halos with MW-like masses are indeed distributed along a special direction that resembles the PoS of the MW. The distribution of the satellites does not necessarily trace that of dark matter in the halo. Rather, it is either close to that of the subhalos with the most massive progenitors in the accretion process or close to the major axis of the host halo \citep{Libeskind:2005,Zentner:2005,Agustsson:2006,Libeskind:2007,Libeskind:2009,Deason:2011,Wang:2013}. 

 However, estimates of the likelihood of reproducing a Milky Way{-like} PoS satellite configuration vary significantly depending on the adopted definition of a “MW-like” system. {  For example, \cite{Wang:2013} find} that only a few percent ($\sim5$--$10\%$) of $\Lambda$CDM satellite systems are as spatially flattened as the PoS of the Milky Way. In contrast, studies focusing on the orientation of the satellite system relative to the host galaxy disk find higher probabilities. {  For instance, \cite{Libeskind:2009} find}  a PoS probability of $\sim17$--$35\%$ depending on the adopted orbital alignment thresholds, {  while \cite{Deason:2011} report} that $\sim20\%$ of { PoS} satellite systems are arranged in a plane roughly perpendicular to the host galaxy disk. The coherent or rotational PoS motion has also been {  studied by \cite{Libeskind:2007}, \cite{Lovell:2011}, and \cite{Deason:2011}. They} have found that alignment of the PoS angular momentum with the host galaxy is possible but also relatively rare ($< 10\%$).

Recent developments in numerical simulations of galaxy formation and evolution in a cosmological context, especially {simulations including} baryonic physical processes, along with new observational data of the satellite systems of MW-like galaxies in the nearby universe, have made it possible to study the PoS phenomenon in much greater detail not only on its rarity but also on the physical processes that might create MW-like satellite planes.

On the rarity of PoS systems, some recent studies found that a MW-like PoS is exceedingly rare in simulations with a $\Lambda$CDM cosmology, especially {when the orbital pole alignment is taken into consideration}. Using Gaia proper motion data of the 11 classical satellites of the MW, \cite{Pawlowski:2020} found that there is an alignment of the orbital poles of 7 of the 11 classical satellites. By analyzing the TNG100 simulation data, they claimed that such an alignment is extremely rare (0.1\%) for their adopted selection and orbital-pole statistic. Similarly, \cite{Seo24} concluded that the odds of finding PoS systems like the MW PoS are very small (zero for certain halo selection methods) from an analysis of the TNG50 simulation.

Regarding the PoS around the Andromeda galaxy (M31), by analyzing Millennium-II simulation data and comparing with the observations of M31, \cite{Ibata:14} claimed that only 0.04\% of the M31 like host galaxies in the simulation display a satellite alignment like the one of M31. By studying the proper motion data of NGC 147 and NGC 185, \cite{Pawlowski:2021} showed that only $\sim 0.1\%$ of randomly drawn on-plane satellite pairs match the observed joint orbital-pole alignment for the most-likely proper motions. Also, \cite{Muller:2018} claimed that finding a kinematically coherent thin plane, such as the one in the Centaurus A galaxy, is extremely unlikely $\le$0.5\% in cosmological simulations from the kinematical data of satellite galaxies around Centaurus A.

However, some other studies found that the {probability of finding} a MW-like PoS are at the percent level. In \cite{Samuel:2021}, FIRE-2 simulations \citep{Wetzel:2023} were utilized to show that spatially thin and/or kinematically coherent satellite planes in Milky~Way-like galaxies occur at the $\sim 1\%$ level and the presence of a Large~Magellanic~Cloud (LMC)--like satellite improves the likelihood of finding such systems. \cite{Sawala:2022a} showed that thin planes of satellites like that of the MW are more common ($5.5\%$) in the zoom-in constrained simulations from the SIBELIUS project \citep{Sawala:2022} if numerically disrupted satellites are included. They also analyzed the MW satellite proper motion data from Gaia EDR3 and showed that the thin plane of satellites around the MW is transitional and thus not rotationally supported. 

Using a new satellite plane finding algorithm on the Magneticum Pathfinder simulation data, \cite{Forster:2022} found that a thin plane of satellites existed in almost all of the simulated galaxy systems even in galaxy clusters, although \cite{Gu:2022} found that the odds are $13.1\%$ for the MW-like systems and $4.7\%$ for the galaxy clusters using the Sloan Digital Sky Survey and the Millennium simulation data. Using ELVIS and Caterpillar simulation suites, \cite{Pham:2023} showed that the  PoS is sensitive to the radial distribution of satellites and that the satellite system of the MW is unusual. However, they conclude that the MW  PoS is consistent with  $\Lambda$CDM cosmology at the $2- 3\sigma$ level. \cite{Xu:2023} was able to find one satellite system that closely resembles the MW PoS both spatially and dynamically among 231 MW-like candidates they selected in the TNG50 simulations. However, the satellite plane was found to be transient in nature.

 Using the same TNG50 simulation, \cite{Sawala:2025} examined the PoS from different reference-frames and showed that while the host-centered satellite plane itself is short-lived, the galaxies that define the plane at z=0 can belong to a spatially coherent group that persists for several billion years prior to infall. Closely related work has focused on kinematically persistent planes (KPPs), defined as fixed subsets of satellites that maintain coherent orbital poles over extended periods of time \citep{Santos23, Gamez-Marin24, Gamez-Marin25}. In particular, \cite{Gamez-Marin25} analyzed the frequency and properties of these structures in MW/M31-like systems using TNG50, finding that such kinematically coherent subsets are not uncommon in $\Lambda$CDM cosmology. In this paper, we focus on the present-day properties of the PoS and on the implications these have for its possible origins. A detailed dynamical study including the time evolution of the PoS and whether this present-day structure reflects long-term coherence or arises from time-dependent orbital evolution will be presented in a forthcoming paper.

The origin of the PoS, and especially the coherent motion of the satellites in the plane, has long been debated. The early claim was that it may be from an anisotropic accretion process along the minor filaments in the cosmic web \citep{Libeskind:2005,Libeskind:2011,Lovell:2011}. However, by analyzing the angular momentum data of the satellites in the Aquarius simulations \citep{Springel:2008} and the Via Lactea simulations \citep{Diemand:2007,Diemand:2008}, \cite{Pawlowski:2012a} have strongly argued that this cannot be the case because the probability is too low (0.5 percent). Also, by studying the observed proper motion of 4 classical satellites of the MW, \cite{Angus:2011} argued that this is unlikely to be the case based on the observational data. However, with new numerical simulations and observational data, recent studies \citep{Wang:2020,Xia:2021,Wang:2021,Dupuy:2022,Forster:2022,Mezini:2025,Madhani:2025} have suggested that it may be the most likely process to form a PoS. In this manuscript we describe a  study of this process in great detail using Milky~Way--like galaxies in the IllustrisTNG simulations \citep{Weinberger:2017,Pillepich:2018}.

Another physical process that can possibly create a PoS is the group accretion of satellites. \cite{Li:2008} and \cite{DOnghia:2008} suggested that the coherent motion of satellites may come from the fact that the satellites have accreted to the main galaxy as a group.  However,  it has been argued that the group will be too extended and short-lived \citep{Metz:2009a,Klimentowski:2010,Nichols:2011,Rocha:2012,Kanehisa23} to account for the current PoS of the MW. Nonetheless, some recent studies [e.g. \cite{Nadler:2020};\cite{Garavito-Camargo:2021};\cite{Samuel:2021};\cite{Mao:2024};\cite{Garavito-Camargo:2024}] suggested that the accretion of Large~Magellanic~Cloud (LMC)-- and Small~Magellanic~Cloud (SMC)--like satellites and/or satellites associated with the LMC and SMC may have played an important role in the formation of the PoS. In the present work we examine the role of LMC and SMC-like galaxies in the formation of the PoS.

An alternate route to form a PoS is the alignment of tidal dwarf galaxies (TDGs) that formed during a close encounter between two galaxies \citep{Zwicky:1956,Lynden-Bell:1976,Kroupa:1997,Kroupa:2005,Metz:2007a,Metz:2008}. \cite{Pawlowski:2011,Pawlowski:2012a} and \cite{Fouquet:2012} have shown that TDGs can have a spatial distribution and orbiting pole alignment similar to MW's PoS. However, TDGs mostly contain only baryonic matter \citep{Barnes:1992,Bournaud:2010}. So, they cannot have a very high mass-to-light ratio as currently observed \citep{Simon:2007,Simon:2011}) and match current observations well without using some modified gravity theory.

In this work we study the PoS of Milky~Way-mass galaxies in IllustrisTNG simulations. The relatively high-resolution simulations and large-size samples of galaxies make it possible for us to identify and study 9 ``Milky~Way--like'' satellite { PoS} systems in detail, especially for their possible formation processes based on their environment. We will also compare the results from the simulations with recent observational data in \cite{Nadler:2020}, ELVES Survey \citep{Carlsten:2022} and SAGA Survey DR3 \citep{Mao:2024}.


{ A new aspect of this work is that we also consider the impact of possible new determinations of the virial mass of the Milky Way based on Gaia DR3. These studies suggest a MW mass somewhat below  the  traditional virial mass  $ 1.28^{+.97}_{-0.48} \times 10^{12}$ M$_\odot$ \citep{Watkins19}. The newer data infer a much lower MW  virial mass of $1.81^{+0.06}_{-0.05} \times 10^{11}$ M$_\odot$ \citep{Ou:2024}, $2.06^{+0.24}_{-0.13} \times 10^{11}$ M$_\odot$ \citep{Jiao23}, $0.64^{+0.15}_{-0.14} \times 10^{12}$ M$_\odot$ \citep{Roche24} and $7.32^{+1.98}_{-1.53} \times 10^{11}$ M$_\odot$ \citep{Ou25}.
We will present results for both the traditional (high) and the new (low) MW virial mass ranges.}

This paper is organized as follows: in Section~\ref{sec_methods}, we describe the simulations and the selection criteria of the MW-like galaxy sample. We also define several characteristic parameters that we will use to study the Plane of Satellites. {We also present our method for estimating the local filament spine direction from the dark matter density field surrounding each host halo.} We present our results in Section~\ref{sec_selection} and \ref{sec_Analysis}: in Section~\ref{sec_Spatial_Dis} and \ref{sec_Orbital}, we analyze the spatial distribution and orbital-pole alignment of the satellite { PoS} systems in the galaxy sample and select galaxies with a MW-like PoS. We then study the Gini coefficient, radial distribution, and bulk velocity of the MW PoS and MW-like PoS systems in the simulations.  These are compared with the whole MW-like galaxy sample in Section~\ref{sec_other_properties}. In Section~\ref{sec_orientation} and \ref{sec_lmc_smc}, we study the possible connection between the satellite plane and local filaments and the impacts of MC-like satellites. {We study the time evolution of the PoS systems in Section~\ref{sec_time_evo}. }We present our summary and conclusions in Section~\ref{sec_conclusions}.



\section{Methodology}
\label{sec_methods}
\subsection{IllustrisTNG simulations}
In order to have a statistically significant sample size of  simulated    Milky~{ Way-mass} galaxies, we utilize 'The Next Generation' IllustrisTNG cosmologcal simulations \citep{Weinberger:2017,Pillepich:2018} for our study. Mainly, we consider the highest-resolution simulation, TNG50 \citep{Nelson:2019,Pillepich:2019}, but we also consider the larger volume simulation, TNG100 \citep{Marinacci:2018,Naiman:2018,Nelson:2018,Pillepich:2018a,Springel:2018}.  

The IllustrisTNG simulations have been set up with initial conditions derived from  a $\Lambda$CDM cosmology with cosmological parameters deduced from the \cite{Planck-Collaboration:2016}, i.e. $\Omm=0.3089$ in which $\Omdm=0.2603$ and $\Omb=0.0486$, $\Oml=0.6911$, $H_0=100 h$ km s$^{-1}$ Mpc$^{-1}$ with $h=0.6774$, $\sigma_8 = 0.8159$ and $n_s = 0.9667$. The simulations started at $z=127$ with the chosen initial conditions and were evolved to the current epoch at $z=0$. The simulations use periodic boundary conditions in co-moving coordinates with a box size of 35 Mpc/$h$ (51.7 Mpc) for the TNG50 and 75 Mpc/$h$ (110.7 Mpc) for the TNG100.  

The simulations were performed with the massive parallel code AREPO \citep{Springel:2010} which is capable of a finite-volume moving-mesh method for the hydrodynamics and a tree-particle-mesh
(TreePM) method for gravity. The AREPO code incorporates important baryonic physical processes in galaxy formation and evolution that include gas cooling and heating, gas metallicity enrichment, star formation and evolution, supernova feedback, as well as black hole formation, growth and feedback.

In order to identify galaxies and their associated satellite systems in the simulations, a friends-of-friends (FOF) algorithm was utilized to link particles based upon their mean particle spacing. The linked particle groups were then treated as galaxy systems that include central galaxies and satellites. To distinguish the central host galaxy and satellites in each galactic system, the SUBFIND algorithm \citep{Springel:2001a} was used to find gravitationally bound particle groups in the systems. The dominating group was categorized as the central host galaxy and the rest of the groups were treated as satellites.  

\subsection{Identifying Milky~Way-mass galaxies}
We identify simulated Milky~Way-mass galaxies on the basis of their virial mass without any other additional conditions. As noted in the introduction, we consider two possible  virial mass ranges for MW-like galaxies.  For the canonical high-mass range, we select host galaxies with a virial mass in the range of $0.8 - 3.0 \times 10^{12}\Msun$. This is the same mass criterion as that used by \cite{Pawlowski:2021}.  For the low-mass range, we adopt $0.1 - 0.8 \times 10^{12} \Msun$, consistent with \cite{Ou:2024, Ou25}, \cite{Jiao23}, and \cite{Roche24}.
Based upon these criteria,  we have identified 186  simulated MW-like galaxies in TNG50 and 1,741 in TNG100 in the high-mass range.  For the low-mass limits, we identify a much larger sample of 1,392 MW-like galaxies in the TNG50 simulation.   We list some key parameters of the simulations in Table \ref{tab_sim_par}.

\begin{table*}
    \centering
    \caption{Characteristic parameters of the TNG50 and TNG100 simulations and the number of identified Milky Way-{mass} galaxies at redshift $z=0$ for the high and low MW mass ranges.}
    \label{tab_sim_par}

    \begin{tabular}{lcccccc}
        \hline
        Simulation &
        Box Size\textsuperscript{a} &
        $m_{\mathrm{gas}}$\textsuperscript{b} &
        $m_{\mathrm{DM}}$\textsuperscript{b} &
        $\epsilon_{\mathrm{gas,min}}$\textsuperscript{c} &
        $\epsilon_{\mathrm{DM,stars}}$\textsuperscript{d} &
        $N_{\mathrm{MW}}$\textsuperscript{e} \\
        &
        [Mpc\,$h^{-1}$] &
        [$10^{4}\,M_\odot$] &
        [$10^{5}\,M_\odot$] &
        [pc] &
        [pc] &
        \\
        \hline
        \multicolumn{7}{l}{\textbf{$M_{\mathrm{virial}} = 0.8$--$3.0\times10^{12}\,M_\odot$}} \\
        TNG50  & 35 & 8.5  & 4.5 &  74 & 288 &  186 \\
        TNG100 & 75 & 140  & 75  & 185 & 740 & 1741 \\
        \hline
        \multicolumn{7}{l}{\textbf{$M_{\mathrm{virial}} = 0.1$--$0.8\times10^{12}\,M_\odot$}} \\
        TNG50  & 35 & 8.5  & 4.5 &  74 & 288 & 1392 \\
        \hline
    \end{tabular}

    \vspace{3pt}
    \parbox{\textwidth}{
    \textbf{Notes.}  
    \textsuperscript{a} Size of the periodic simulation box.  
    \textsuperscript{b} { Mean initial mass of the gas cells and dark matter particles. Note that the star-particle mass diminishes with time due to mass loss during stellar evolution.} 
    \textsuperscript{c} Minimum gravitational softening length for the gas cells.  
    \textsuperscript{d} Softening length of the dark matter and star particles at $z=0$.  
    \textsuperscript{e} Number of Milky { Way-mass} galaxies in the simulations within the adopted virial mass limits.
    }
\end{table*}

Among the galaxies that satisfy the high- and low-mass criterion, we next selected the 11 most luminous (mass in stars M$_* \ge 10^5$ M$_\odot$) satellites within 300 kpc as a proxy for the 11 most luminous satellites observed in the MW. We also require a non-zero \text{subhalo\_flag} parameter for each satellite. This ensures that only satellites of cosmological origin are included.

Of particular interest for the present study is the recent SAGA DR3 census \citep{Mao:2024} of 101 satellite systems around Milky-Way mass galaxies within a distance of 25-40.75 Mpc.   The selection criterion in the SAGA survey for MW-like galaxies was primarily based upon distance, K-band luminosity, and the local environment \citep{Mao:2021}. The specific selection criteria includes the following:
\begin{itemize}
    \item stellar mass: {$-23 < M_K < -24.6$};
    \item stellar foreground: $|b| > 25^\circ$;
    \item stellar foreground: $H_p^{\rm brightest~star} (<300~{\rm kpc}) > 5$;
    \item environment: $K < K_{\rm brightest~gal.\ }^{(<300~{\rm kpc})} - 1.6$;
    \item environment: $M_{\mathrm{halo}} < 10^{13}\, M_\odot$;
    \item distance: $25\text{--}40.75~\mathrm{Mpc}$; and
    \item distance: $|v_{\rm helio}| > 1400~\mathrm{km\,s^{-1}}$.
\end{itemize}

Here, $M_K$ is the absolute K-band magnitude, $\vert b \vert$ is the angle above the Galactic plane, and $H_p$ is the broad
optical Hipparcos magnitude. {The K-band magnitude range is adopted as a proxy for host stellar mass and corresponds to central galaxies with stellar masses of approximately $10^{10}$--$10^{11} M_\odot$. The criterion requires that there be no galaxy within a projected radius of 300 kpc satisfying
$K < K_{\rm host}-1.6$.} The magnitude difference of
1.6 mag was chosen to ensure that no two MW host galaxies in the K-band magnitude range could be found inside the same
virial volume. 

{The SAGA DR3 sample contains 378 confirmed satellites around 101 MW-mass host systems, corresponding to an average of 4 satellites per host. The survey completeness is highest for satellites with stellar masses above approximately $10^{7.5} M_\odot$.}

 For the high-mass range in the simulations studied here, nearly every host galaxy (185 out of 186) had 11 satellites satisfying the selection criterion of 11 satellites within 300 kpc. This is a higher fraction than in the SAGA DR3 sample.
For the low-mass range criterion, we identified 173 systems with 11 satellites within 300 kpc among 1392 MW-like systems.  This seems more or less consistent with the SAGA DR3 sample, where only about 10\% of the 101 complete host galaxies (cf. inset in  Fig. 8 of that paper) had 11 or more satellites within 300 kpc, although the 25\% most compact had a similar radial distribution to that of the MW. However, a direct comparison is not straightforward since the SAGA survey is probably incomplete for less luminous  lower-mass satellites.

\subsection{Characterization of spatial and orbital pole alignment of satellites}
Having identified MW-like systems with 11 selected satellites, the next step was to quantify the aspect ratio of the distribution of the satellite galaxies around the central galaxy.  It is common practice (e.g. \cite{Sawala:2022a}) to diagonalize a  second moment tensor of the  spatial coordinates of the satellites relative to the geometric center of the satellite distribution, i.e.
\begin{equation}
\label{moment_tensor}
I_{ij} = \sum_n x_{i,n}x_{j,n}~~,
\end{equation}
where $x_{i,n}$ and $x_{j,n}$ are spatial coordinates 
of the $n$-th satellite and the indices $i$ and $j = 1,2,3$ run over the 3 spatial dimensions.  Note, this is not to be confused with the moment of inertia tensor nor the reduced inertia tensor.  It is an often employed measure of the geometrical distribution of the satellites.

The three eigenvectors of the tensor can be interpreted as the three principal axes $\vec a$, $\vec b$ and $\vec c$ of an ellipsoid for which the magnitudes $a>b>c$ are the lengths of the semi-axes. In this parameterization, satellites have the least spatial extension in the $\vec c$ direction.  Hence,  they are closer to the plane formed by the vectors $\vec a$ and $\vec b$ and the ratio $c/a$ can be used to characterize the aspect ratio of the satellite distribution. This plane is defined as the "Plane of Satellites (PoS) ". For our comparisons, we adopt a value of $c/a = 0.182 $ for the MW PoS as deduced in \cite{Pawlowski:2014}.

The Plane of Satellites can be represented with the Hesse normal form,
\begin{equation}
\vec{n}\cdot\vec{x} - d =0,
\end{equation}
where $\vec{n}$ is the unit normal vector of the plane. It is in the same direction as the $\vec c$ axis of the PoS.  The vector  $\vec{x}$ is the position of any point in the plane  and $d$ is the distance from the plane to the coordinate origin (host galactic center).

 To characterize the clustering of orbital poles of the PoS galaxies, we use the formulation introduced by \cite{Metz:2007} and \cite{Pawlowski:2020}.   That is, we calculate the average direction $\langle \vec n\rangle$ of a sample of $k$  satellite orbital poles on the unit sphere.  We then determine  the orbital pole dispersion $\Delta_{k}$ about this average direction, given by:
\begin{equation}
\label{pole_disp}
\Delta_{k}  = \sqrt{\frac{1}{k}\sum_{i=1}^{k} \theta_i^2} = \sqrt{\frac{1}{k}\sum_{i=1}^{k} [\cos^{-1}(\langle \vec n \rangle \cdot \vec n_i)]^2}~~,
\end{equation}
where $\theta_i$ is the angle between the orbital pole of the $i^{\, th}$ satellite and the direction of the mean orbital pole of the total satellite sample. The direction of the orbital pole is calculated as the direction of the angular momentum without mass weighting.  This parameter gives a measure of the degree of alignment of the PoS.  That is, a small value of $\Delta_k$ corresponds to a close alignment of the rotation of the individual satellites with the PoS.  For reference, we adopt values for the Milky Way PoS of $\Delta_k = 56^o$ \citep{Pawlowski:2020} (with $k=11$ for the whole sample of the 11 most luminous satellites), and $\Delta_{k7} = 18.9^o$ \citep{Sawala:2022a} (with $k=7$ for the 7 most aligned satellites in the whole sample). 

\subsection{Gini Coefficient}
\cite{Sawala:2022a} and \cite{Pham:2023} have suggested that the radial distribution of satellites around Milky Way-mass galaxies may have a crucial impact on the formation of the PoS. One way to study this impact quantitatively is to use the  Gini coefficient, $G$ as introduced by \cite{Sawala:2022a}. For a sample of $k$ satellites, $G$ is defined by:
\begin{equation}
\label{gini_coef}
    G = \frac{\sum_{i=1}^{k} (2i - k -1) r_i^2}{(k-1)\sum_{i=1}^{k}r_i^2}~~.
\end{equation}
The square of the radius $r_i$ is a measure of the contribution of each satellite to the second moment tensor.  The Gini coefficient quantifies the inequality of these contributions. That is, the symmetric nature of the numerator forces the Gini coefficient to be 0 for a uniform radial contribution of all satellites to the  second moment tensor. This would occur, for example, if all satellites formed a ring at fixed radius. However, a near unity value of the Gini coefficient corresponds to a large contribution from only one or a few distant satellites, implying a very uneven radial distribution. We calculated $G = 0.645$ for the Milky Way PoS using the data listed in Table \ref{tab_obs_par}. As noted in \cite{Sawala:2022a}, both the $c/a$ values and the Gini coefficient also depend on the orbital phase.

\begingroup
\setlength{\tabcolsep}{2pt}
\begin{table}
    \centering
    \caption{Characteristic parameters of the 11 satellites in the plane of satellites of the Milky Way.}
    \label{tab_obs_par}

    \begin{tabular}{lccccc}
        \hline
        Name &
        $M_\star$ ($M_\odot$)\textsuperscript{a} &
        {Dist.} (kpc)\textsuperscript{b} &
        $v_x$ (km\,s$^{-1}$)\textsuperscript{c} &
        $v_y$ (km\,s$^{-1}$)\textsuperscript{c} &
        $v_z$ (km\,s$^{-1}$)\textsuperscript{c} \\
        \hline
        Sagittarius & $3.4\times10^{7}$   & $16.0\pm2.0$   & $233\pm2$    & $-8\pm4$     & $209\pm4$ \\
        LMC         & $1.1\times10^{9}$   & $50.2\pm2.2$   & $-42\pm6$    & $-223\pm4$   & $231\pm4$ \\
        SMC         & $3.7\times10^{8}$   & $56.9\pm2.2$   & $6\pm8$      & $-180\pm7$   & $167\pm6$ \\
        Ursa Minor  & $5.4\times10^{5}$   & $68.1\pm3.0$   & $7\pm12$     & $56\pm9$     & $-154\pm9$ \\
        Sculptor    & $3.9\times10^{6}$   & $79.2\pm4.0$   & $31\pm7$     & $184\pm7$    & $-97\pm1$ \\
        Draco       & $3.2\times10^{5}$   & $82.0\pm6.0$   & $62\pm3$     & $14\pm2$     & $-166\pm3$ \\
        Sextans     & $7.0\times10^{5}$   & $89.2\pm4.0$   & $-221\pm13$  & $81\pm10$    & $59\pm10$ \\
        Carina      & $3.8\times10^{5}$   & $102.7\pm5.0$  & $-46\pm16$   & $-39\pm7$    & $143\pm16$ \\
        Fornax      & $2.4\times10^{7}$   & $140.1\pm8.0$  & $17\pm18$    & $-140\pm18$  & $94\pm8$ \\
        Leo II      & $1.2\times10^{6}$   & $207.7\pm12.0$ & $-24\pm34$   & $86\pm36$    & $36\pm14$ \\
        Leo I       & $4.9\times10^{6}$   & $254.0\pm30.0$ & $-167\pm29$  & $-28\pm28$   & $100\pm21$ \\
        \hline
    \end{tabular}

    \vspace{3pt}
    \parbox{\columnwidth}{
    \textbf{Notes.}
    \textsuperscript{a} Stellar mass (Garrison et al. 2019).\\
    \textsuperscript{b} Distance from the Galactic centre (Metz et al. 2007).\\
    \textsuperscript{c} Velocities in Galactic Cartesian coordinates (Pawlowski et al. 2020).
    }
\end{table}
\endgroup

\subsection{PoS Bulk Velocity}
In order to study the motion of the satellites in the PoS relative to the main galaxy as a group,  we define a characteristic PoS velocity parameter, as
\begin{equation}
\label{v_pos}
    V_{PoS} = \Bigg| \sum_{i=1,11} (\vec v_i - \vec v_{host}) \Bigg| ~~.
\end{equation}
 Here, $\vec v_{host}$ is the center–of–mass velocity of the host galaxy relative to the simulation box.  This is computed in the simulation as the mass-weighted sum of the velocities of all particles associated with the host.  $V_{PoS} = 668$ km s$^{-1}$ for the MW system as can be deduced directly from the data listed in Table \ref{tab_obs_par}. 
 (Note that this parameter differs from  the average PoS velocity which would be obtained from a mass-weighted sum of satellite velocities divided by the number of satellites.)

\subsection{Estimation of Local Filament Spine Directions}
\label{filament_spine_tensor}
To study the environmental effect of the surrounding large-scale structure on the PoS systems, we determine the direction of the local filament in which each halo resides. In this work, the filament spine direction is estimated directly from the dark matter density field in the vicinity of each halo using sub-volumes extracted from the TNG50 simulation.

For each system, we analyze a cubic region of width $5\,\mathrm{Mpc}/h$ centered on the halo, within which the filamentary structure of the cosmic web is identified and its local orientation is estimated.
The filament direction is obtained from the local curvature of the density field using a multiscale Hessian-based method. The filament-identification pipeline follows \citet{Alithesis2015}; an updated version will be presented in a forthcoming paper.

In practice, we apply a filament-identification algorithm to the Gaussian-smoothed logarithmic dark matter density field within each sub-volume. At each location, the Hessian matrix provides three orthogonal eigenvectors corresponding to the principal directions of density variation. A dimensionless filament morphology response, denoted as $R_{fil}$, is derived from the local Hessian eigenvalue configuration and density contrast. A filament mask is then constructed by selecting voxels with a significant filamentary response.

In regions identified as filamentary, the eigenvector associated with the eigenvalue of smallest absolute magnitude traces the direction along which the density varies least, and we adopt this as the local filament spine direction. The density field is analyzed over multiple smoothing scales to account for the varying physical widths of filaments. At each point, the scale producing the strongest filamentary response is retained as the characteristic local filament scale and denoted by $\sigma_{\rm win}$. The filament spine direction is subsequently estimated from this multiscale filament reconstruction, which is applied consistently to all systems in the analysis.

To obtain a stable estimate of the filament orientation at each halo position, we combine information from neighboring filamentary regions within a spherical volume around the host halo center of each system. Because the multiscale analysis produces significant filamentary responses over a finite range of scales around $\sigma_{\rm win}$, the effective filament width is not uniquely defined. In practice, the neighborhood radius $R$ is chosen within the range $\sigma_{\rm win}$ to $2\sigma_{\rm win}$ according to the local filament morphology, such that the filament cross-section is sufficiently covered without introducing excessive contamination from nearby structures. 

For a given location $\mathbf{x}_0$, we consider all filamentary voxels within a radius $R$. Since filament spine directions represent unoriented axes rather than directed vectors, neighboring orientations are combined using dyadic products, which are invariant under the transformation $\mathbf{e}_j \rightarrow -\mathbf{e}_j$. We therefore construct a weighted orientation tensor,
\begin{equation}
\mathbf{T}(\mathbf{x}_0) = \sum_{j} w_j \, \mathbf{e}_j \otimes \mathbf{e}_j~~,
\end{equation}
where $\mathbf{e}_j$ is the local filament direction at position $\mathbf{x}_j$, and the weights are defined as
\begin{equation}
w_j = R_{fil}^j \, \exp\left(-\frac{|\mathbf{x}_j - \mathbf{x}_0|^2}{2R_{\rm w}^2}\right)~~,
\end{equation}
where voxels with weaker filamentary signatures or larger separations from $\mathbf{x}_0$ contribute less to the orientation estimate. The weighting scale is set to $R_{\rm w} = 0.5\sigma_{\rm win}$, which suppresses contributions from peripheral regions within the sampling volume and reduces the influence of nearby filament branches and background fluctuations. 

The final spine direction at $\mathbf{x}_0$ is taken to be the principal eigenvector of $\mathbf{T}$, corresponding to the dominant orientation within the neighborhood.
In regions where the contributing filamentary voxels have no dominant shared orientation, the resulting tensor becomes less anisotropic, indicating a less well-defined local spine direction. Therefore, the tensor anisotropy is used to assess the coherence of the local orientation estimate. We quantify this using the eigenvalues of $\mathbf{T}$ by defining a dimensionless confidence parameter,
\begin{equation}
\label{c_para}
C = \frac{\mu_1 - \mu_2}{\mu_1 + \mu_2 + \mu_3},
\end{equation}
where $\mu_1 \geq \mu_2 \geq \mu_3$ are the eigenvalues of $\mathbf{T}$. Since $\mathbf{T}$ is constructed from weighted dyadic products, its eigenvalues are nonnegative by construction. Larger values of $C$ correspond to a more coherent local orientation distribution and thus a more well-defined local filament spine direction.

\begin{figure*}
\centering

\includegraphics[trim=0cm 0cm 0cm 0cm, clip, width=\textwidth]{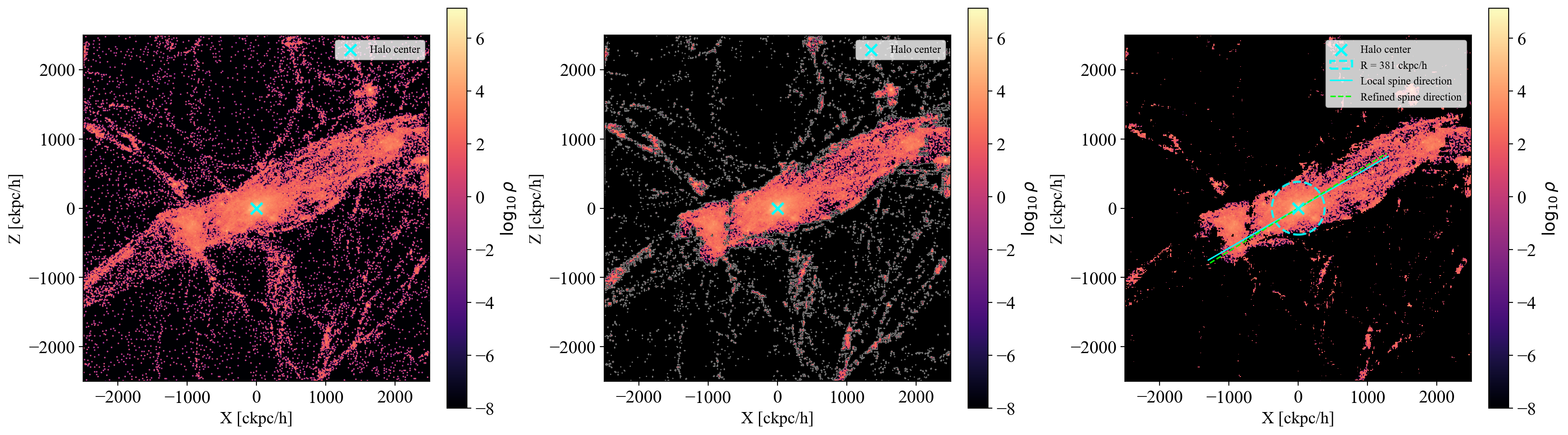}

\caption{Local filament environment of Group~289 in the Cartesian $y$-slice passing through the halo center, marked by the cyan cross. Left: dark matter density slice within the $5~ \mathrm{Mpc}/h$ sub-volume centered on the halo. Middle: density field filtered by the filament mask. The contour indicates the boundary of the filament mask. Right: the same filament density field overlaid with the local spine-direction estimates at the halo position. The solid cyan line shows the local filament direction obtained directly from the Hessian eigenvector at the halo position, while the dashed green line shows the final spine direction derived from the weighted orientation tensor. The dashed circle denotes the neighborhood used for the tensor analysis.}

\label{group289_filament}
\end{figure*}

{As an illustrative example, Figure~\ref{group289_filament} shows the local filament environment identified around Group~289, one of the MW-like systems analyzed in this work. This system resides in a particularly prominent filamentary structure. For visualization purposes, we show the halo-centered slice along the Cartesian direction in which the filamentary morphology is most clearly visible. }

{ The left panel shows a slice of the dark matter density field within the $5~\mathrm{Mpc}/h$ sub-volume centered on the halo. The middle panel shows the density field filtered by the filament mask obtained from the Hessian-based segmentation algorithm with the mask boundary indicated by the contour. This visualization confirms that the halo resides within a coherent filamentary structure rather than being associated with a projection effect. The right panel shows the filament density field together with the spherical neighborhood used to construct the orientation tensor. The solid cyan line indicates the local filament direction obtained directly from the Hessian eigenvector associated with the eigenvalue of smallest absolute magnitude at the halo position, while the dashed green line shows the final spine direction derived from the weighted orientation tensor. The close agreement between the two estimates reflects the high degree of local filament coherence in this system.}

\section{Milky Way-Like Planes of Satellites formed in the simulations}
\label{sec_selection}
We first present our result on the spatial alignment of satellite galaxies around Milky { Way-mass} hosts in the TNG simulations. Then, we study the dynamical properties of the Plane of Satellites formed in the simulations based upon the orbital-pole alignments, Gini coefficients, and PoS bulk velocities.

\subsection{Spatial Distribution}
\label{sec_Spatial_Dis}
First, we compare the spatial distribution of the satellite galaxies with that of the dark matter sub-halo distribution and the total dark matter around the main galaxy.  This is to see whether luminous satellite galaxies have their own spatial distribution or simply trace the sub-halos or dark matter halo around the main host galaxy.

\begin{figure}
\centering
\includegraphics[trim=3cm 2cm 3cm 3cm, clip, width=\columnwidth]{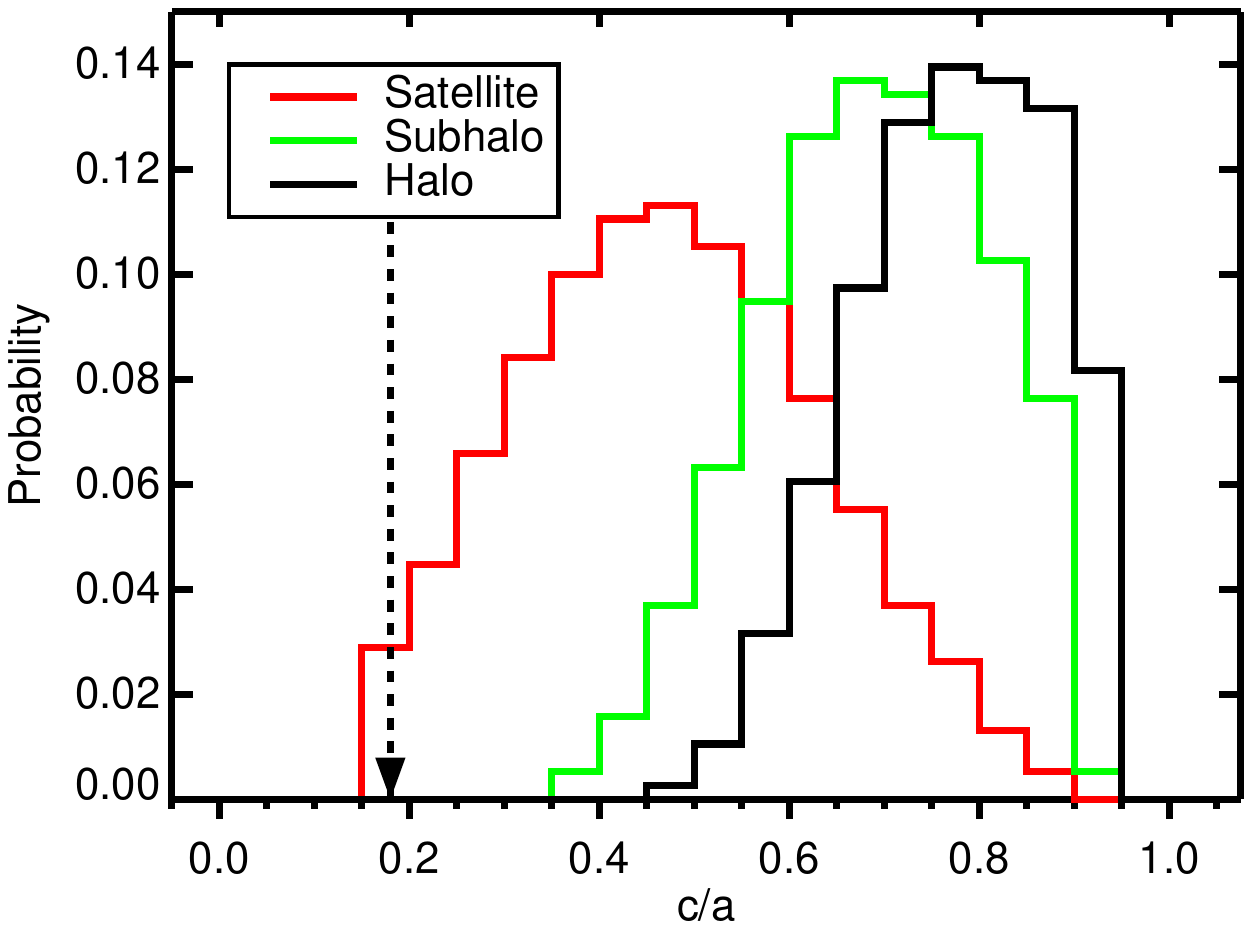}
\caption{Probability distribution of the $c/a$ aspect ratio for the plane of the 11 selected satellites (red line).  The green line gives the $c/a$ ratio for  the total sum of all subhalos including these satellites.  The same ratio for the dark matter halos of the main host galaxies is given black line. The dashed arrow shows the adopted ratio for the MW PoS, $c/a = 0.182$.}
\label{fig_ca_ratio_comp}
\end{figure}

Figure \ref{fig_ca_ratio_comp} shows the probability distribution of the fitted $c/a$ aspect ratios of the spatial distribution of the 11 selected satellites {for the simulated MW-like PoS systems (red histogram), the total of all dark matter subhalos (green histogram), and the dark matter halo of the main galaxy \citep{Anbajagane:2022} in the simulated MW-like systems within the high-mass range of the TNG100 simulations (black histogram).}  By total subhalos, we mean all surrounding {structures} (not just the 11 brightest) associated with the host galaxy. The shape of the dark-matter halo is never a perfect sphere (i.e. $c/a<1$) in the system. 

The distribution of all subhalos roughly traces the distribution of dark matter in the halo as seen in Figure \ref{fig_ca_ratio_comp}.  That is, the green line has a similar shape to that of  the black line but shifted to a slightly lower $c/a$ ratio due to the finite number of subhalos. An important feature is that the full set of subhalos can never form a thin plane with a $c/a$ ratio of 0.182 as that of the Milky Way. 

Previous studies based upon dark-matter only simulations (e.g. \cite{Gu:2022, Sawala:2022a,Pham:2023}) chose a subset based upon the largest circular velocities ($V_{peak}$ constraint \cite{Sawala:2022a}) or a mass constraint \citep{Pham:2023} to select the 11 subhalos. They concluded that the  subset might have a very small probability to form a thin plane with a $c/a$ ratio equal to or less than that of the Milky Way value of 0.182. However, we chose the 11 most luminous satellites within 300 kpc of the host in the present study. This is, perhaps, the most realistic choice, as this is how the MW PoS was originally identified from observation.

The probability distribution of the $c/a$ ratios of the 11 selected satellites exhibits a distinct difference from that of the {total subhalos} and the DM halo. It spans a much broader range and its peak is shifted to a lower value. Most importantly, about 3\% of the systems can have a $c/a$ ratio as low as that of the Milky Way. This is a small but non-negligible probability. We believe that there are two main reasons behind this low $c/a$ ratio.

First, we only have a small number, 11, of satellites that can be treated as the sampling points of the total subhalo distribution, which is  shown by the green line. The fewer the sampling points, the more likely the subset will have a low $c/a$ ratio. For example, if there are only three points in the subset, it is guaranteed to form a plane with a $c/a$ ratio of 0. 

Second, the likelihood that a subhalo can have a significant stellar component and thus form a satellite is dependent upon several factors, i.e. the mass of the subhalo, its environment (i.e. proximity to filaments and availability of star-forming gas), and its history (e.g. a recent major merger). Hence, the distribution of luminous satellites is not completely random, but may be impacted by the larger cosmic structure environment containing the main galaxy and its evolution history. However, no matter what the cause is, the occasional formation of a system with a small $c/a$ ratio is possible.

\subsection{Orbital Pole Alignment}
\label{sec_Orbital}
\begin{figure*}
\centering
\includegraphics[width=0.45\textwidth]{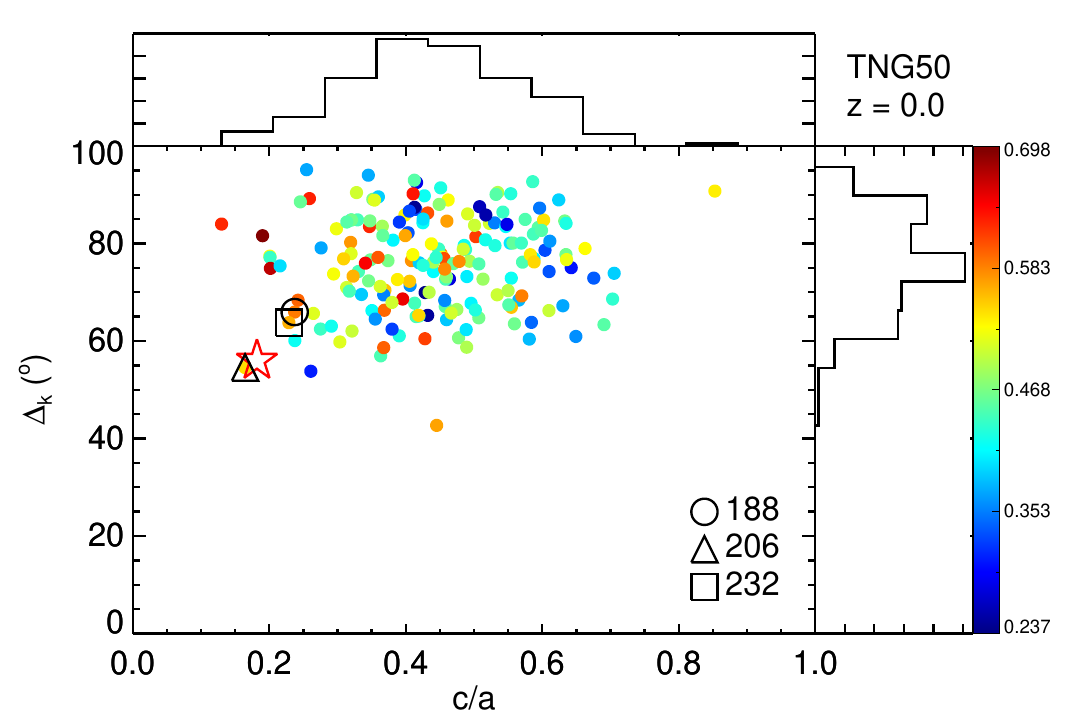}
\includegraphics[width=0.45\textwidth]{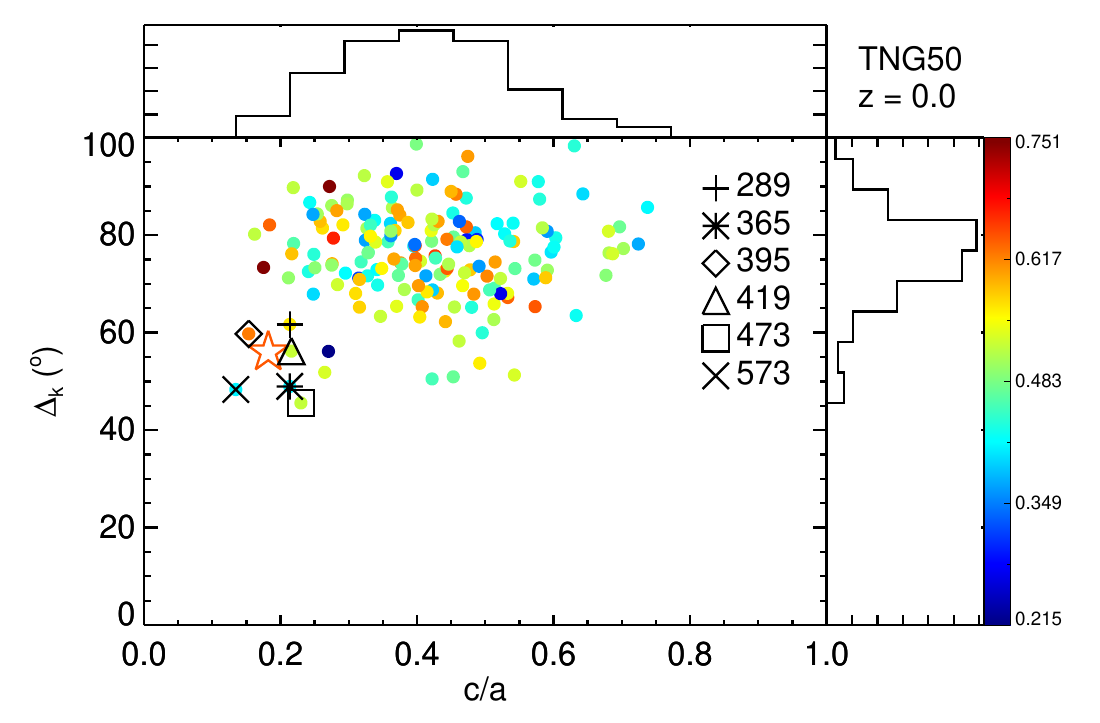}

\includegraphics[width=0.45\textwidth]{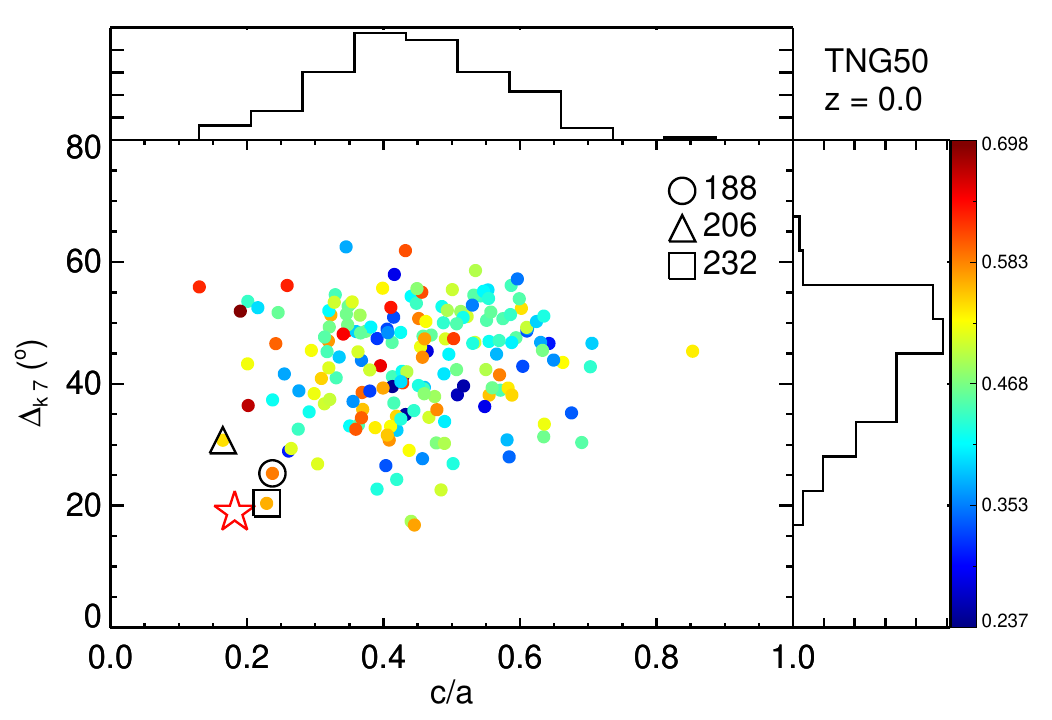}
\includegraphics[width=0.45\textwidth]{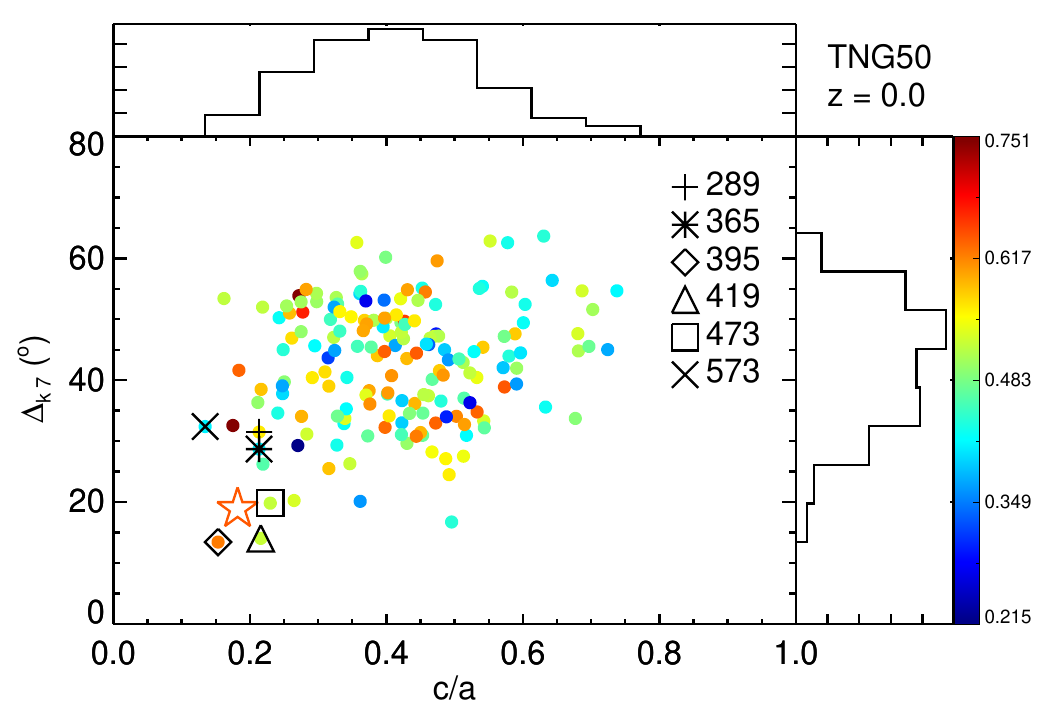}

\includegraphics[width=0.45\textwidth]{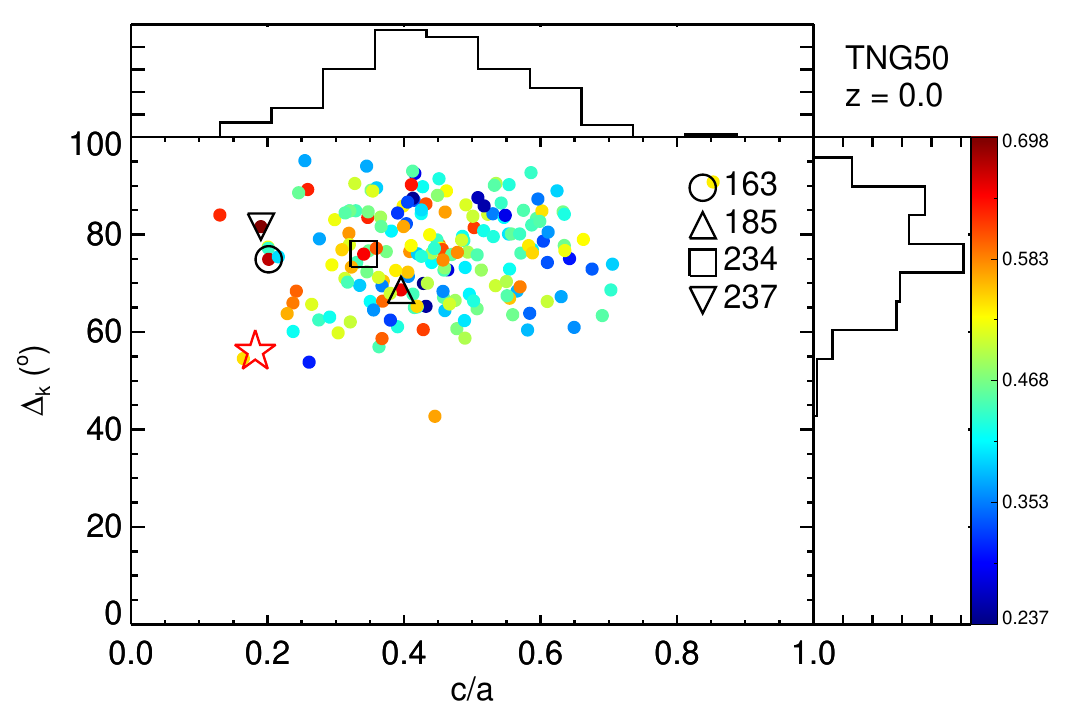}
\includegraphics[width=0.45\textwidth]{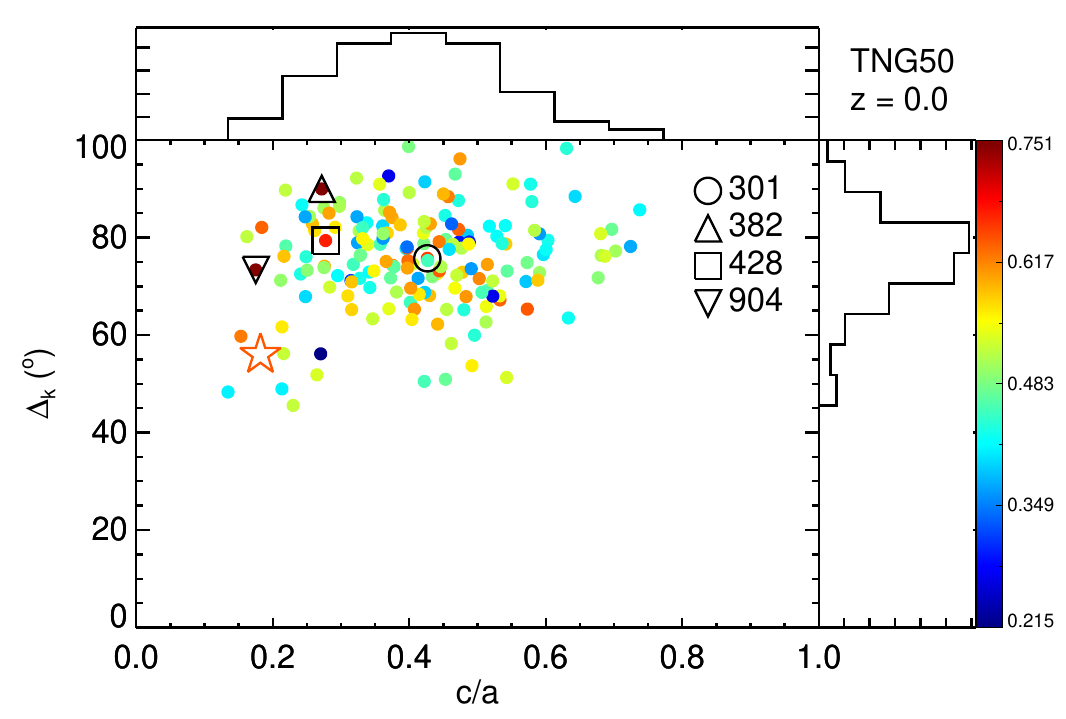}

\includegraphics[width=0.45\textwidth]{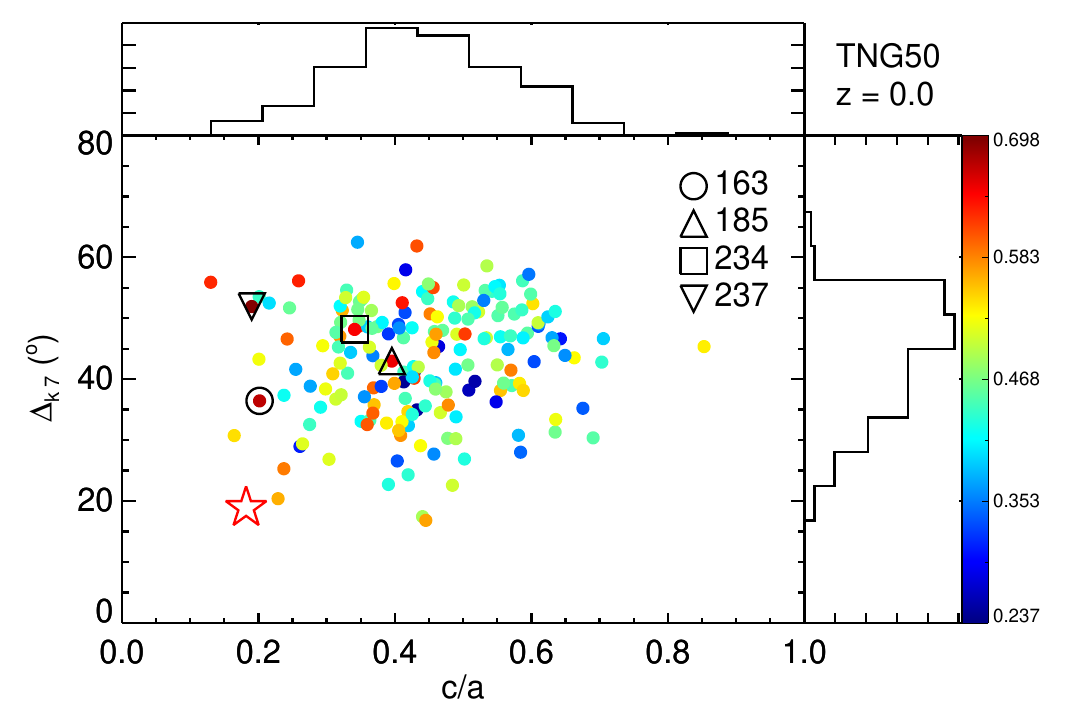} 
\includegraphics[width=0.45\textwidth]{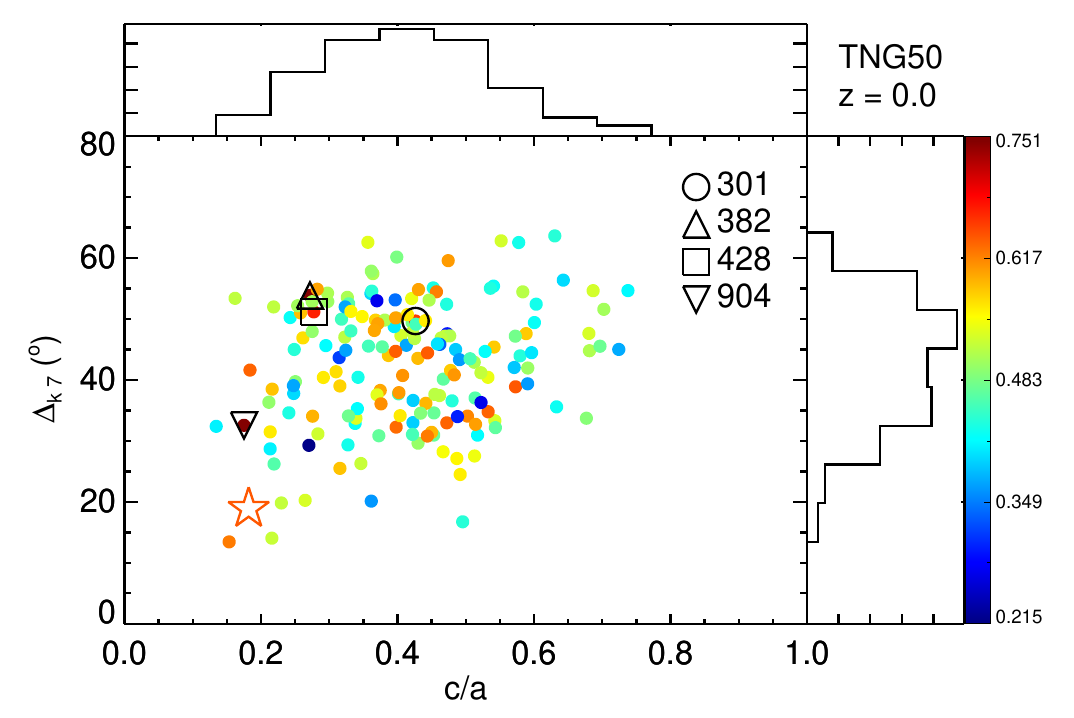} 

\caption{The two orbital pole dispersions ($\Delta_k$ and $\Delta_{k7}$) as a function of the spatial distribution aspect ratio ($c/a$) for the MW-like satellite systems at redshift $z=0$. {The left and right columns correspond to high and low mass ranges, respectively.} The color of each symbol indicates the Gini coefficient of the PoS system. The 11 most luminous satellites are shown in the top panels; a subsample of the 7 most aligned satellites is shown in the second row. The MW is represented by the star symbol, color coded to indicate its high Gini coefficient. Satellite systems with a MW-like PoS are labeled by group number in the TNG50 simulations. The two bottom rows are the same as the top two, except the group numbers identify systems with Gini coefficients higher than that of the MW.}
\label{fig_delta_k_gini}
\end{figure*}

Apart from the spatial distribution, it has been argued \citep{Metz:2007,Pawlowski:2020} that the possible alignment of the orbital poles of the satellite galaxies of the MW makes such a system extremely rare. In order to study this phenomenon and select candidate satellite systems similar to the MW for further study, we plot the pole dispersion defined in Equation (\ref{pole_disp}). Figure \ref{fig_delta_k_gini} shows the orbital pole dispersion $\Delta_k$ versus the spatial $c/a$ ratio for  the 11
most luminous satellites, and as in \cite{Sawala:2022a}, we also plot the 7 most aligned satellites among the 11 
in each MW-like system in both high and low mass ranges in the TNG50 simulations. Henceforth, we concentrate on the TNG50 results, as these best resolve the satellites of the PoS.

We compare the parameters of the simulated systems with those of the MW ($c/a = 0.182, \Delta_k = 56^o$, and $\Delta_{k7} = 18.9^o$). We define a satellite system to have a MW-like PoS by requiring its $c/a \le 0.25, \Delta_k \le 66^o$, and $\Delta_{k7} \le 35^o$. The reason that we did not use the most strict criteria possible ($c/a \le 0.182, \Delta_k \le 56^o$, and $\Delta_{k7} \le 18.9^o$) is that we want to find a reasonable sized sample of galaxies with satellite systems that closely resemble the MW PoS. This enables a study of the physical processes behind the formation of such satellite planes. 

{ We caution that  finding a perfect match to the MW PoS with an increasing set of parameters would inevitably make such matches exceedingly rare.   This could restrict  artificially recovered sample to only the closest matches to the MW, possibly ignoring other systems that are meaningfully analogous.  Nevertheless, within the parameter space of aspect ratio and orbital pole dispersion we find interesting analogies to the MW that we wish to examine.}

The systems with a group number in Figure \ref{fig_delta_k_gini} identify the PoS systems most similar to that of the MW.   Note that systems with a small aspect ratio and orbital pole dispersion like that of the MW appear as a class of outliers separated from the main distribution.

If we use these criteria, the probability of finding such a system is 1.6 percent (3 out of {186}) in the high-mass range and {0.4 percent} (6 out of {1392}) in the low-mass range in the TNG50 simulations. This is in general agreement with other recent studies such as \cite{Samuel:2021} and \cite{Sawala:2022a}. A MW-like PoS is {therefore} a rare outlier, but a natural occurrance in $\Lambda$CDM cosmology even if we require a satellite system to satisfy both spatial and dynamical criteria. For an easy comparison to the MW, we list in Table~\ref{tab_pos_par} some characteristic parameters of the MW and MW-like PoS systems in the TNG50 simulation. 

{We note that the lowest-mass satellites in our sample are represented by relatively few star particles in the TNG50 simulation. However, the plane analysis presented here is based on the subhalo positions, which are defined by the position of all  particles (stars, gas, and DM) with respect to the minimum gravitational potential of the bound subhalo. Hence, all identified satellites contain a large number of particles ($\sim 100$ to 1000's).   

{  To allow readers to assess the numerical resolution of the analyzed satellites, Tables~\ref{tab_resolution_188}--\ref{tab_resolution_573} in the Appendix~\ref{sec_appendix} give the stellar mass, dark matter} mass, number of star particles and total number of bound particles for every satellite belonging to the nine MW-like systems listed in Table~\ref{tab_pos_par}.}

\begin{table}
    \centering
    \caption{Characteristic parameters of the MW plane of satellites (PoS) and the MW-like PoS systems in the TNG50 simulation.}
    \label{tab_pos_par}

    \begin{tabular}{lccccc}
        \hline
        Group ID &
        $c/a$ &
        $\Delta_k\ (^\circ)$ &
        $\Delta_{k7}\ (^\circ)$ &
        $G$ &
        $V_{\mathrm{PoS}}\ (\mathrm{km\,s^{-1}})$ \\
        \hline
        MW   & 0.182 & 56.0 & 18.9 & 0.645 & 668 \\
        188  & 0.237 & 65.9 & 25.3 & 0.589 & 1711 \\
        206  & 0.164 & 54.5 & 30.7 & 0.544 & 1782 \\
        232  & 0.229 & 63.8 & 20.4 & 0.566 & 972 \\
        289  & 0.214 & 61.7 & 31.5 & 0.566 & 630 \\
        365  & 0.213 & 48.9 & 28.7 & 0.412 & 1361 \\
        395  & 0.153 & 59.7 & 13.4 & 0.625 & 156 \\
        419  & 0.216 & 56.2 & 14.0 & 0.525 & 683 \\
        473  & 0.230 & 45.6 & 19.8 & 0.524 & 795 \\
        573  & 0.134 & 48.3 & 32.4 & 0.411 & 959 \\
        \hline
    \end{tabular}
\end{table}

\subsection{Gini Coefficient, Radial Distribution, and Bulk Velocity}
\label{sec_other_properties}
\subsubsection{Gini Coefficient}
One aspect that may contribute to the rather rare MW-like PoS is that 9 of the 11 MW classical satellites are relatively close to the MW central galaxy compared to the other two more distant satellites. It has been  claimed \citep{Sawala:2022a} that such a non-uniform radial distribution (a Gini coefficient $G = 0.645$ as defined in Equation (\ref{gini_coef})) may be extremely rare for the satellite systems formed in the simulations. 

In order to further study this claim, we color-code the satellite symbols with their Gini coefficients in Figure \ref{fig_delta_k_gini}.  This figure shows that there is no direct correlation between the Gini coefficient and the ($c/a$) ratio or the pole dispersion. 
Also, \cite{Sawala:2022a} found no satellite system with a Gini coefficient as high as that of the MW using the SIBELIUS constrained simulations. However, we find that there are 4 satellite systems in both the high-mass and low-mass range (8 in total) that have Gini coefficients higher than $G = 0.645$ of the MW PoS.  Also, 3 of them have a ($c/a$) ratio close to that of the MW. However, their pole dispersions are not comparable to that of the MW as shown in the bottom two rows of Figure \ref{fig_delta_k_gini}. Also, the candidate MW-like PoS systems selected in the simulations do not necessarily have a high Gini coefficient. So, the Gini coefficient is not a crucial indicator of a MW-like PoS. Nevertheless, satellite systems with a very high Gini coefficient tend to have a lower ($c/a$) ratio as is evident in Figure \ref{fig_delta_k_gini}.  This is simply because a few distant satellites in a high Gini coefficient system tends to lead to a small $c/a$ ratio.

\subsubsection{Radial Distribution}

In addition to uniformity, it is possible that other aspects of the radial distribution of satellites may  impact their spatial distribution. \cite{Carlsten:2020} find that satellites around MW like galaxies in the Local Volume are significantly more centrally concentrated than the similar simulated systems. They suggested that this can be partially due to artificial disruption in the simulations. The updated analysis in \cite{Carlsten:2022} using the full ELVES sample shows that it could be due to the incomplete sample used in the first study. Using the SAGA Survey DR3 data, \cite{Mao:2024} found that the radial distribution of satellites around the MW is substantially more concentrated (top 25\%) than that of most of the MW-like hosts in the SAGA Survey. In fact, the MW is one of the most radially concentrated satellite systems in observed MW analogs analyzed by \cite{Patel:2024} for the ages between now and 2 Gigayears ago.

There are also recent studies that compare the radial distribution of the satellites of the MW to that of MW-like galaxies in simulations. \cite{Samuel:2020} found that the radial distribution of satellites of 12 MW like hosts in the FIRE-2 simulations generally agreed with that of Local-Group galaxies. However, again the MW has a more concentrated satellite distribution than the simulated systems. More massive host galaxies in the simulations have fewer satellites at distances less than 100 kpc possibly due to tidal disruption by the host. \cite{Hu:2025} found that the radial density profile of the MW satellites  matches that of MW analogs in the TNG50 simulation. However,  their satellite samples may have included non-cosmological objects (i.e. clusters formed within the host galaxy during its evolution).

\begin{figure*}
\centering
\includegraphics[width=0.48\textwidth]{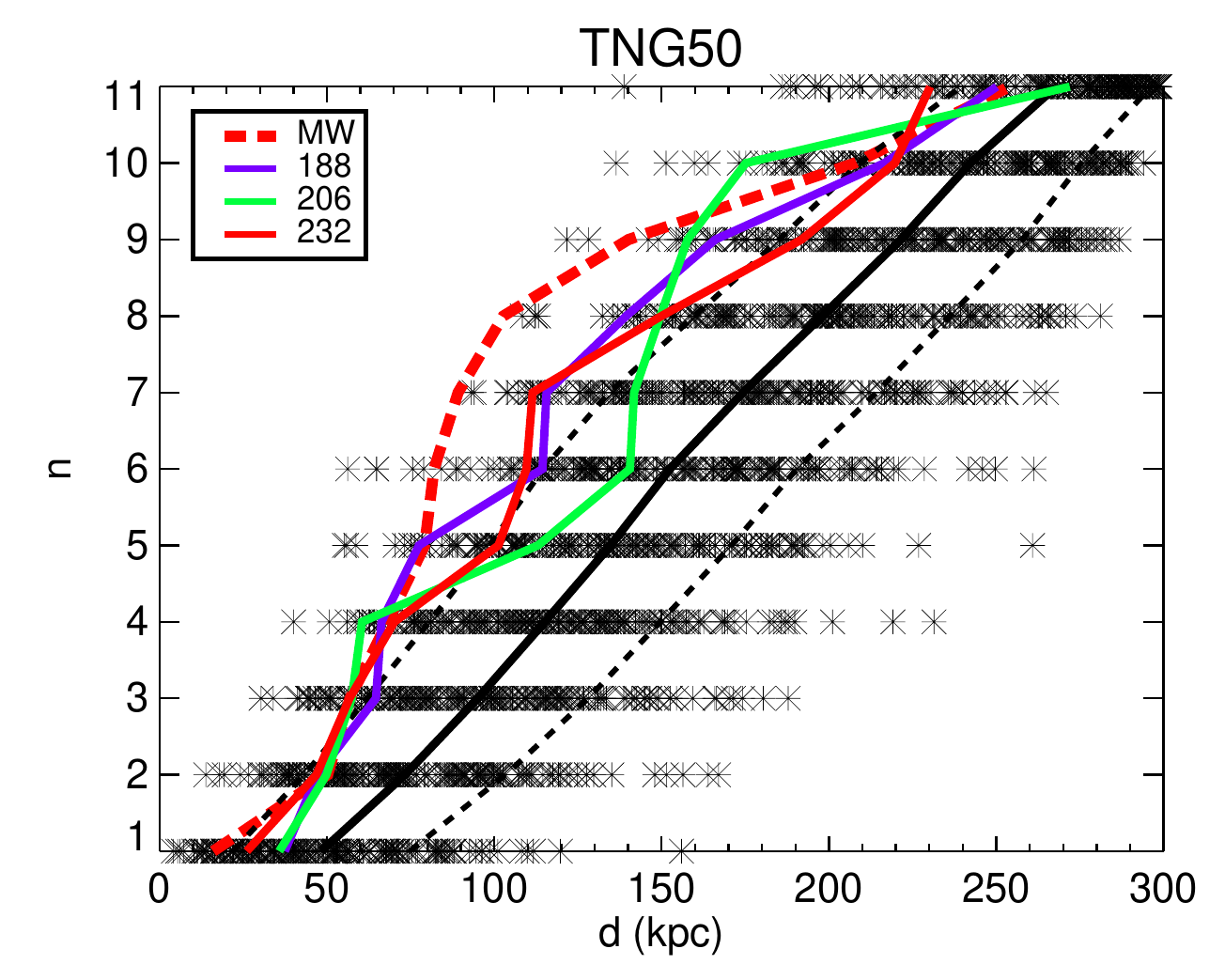}
\includegraphics[width=0.48\textwidth]{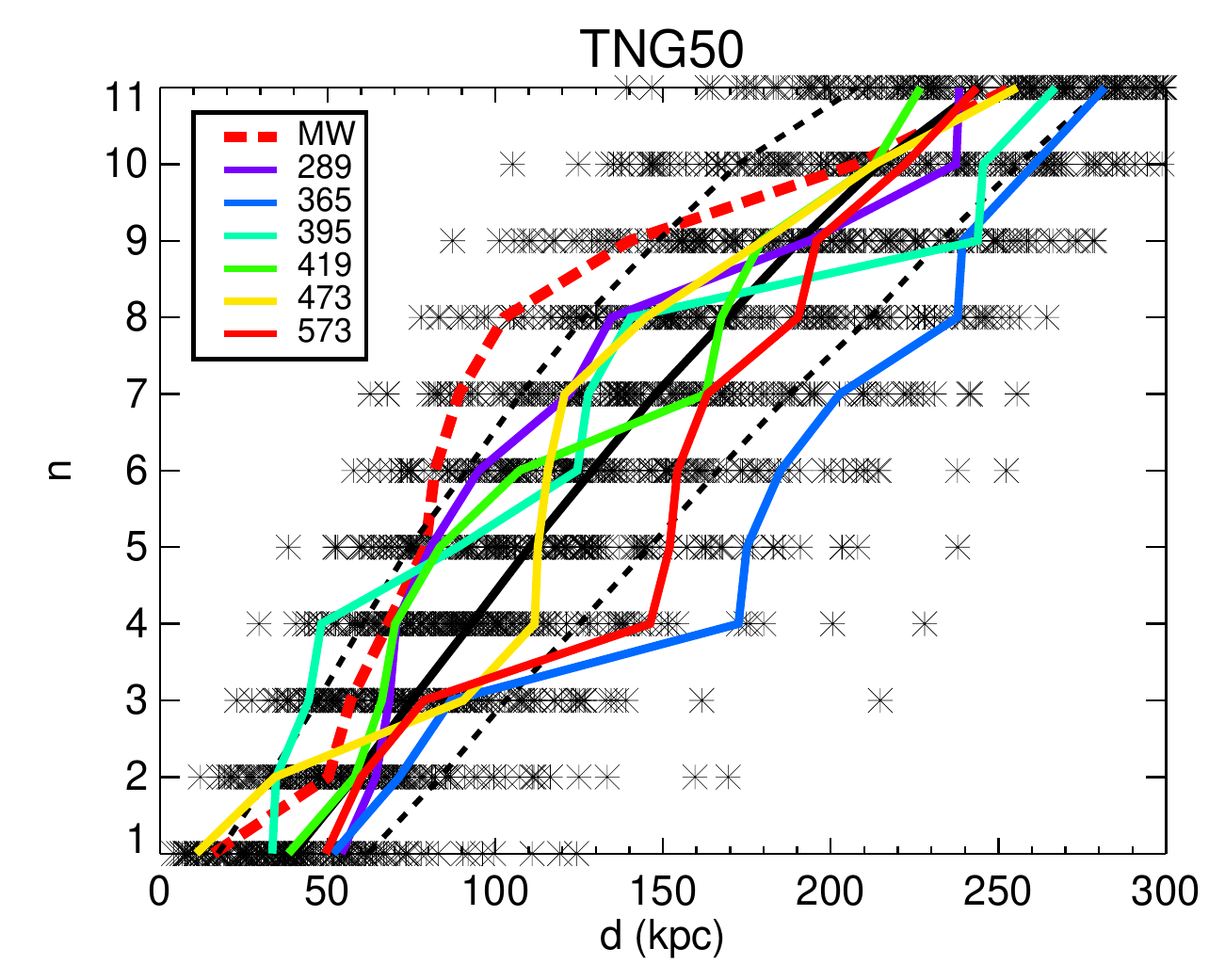}

\includegraphics[width=0.48\textwidth]{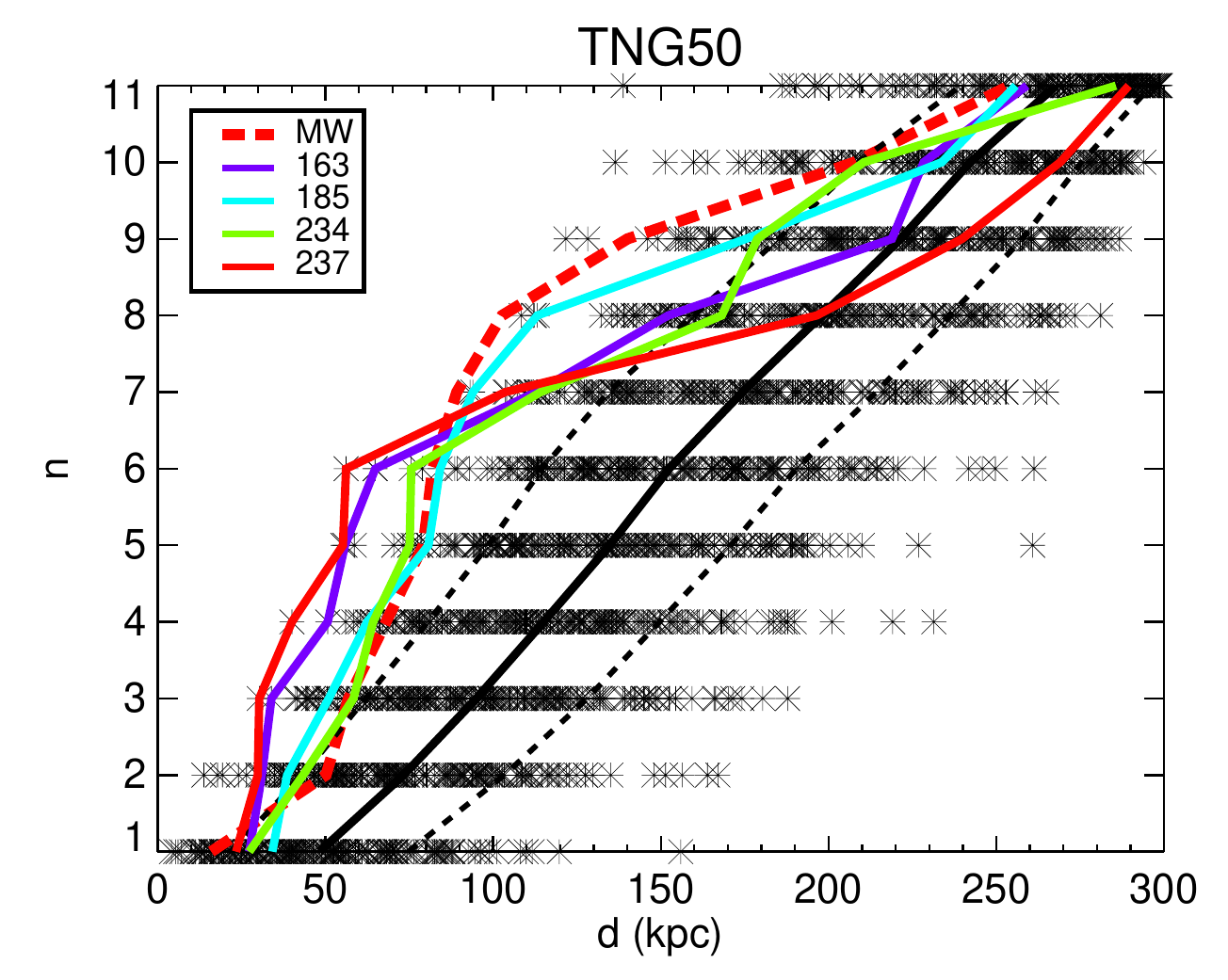}
\includegraphics[width=0.48\textwidth]{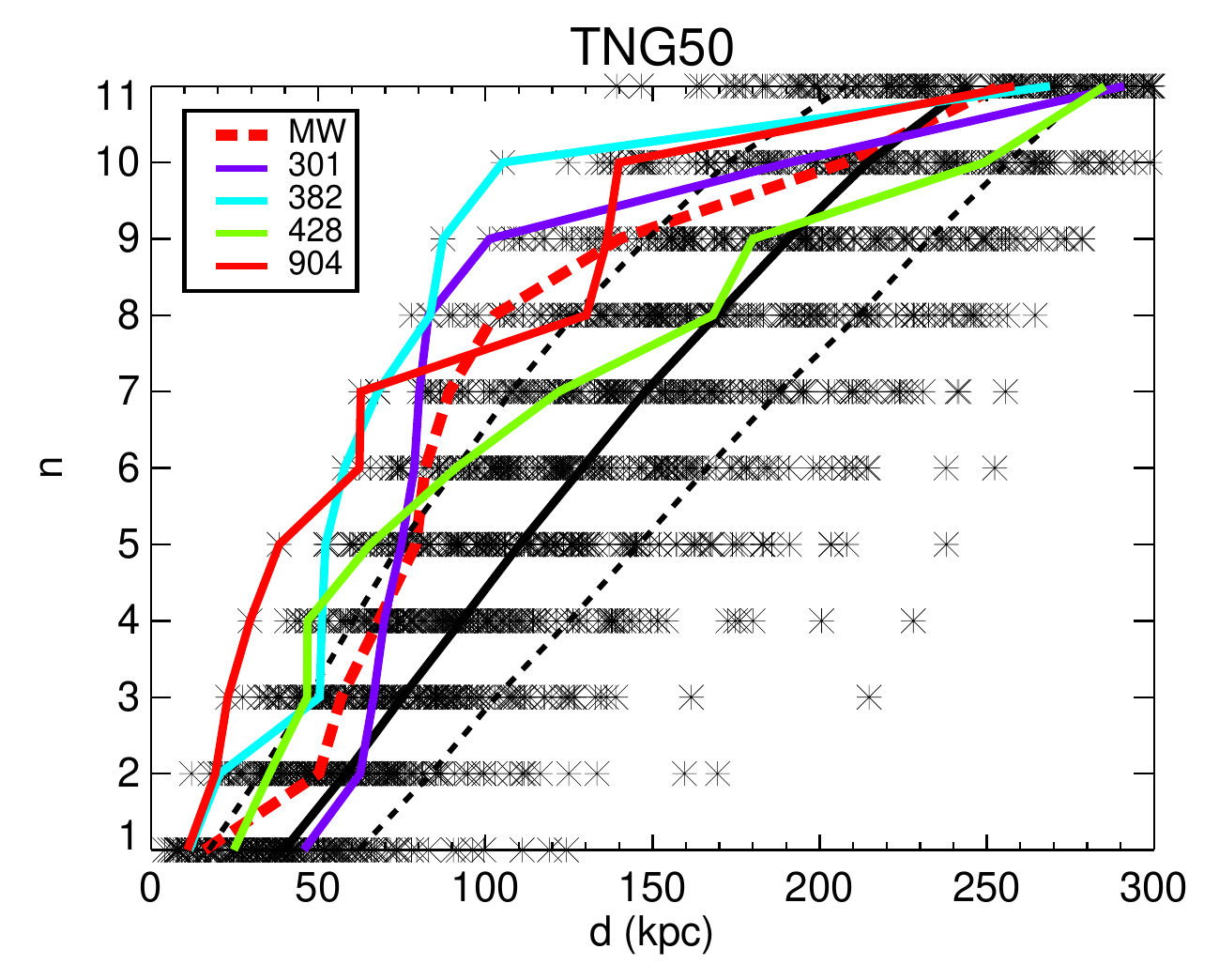}

\caption{Distance of the 11 most luminous satellites to the host galaxy center in MW-like systems. Results are shown for the high-mass (left column) and low-mass (right column) ranges, ordered from the closest ($n=1$) to the most distant ($n=11$) satellite. Black asterisks show the satellite distance distribution for all MW-like systems. The red dashed line marks the 11 luminous satellites of the Milky Way PoS. Colored solid lines identify satellite systems with a MW-like PoS (top row) or with a Gini coefficient higher than that of the MW (bottom row). The black solid line is the mean distance of the $n$th satellite, and the black dashed lines indicate the standard deviation.}
\label{fig_radial_dist}
\end{figure*}

In Figure \ref{fig_radial_dist}, we plot the distance of the 11 most luminous satellites from the center of the MW and from the MW-like host in the TNG50 simulations. This provides a {well-posed} study of the radial distribution of satellites around MW-like galaxies and its connection to the formation of a MW-like plane of satellites in the simulations.  We chose the 11 most luminous satellites in each case, rather than the whole satellite population, to make a clean comparison with the Milky Way.

From Figure \ref{fig_radial_dist}, it is very clear that among all the systems plotted, the MW satellite system is the most radially concentrated system with 8 satellites at a distance of 100 kpc or less from the center of the MW. This is consistent with the finding in \cite{Mao:2024} based upon the SAGA DR3 data. This concentration may be due to the fact that the LMC is only about 50 kpc from the MW center (\cite{Nadler:2020}). We further discuss the impacts of LMC- and SMC-like satellites in Section \ref{sec_lmc_smc}. 

As we have discussed, the Gini coefficient for the MW is higher than that of most galaxies.  From Table \ref{tab_pos_par}, the Gini coefficients of 8 MW-like PoS systems in the simulation are all lower than that of the MW. However,  the bottom panels of Figure \ref{fig_radial_dist}  show the radial distribution for 4 galaxies in each mass range with a high Gini coefficient. Their satellites are concentrated similarly to those of the MW. However, they do not have a MW-like $c/a$ and $\Delta_k$ as seen in Figure~\ref{fig_delta_k_gini}. Hence, a high Gini coefficient and/or a high concentration is not a sufficient condition to identify a MW-like PoS.


It is very clear in the upper left panel of Figure \ref{fig_radial_dist} that the MW and MW-like PoS systems as a whole are more radially compact than most of the galaxies identified in the high-mass range ($0.8 - 3.0 \times 10^{12} \Msun$). The thick black diagonal solid lines on Figure \ref{fig_radial_dist} show the mean distance of each satellite from the center of its host for all MW-like galaxies.  The black diagonal dashed lines show the $\pm 1 ~\sigma$ deviation. For the MW-like PoS systems, their radial distances are all roughly about 1 $\sigma$ lower than the mean of the distribution. Hence, the radial compactness may play an important role in the formation of a MW-like PoS, at least for the high-mass range. How these galaxies ended up in a more radially concentrated configuration may depend on their environment and evolution history. Whether this is because artificial tidal disruption in the simulations makes it difficult for satellites to form close to the stellar disk of the main galaxy or this is because baryonic processes realized in the simulation make more massive galaxies less compact needs further study. If the latter is the case, the Milky Way is indeed an outlier on the radial distribution of satellites in the high-mass range.

For the low-mass range ($0.1 - 0.8 \times 10^{12} \Msun$) shown in the upper right panel of Figure \ref{fig_radial_dist}, the MW and MW-like galaxies are not significant outliers in radial compactness compared to the rest of the galaxies. This is because lower-mass galaxies tend to have satellites closer to the center of their host. However, the satellite distances from the MW are still below the mean distance of {all simulated satellites}. Based upon this, and if the radial distribution of satellites of galaxies in this mass range is not significantly impacted by artificial tidal disruption, one may favor a MW in the low-mass range. However, future studies on this point are needed.

{ As a caveat, we note that the radial distribution of satellites shown in Figure \ref{fig_radial_dist} could be affected by artificial tidal disruption in the simulations. 
\cite{Bosch:2018} showed that the gravitational softening length and the number of particles in subhaloes are key parameters to resolve the tidal evolution of subhaloes.
To assess the extent of the impact of artificial tidal disruption on the radial distribution, we plot in Figure \ref{fig_radial_dist_100} the radial distribution of the MW-like PoS satellites in the TNG100 simulation  for which there are fewer particles. 

As can be seen in Figure \ref{fig_radial_dist_100}, the mean distance of the 11 selected satellites in the TNG100 simulation is slightly further away from the galactic center than those of the TNG50. So, with higher resolution,  satellites closer to the galactic disk are more easily identified. Nevertheless, the overall shape of the radial distribution is similar, and the MW remains as  a significant outlier in both cases. Future high resolution simulations, in which lower mass subhalos are better resolved, will be needed to address this issue more thoroughly.  Nevertheless, our conclusions regarding the radial distribution remains robust and are unlikely to be affected by artificial tidal disruption.

\begin{figure}
\centering
\includegraphics[width=0.48\textwidth]{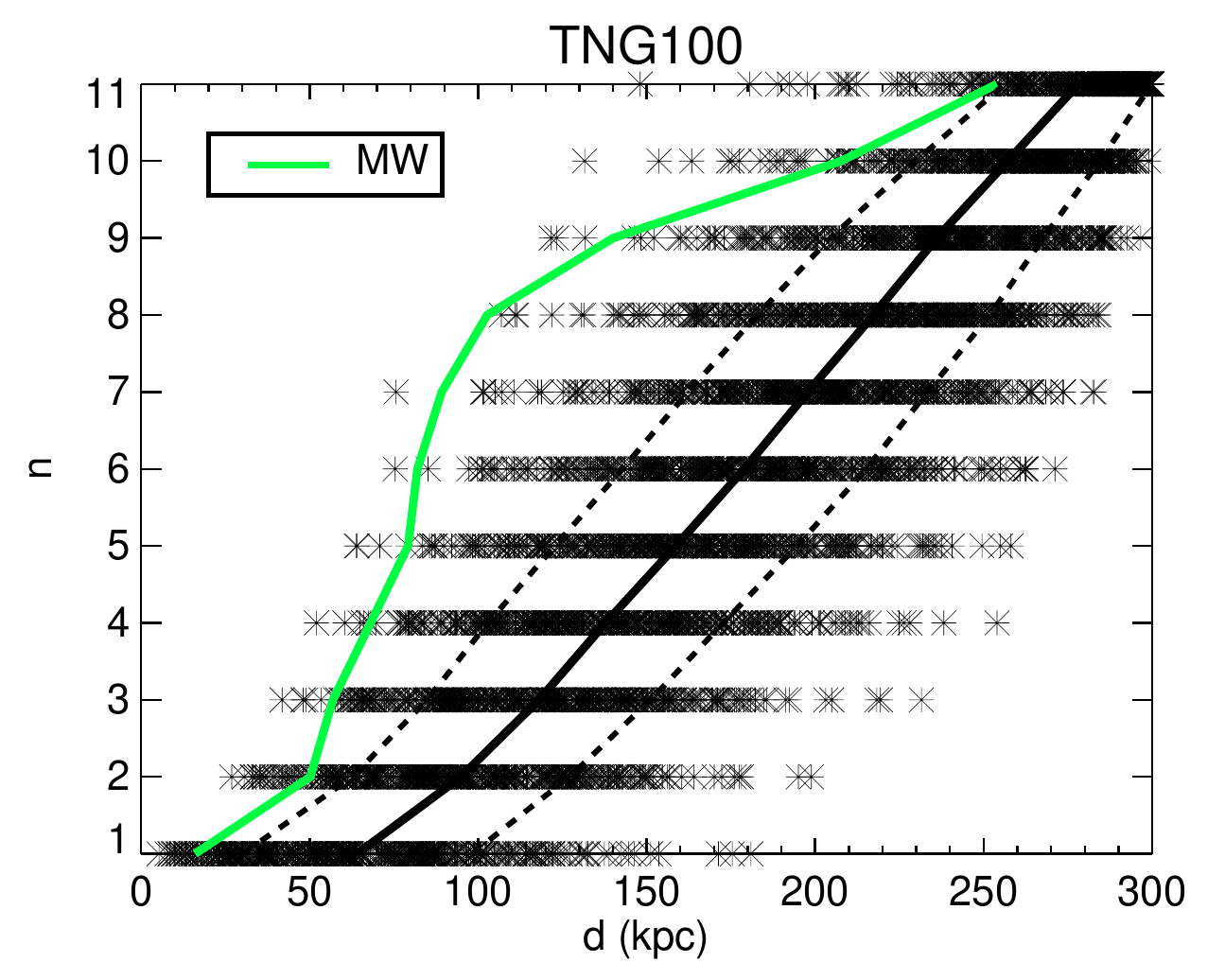}
\caption{{Distance of the 11 most luminous satellites to the host galaxy center in MW-like systems for the high-mass range in TNG100 simulation ordered from the closest ($n=1$) to the most distant ($n=11$) satellite. Black asterisks show the satellite distance distribution for all MW-mass systems. The green line marks the 11 luminous satellites of the Milky Way PoS. The black solid line is the mean distance of the $n$th satellite, and the black dashed lines indicate the standard deviation.}}
\label{fig_radial_dist_100}
\end{figure}

In addition to artificial disruption, it has been shown \citep{Moreno:2025} that different halo-finding algorithms can have significant differences when identifying subhalos very close to the main galactic disk in simulations. This could  affect the simulated  satellite radial distribution around the main galaxy. In particular, SUBFIND can underestimate satellite abundances in the inner region of the system. However, this will mainly impact the region less than 10\% of $R_{200}$.  That is $\sim 30$ kpc from the center of the systems in Figure \ref{fig_radial_dist}.   However, the PoS satellites of the MW are about 50-100 kpc from the center of the galaxy. Also, the host galaxies considered in \cite{Moreno:2025}  were much more massive than the MW mass ranges considered  in the present work. We note that more studies on this issue are needed before the discrepancy between the Milky Way and simulated satellite  distributions can be attributed  to defects of the halo-finding algorithm.  However, we believe that the masses are low enough and the distances large enough that halo-finding algorithm employed here is adequate for the comparisons made in the present work.}

\subsubsection{PoS Bulk Velocity}
\begin{figure*}
\centering
\includegraphics[width=0.48\textwidth]{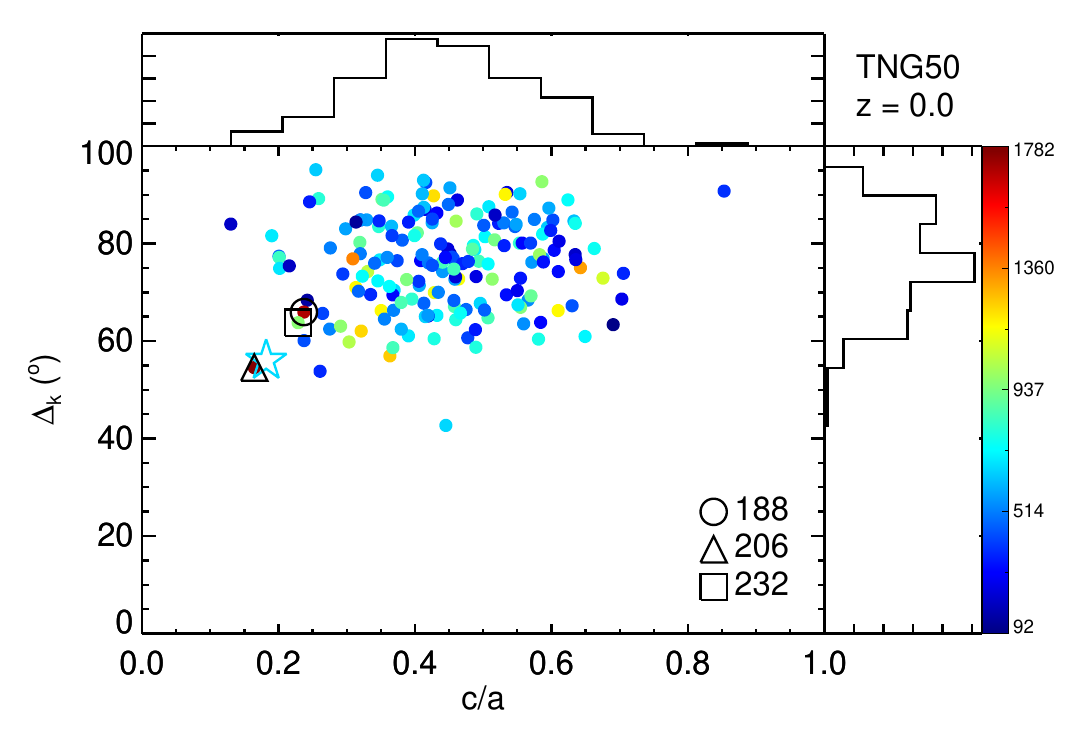}
\includegraphics[width=0.48\textwidth]{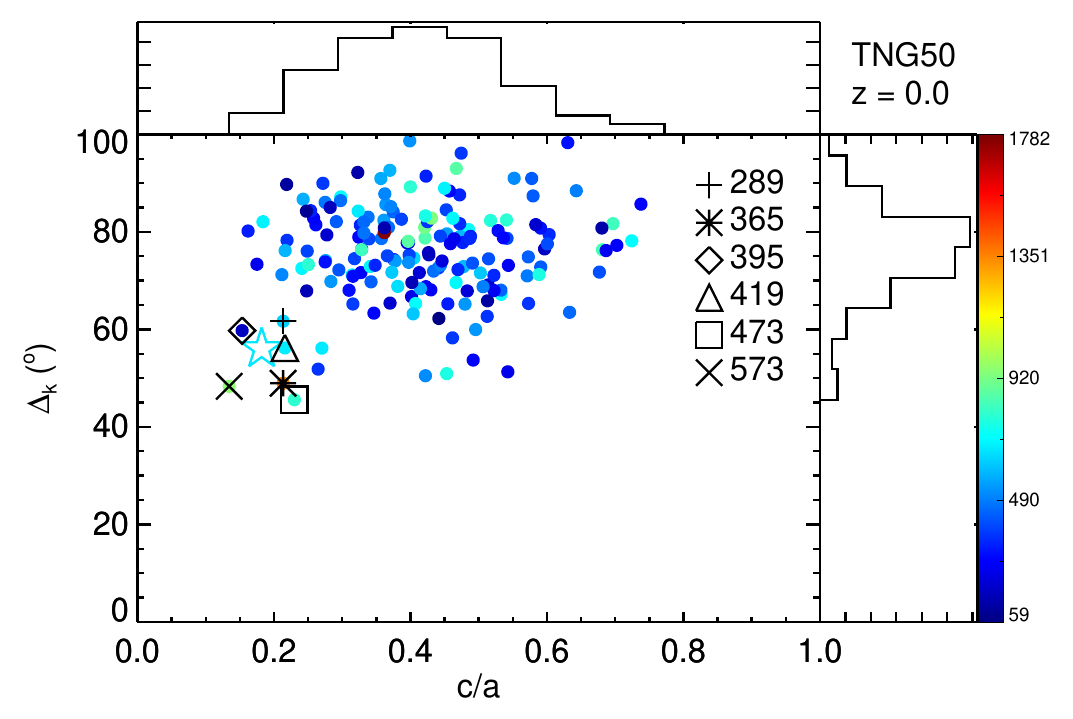} 

\includegraphics[width=0.48\textwidth]{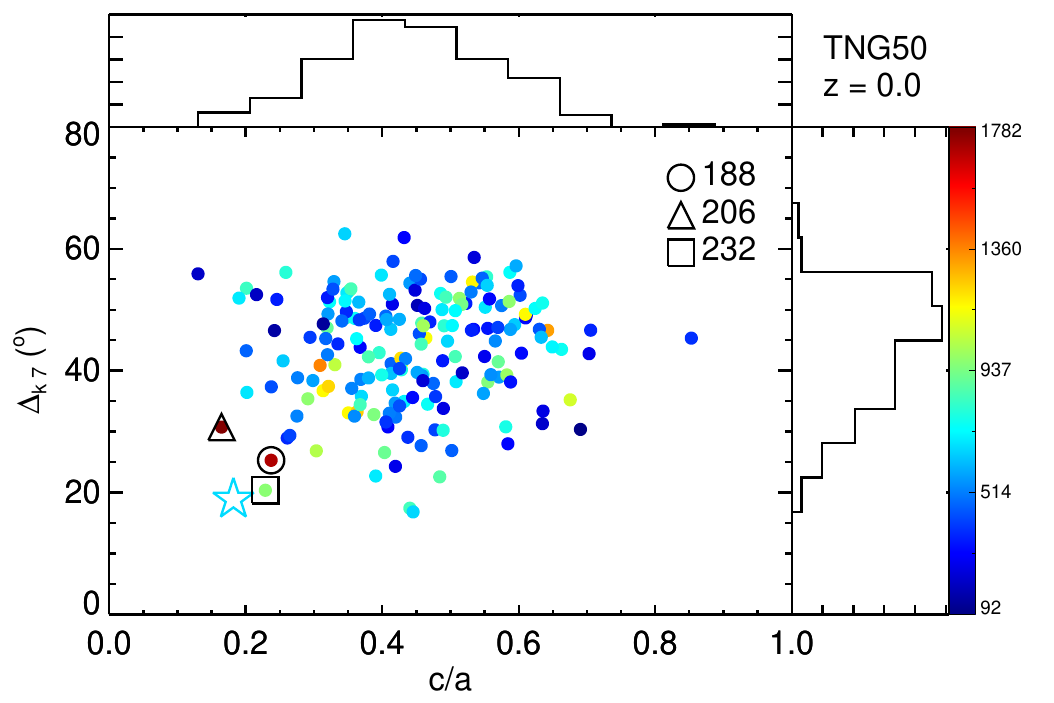}
\includegraphics[width=0.48\textwidth]{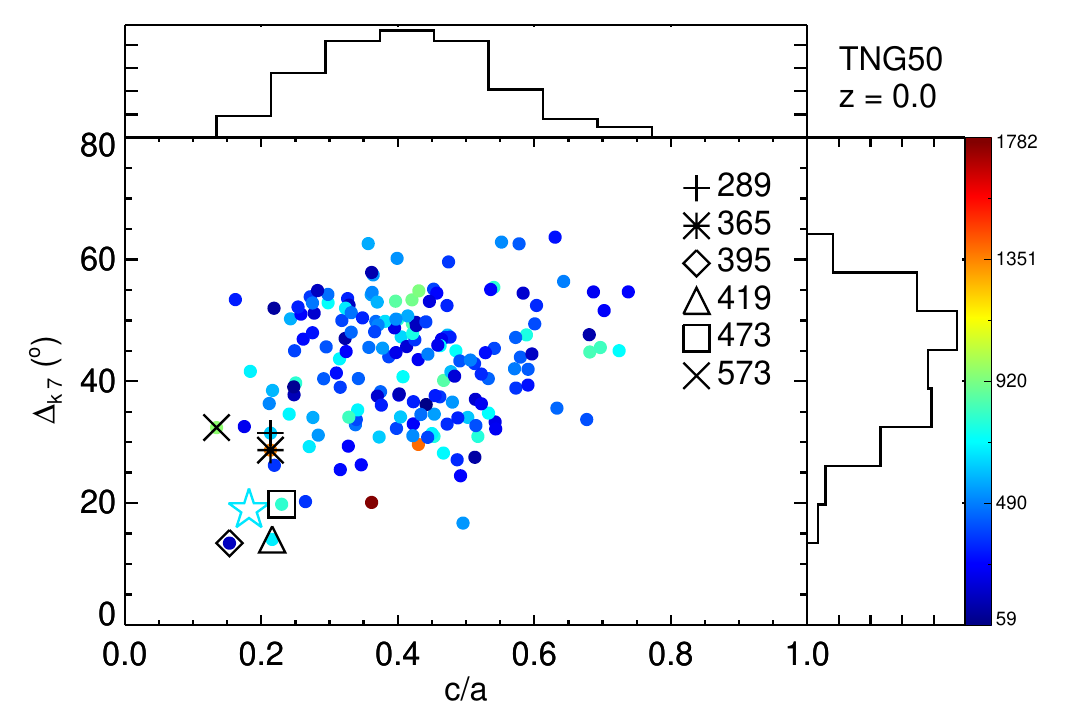} 

\caption{Orbital pole dispersion ($\Delta_k$ {and $\Delta_{k7}$}) versus spatial distribution aspect ratio ($c/a$) for MW-like satellite systems at $z=0$. Symbols are color-coded by the PoS velocity, $V_{\mathrm{PoS}}$, relative to the host galaxy (km s$^{-1}$). The MW is shown by a star symbol, color-coded to indicate a moderate $V_{\mathrm{PoS}}$. Other aspects match the first two rows of Figure~\ref{fig_delta_k_gini}.}
\label{fig_v_pos}
\end{figure*}

In order to study the motion of the satellites as a whole in addition to the orbital motion, we color-code the PoS velocity in Figure \ref{fig_v_pos} relative to the main galaxy, where $V_{PoS}$ is defined in Equation~(\ref{v_pos}). If a satellite system has a high $V_{PoS}$, it is likely that the system is going through a major merger or is near a massive galaxy cluster that causes a significant flow of the intergalactic medium.  On the other hand,  a low $V_{PoS}$ would mean that the system is more virialized and isolated. The MW has a moderate  $V_{PoS} = 668$ km s$^{-1}$ compared to galaxies in the simulations. 

Among the systems with a MW-like PoS in  Table \ref{tab_pos_par}, systems 188, 206, 232, 365, and 573 have a $V_{PoS}$ much higher than that of the MW, while system 395 has a very low $V_{PoS}$. Systems 289, 419, and 473 have $V_{PoS}$ very close to that of the MW. In the next section, we utilize $V_{PoS}$ as a characteristic parameter to analyze the environment of the MW-like systems formed in the simulation and to compare it with that of the MW.

\section{Galaxies with the MW-like Plane of Satellites and their interactions with the environment}
\label{sec_Analysis}
\begin{figure*}
\centering
\includegraphics[trim=0cm 0cm 0cm 0cm, clip, width=0.32\textwidth]{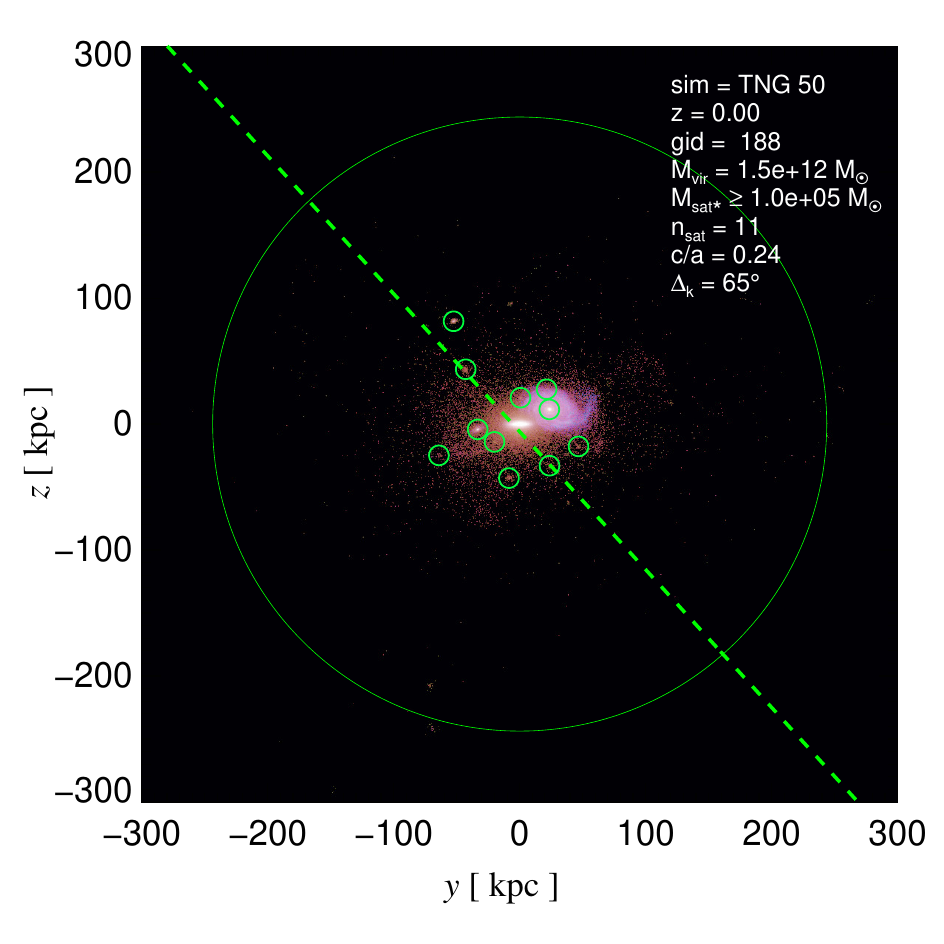} 
\includegraphics[trim=0cm 0cm 0cm 0cm, clip, width=0.32\textwidth]{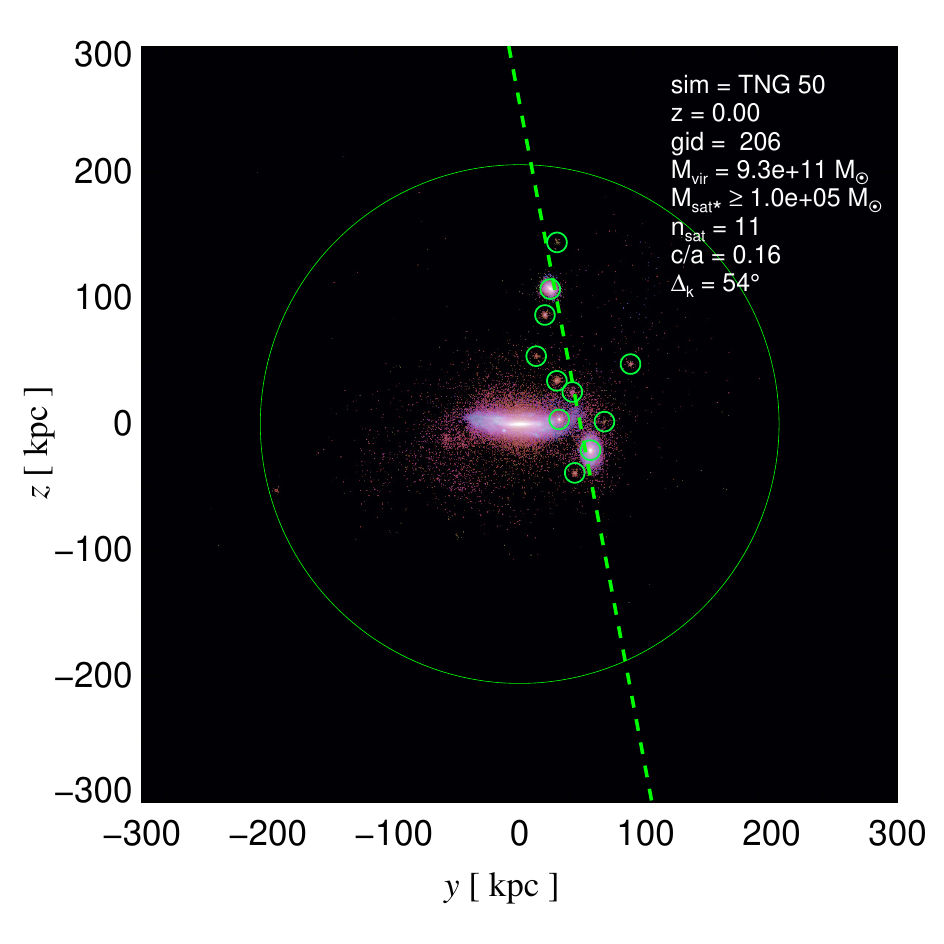} 
\includegraphics[trim=0cm 0cm 0cm 0cm, clip, width=0.32\textwidth]{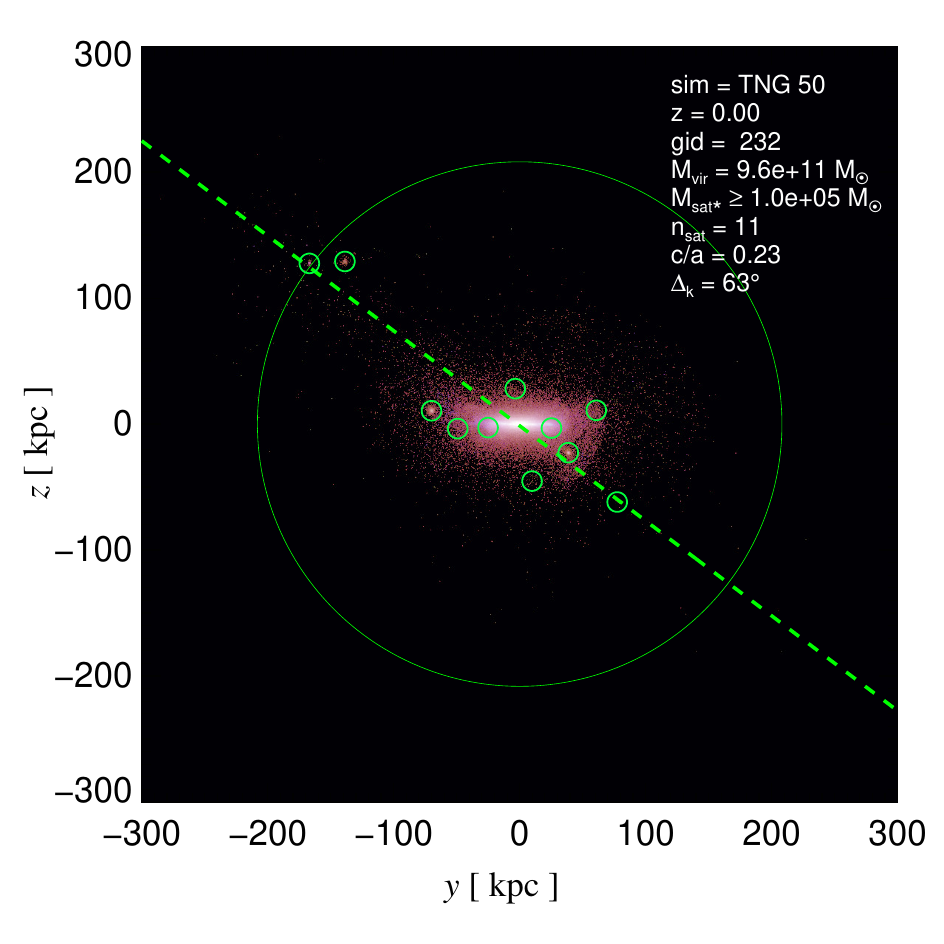} 

\includegraphics[trim=0cm 0cm 0cm 0cm, clip, width=0.32\textwidth]{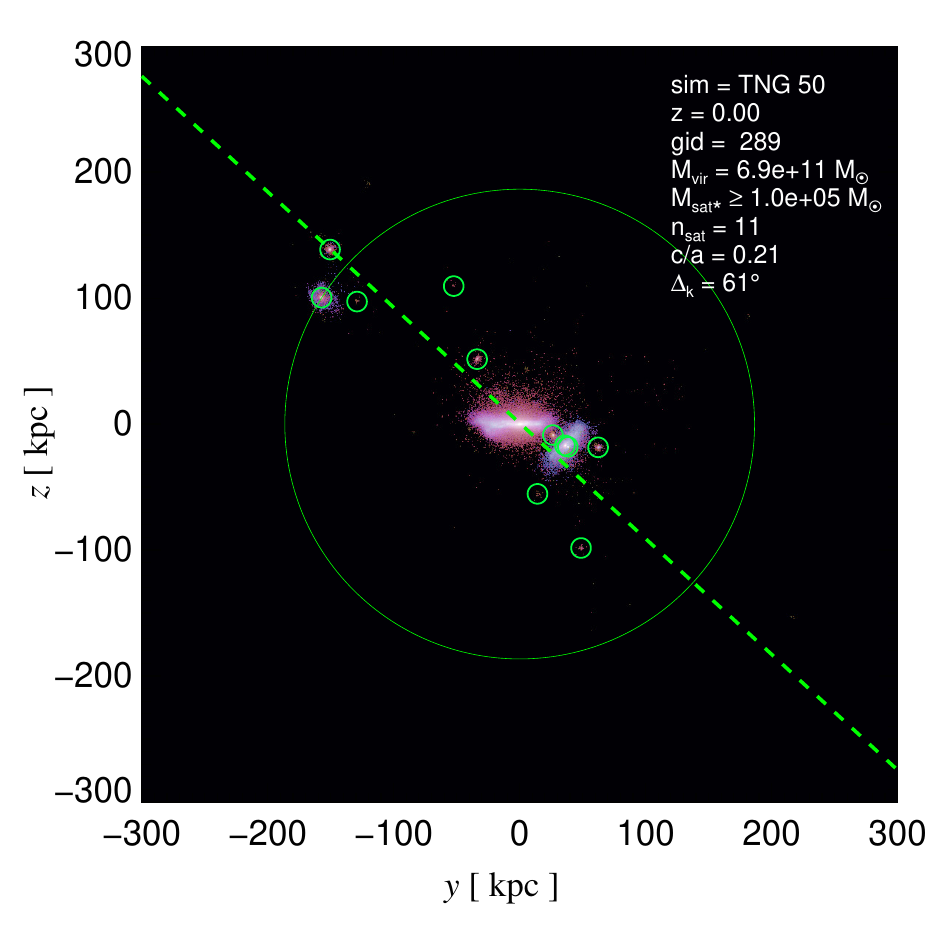} 
\includegraphics[trim=0cm 0cm 0cm 0cm, clip, width=0.32\textwidth]{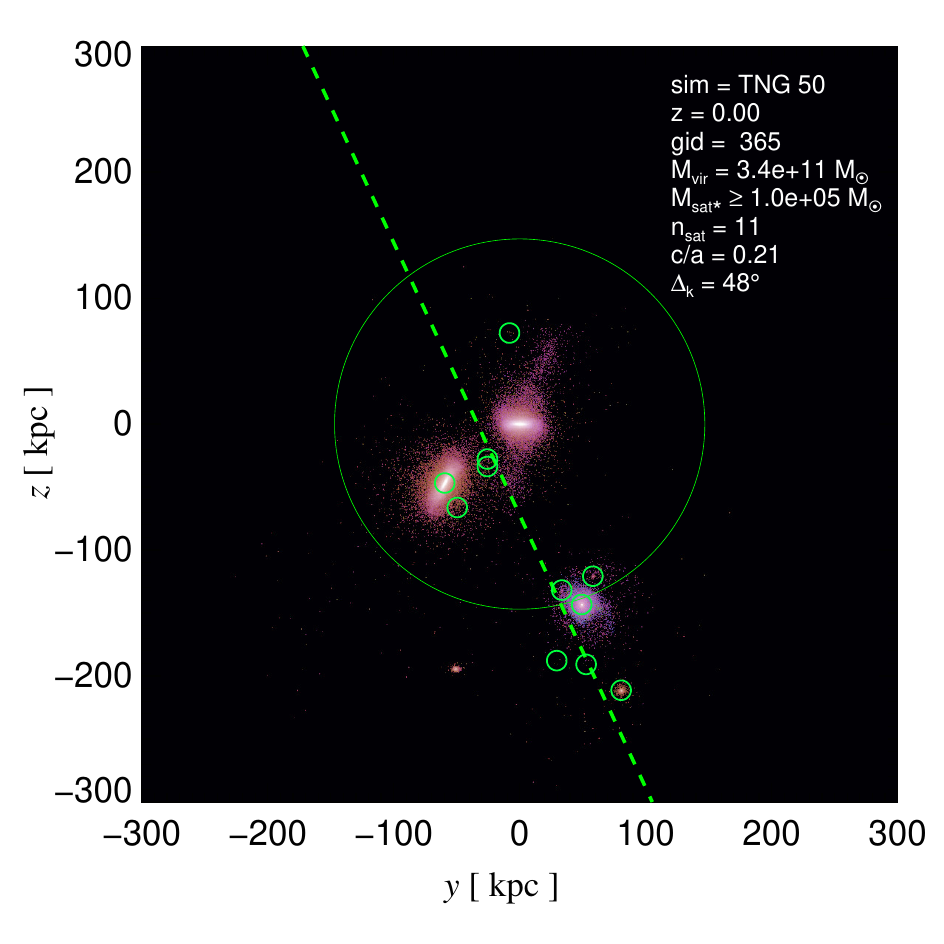} 
\includegraphics[trim=0cm 0cm 0cm 0cm, clip, width=0.32\textwidth]{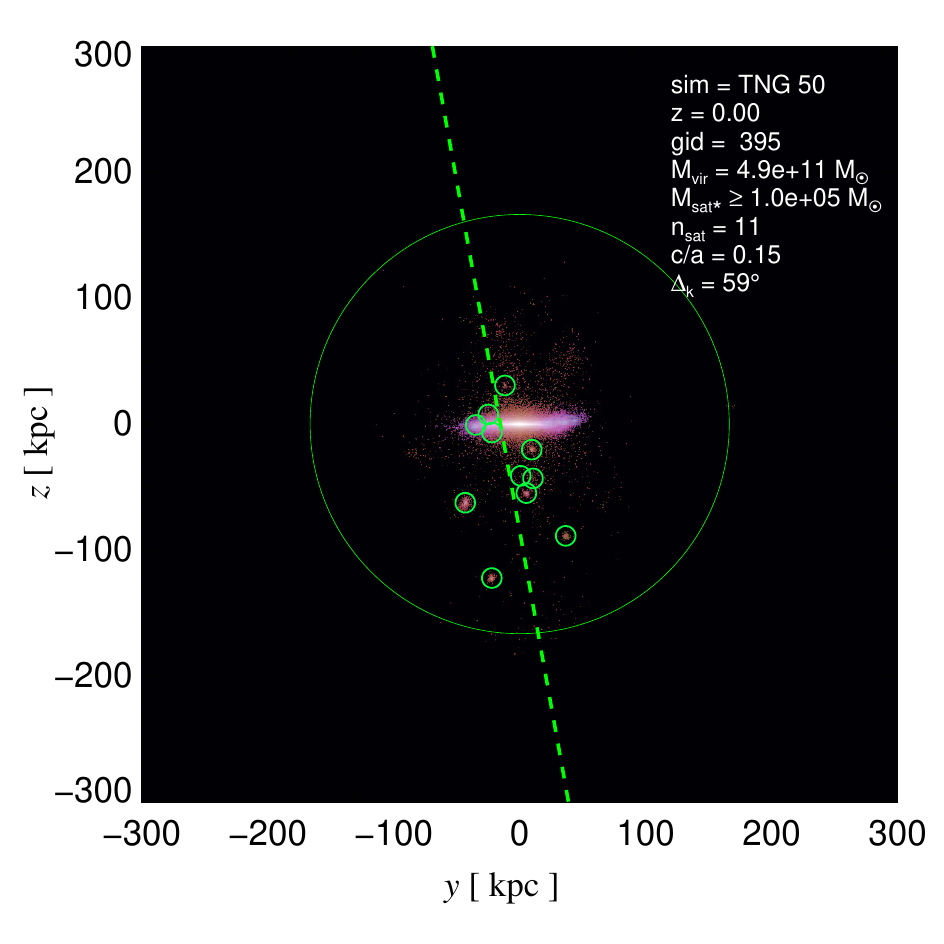} 

\includegraphics[trim=0cm 0cm 0cm 0cm, clip, width=0.32\textwidth]{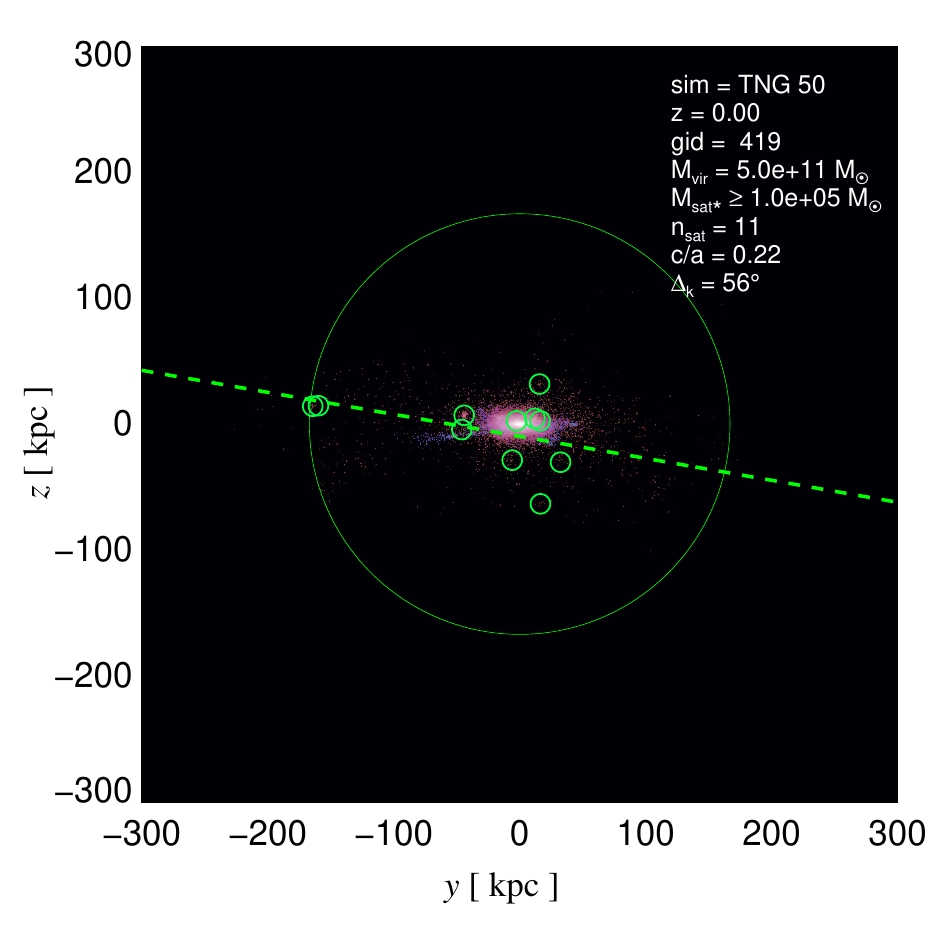} 
\includegraphics[trim=0cm 0cm 0cm 0cm, clip, width=0.32\textwidth]{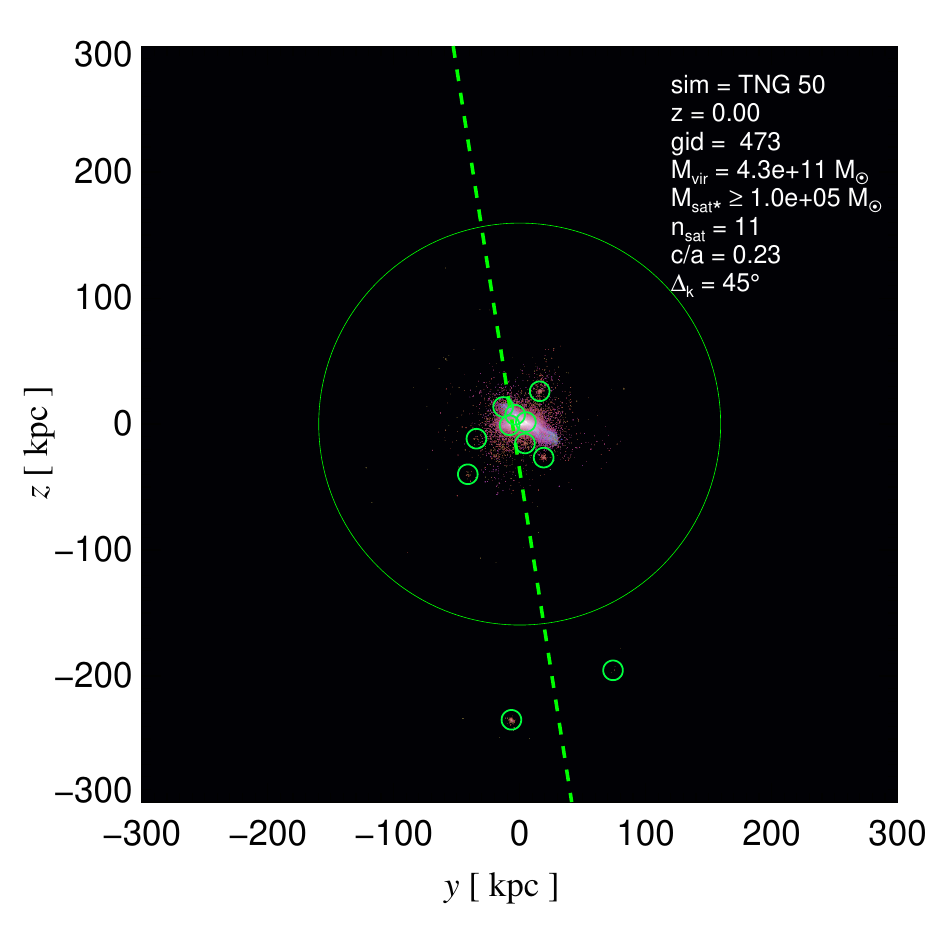} 
\includegraphics[trim=0cm 0cm 0cm 0cm, clip, width=0.32\textwidth]{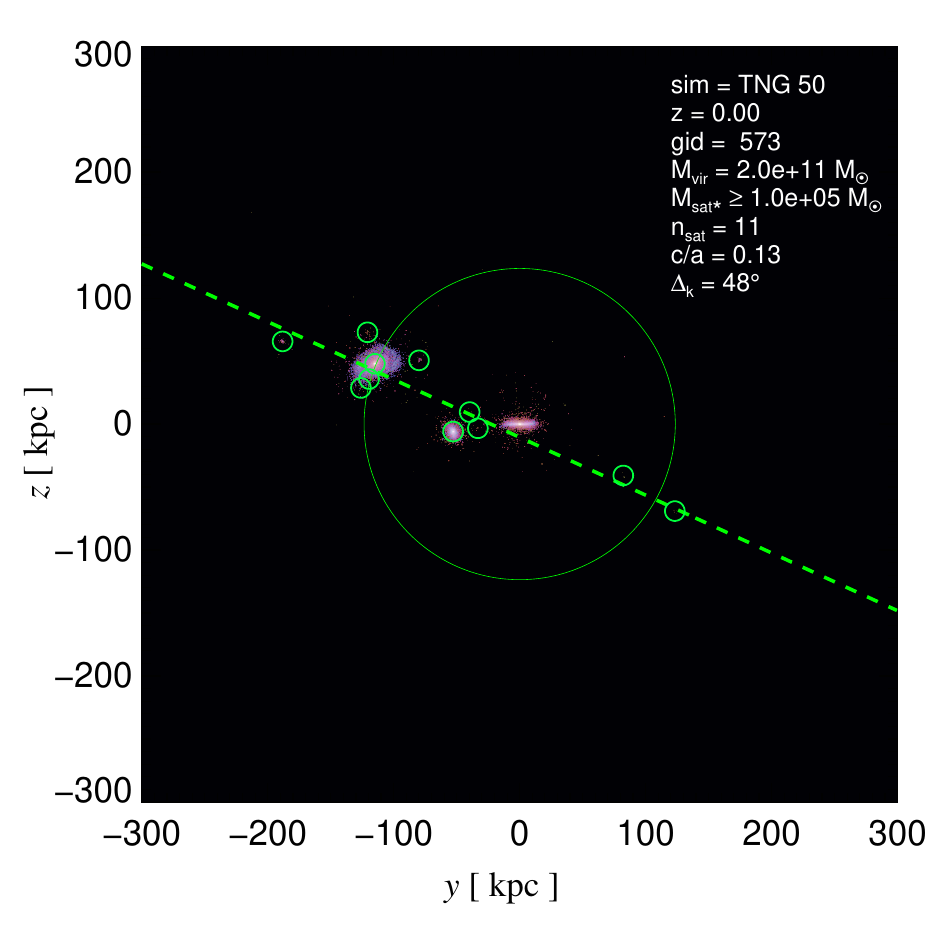} 

\caption{Edge-on views of the plane of the nine MW-like PoS systems in the TNG50 simulation on a 300 kpc scale. The stellar component is shown as 2D column density; color represents stellar age (blue to red/yellow for younger to older). The larger circle indicates the system’s virial radius. Small circles mark each of the 11 most luminous satellites. Each system is oriented so the central galaxy is at the origin with its stellar disk in the $xy$ plane; the fitted PoS is edge-on and marked by a dashed line.  Note that some satellites nearly overlap in the edge-on view as in satellites 1 and 3 of group 289.}
\label{fig_pos_star_300}
\end{figure*}

Figure \ref{fig_pos_star_300} illustrates the stellar component of the main and satellite galaxies and the orientation of the PoS at $z=0$ for the systems listed in Table \ref{tab_pos_par}. The projection is within a cubic space of 300 kiloparsecs in each dimension in the TNG50 simulation.

Table \ref{sat-tab1} summarizes the satellite stellar masses and distances from the host galaxy for the three systems that have a PoS most like that of the Milky Way for the high-MW mass range. These systems are labeled by the TNG50 halo finder ID as 188, 206, and 232.

\begin{table*}
    \centering
    \caption{Satellite stellar mass $M_*$ and distance from the host center $r$ for a high Milky Way mass.}
    \label{sat-tab1}
    \begin{tabular}{ccccccc}
        \hline
        Sat. Num. &
        \multicolumn{2}{c}{\textbf{Group 188}} &
        \multicolumn{2}{c}{\textbf{Group 206}} &
        \multicolumn{2}{c}{\textbf{Group 232}} \\
        &
        $M_\star\ (M_\odot)$ & $r\ (\mathrm{kpc})$ &
        $M_\star\ (M_\odot)$ & $r\ (\mathrm{kpc})$ &
        $M_\star\ (M_\odot)$ & $r\ (\mathrm{kpc})$ \\
        \hline
        Host & $4.6 \times 10^{10}$ & 0.0   & $2.9 \times 10^{10}$ & 0.0   & $4.9 \times 10^{10}$ & 0.0 \\
        1    & $1.1 \times 10^{10}$ & 77.5  & $2.3 \times 10^9$    & 272.0 & $7.5 \times 10^7$    & 111.5 \\
        2    & $2.5 \times 10^8$    & 37.4  & $1.6 \times 10^9$    & 149.6 & $5.2 \times 10^7$    & 46.8 \\
        3    & $5.8 \times 10^8$    & 114.4 & $3.4 \times 10^8$    & 35.7  & $1.05 \times 10^7$   & 26.1 \\
        4    & $8.7 \times 10^6$    & 216.6 & $2.5 \times 10^7$    & 113.0 & $1.05 \times 10^7$   & 191.8 \\
        5    & $6.9 \times 10^6$    & 139.4 & $1.9 \times 10^7$    & 49.8  & $4.5 \times 10^6$    & 230.2 \\
        6    & $6.0 \times 10^6$    & 66.3  & $1.0 \times 10^7$    & 60.5  & $2.2 \times 10^6$    & 219.7 \\
        7    & $5.6 \times 10^6$    & 250.2 & $6.6 \times 10^6$    & 141.8 & $7.5 \times 10^5$    & 56.7 \\
        8    & $4.5 \times 10^6$    & 47.1  & $5.3 \times 10^6$    & 57.4  & $4.6 \times 10^5$    & 70.1 \\
        9    & $2.8 \times 10^6$    & 115.6 & $5.2 \times 10^6$    & 140.5 & $4.5 \times 10^5$    & 149.9 \\
        10   & $1.47 \times 10^6$   & 166.0 & $2.2 \times 10^6$    & 157.7 & $4.2 \times 10^5$    & 101.2 \\
        11   & $1.46 \times 10^6$   & 64.6  & $1.8 \times 10^6$    & 175.1 & $4.0 \times 10^5$    & 109.6 \\
        \hline
    \end{tabular}
\end{table*}

Tables \ref{sat-tab2} and \ref{sat-tab3} summarize the host galaxy and satellite stellar masses and distances from the host galaxy of the six systems most like the Milky Way PoS for the low-MW mass range. These systems are labeled with the TNG50 halo finder ID as 289, 365, and 395 in Table \ref{sat-tab2} and 419, 473, and 573 in Table \ref{sat-tab3}.

\begin{table*}
    \centering
    \caption{Satellite stellar mass $M_*$ and distance from the host center $r$  for the first three MW-like { PoS}  systems assuming a low MW mass.}
    \label{sat-tab2}
    \begin{tabular}{ccccccc}
        \hline
        Sat. Num. &
        \multicolumn{2}{c}{\textbf{Group 289}} &
        \multicolumn{2}{c}{\textbf{Group 365}} &
        \multicolumn{2}{c}{\textbf{Group 395}} \\
        &
        $M_\star\ (M_\odot)$ & $r\ (\mathrm{kpc})$ &
        $M_\star\ (M_\odot)$ & $r\ (\mathrm{kpc})$ &
        $M_\star\ (M_\odot)$ & $r\ (\mathrm{kpc})$ \\
        \hline
        Host & $1.7 \times 10^{10}$ & 0.0   & $1.4 \times 10^{10}$ & 0.0   & $2.5 \times 10^{10}$ & 0.0 \\
        1    & $2.6 \times 10^{9}$  & 64.4  & $1.3 \times 10^{10}$ & 260.7 & $4.5 \times 10^{7}$  & 243.8 \\
        2    & $2.9 \times 10^{8}$  & 238.4 & $2.0 \times 10^{9}$  & 185.0 & $1.5 \times 10^{7}$  & 124.5 \\
        3    & $1.63 \times 10^{8}$ & 68.4  & $6.7 \times 10^{7}$  & 237.8 & $1.22 \times 10^{7}$ & 88.8 \\
        4    & $1.62 \times 10^{8}$ & 237.4 & $3.7 \times 10^{6}$  & 175.0 & $1.18 \times 10^{7}$ & 34.7 \\
        5    & $1.4 \times 10^{8}$  & 54.1  & $1.3 \times 10^{6}$  & 172.5 & $7.9 \times 10^{6}$  & 127.7 \\
        6    & $3.6 \times 10^{7}$  & 70.5  & $7.0 \times 10^{5}$  & 281.7 & $2.3 \times 10^{6}$  & 33.5 \\
        7    & $2.5 \times 10^{7}$  & 80.8  & $3.6 \times 10^{5}$  & 71.4  & $1.5 \times 10^{6}$  & 48.2 \\
        8    & $3.0 \times 10^{6}$  & 134.7 & $2.5 \times 10^{5}$  & 51.7  & $1.0 \times 10^{6}$  & 245.5 \\
        9    & $1.4 \times 10^{6}$  & 193.5 & $1.4 \times 10^{5}$  & 86.8  & $9.8 \times 10^{5}$  & 44.4 \\
        10   & $1.3 \times 10^{6}$  & 95.1  & $1.2 \times 10^{5}$  & 202.7 & $6.3 \times 10^{5}$  & 267.1 \\
        11   & $7.7 \times 10^{5}$  & 122.2 & $1.1 \times 10^{5}$  & 239.3 & $4.5 \times 10^{5}$  & 140.4 \\
        \hline
    \end{tabular}
\end{table*}

\begin{table*}
    \centering
    \caption{{  Same as Table 5, but for Groups 419, 473, and 573.}}
    \label{sat-tab3}
    \begin{tabular}{ccccccc}
        \hline
        Sat. Num. &
        \multicolumn{2}{c}{\textbf{Group 419}} &
        \multicolumn{2}{c}{\textbf{Group 473}} &
        \multicolumn{2}{c}{\textbf{Group 573}} \\
        &
        $M_\star\ (M_\odot)$ & $r\ (\mathrm{kpc})$ &
        $M_\star\ (M_\odot)$ & $r\ (\mathrm{kpc})$ &
        $M_\star\ (M_\odot)$ & $r\ (\mathrm{kpc})$ \\
        \hline
        Host & $1.5 \times 10^{10}$ & 0.0   & $6.1 \times 10^{9}$  & 0.0   & $1.8 \times 10^{9}$  & 0.0 \\
        1    & $6.3 \times 10^{7}$  & 167.4 & $2.6 \times 10^{9}$  & 112.7 & $1.5 \times 10^{9}$  & 195.8 \\
        2    & $3.2 \times 10^{7}$  & 70.2  & $1.1 \times 10^{8}$  & 111.8 & $8.8 \times 10^{8}$  & 60.0 \\
        3    & $1.9 \times 10^{7}$  & 107.2 & $9.3 \times 10^{6}$  & 255.6 & $4.0 \times 10^{6}$  & 243.6 \\
        4    & $1.3 \times 10^{7}$  & 226.9 & $6.4 \times 10^{6}$  & 179.9 & $1.1 \times 10^{6}$  & 154.3 \\
        5    & $1.1 \times 10^{6}$  & 213.1 & $5.3 \times 10^{6}$  & 34.3  & $7.9 \times 10^{5}$  & 152.0 \\
        6    & $6.7 \times 10^{5}$  & 83.5  & $2.2 \times 10^{6}$  & 115.7 & $2.8 \times 10^{5}$  & 49.5 \\
        7    & $6.4 \times 10^{5}$  & 38.3  & $1.8 \times 10^{6}$  & 90.5  & $2.7 \times 10^{5}$  & 146.2 \\
        8    & $6.3 \times 10^{5}$  & 179.4 & $1.0 \times 10^{6}$  & 120.7 & $2.6 \times 10^{5}$  & 190.3 \\
        9    & $5.5 \times 10^{5}$  & 66.2  & $4.1 \times 10^{5}$  & 10.9  & $2.2 \times 10^{5}$  & 222.2 \\
        10   & $1.7 \times 10^{5}$  & 58.1  & $3.4 \times 10^{5}$  & 213.1 & $1.4 \times 10^{5}$  & 163.3 \\
        11   & $1.5 \times 10^{5}$  & 162.8 & $2.5 \times 10^{5}$  & 144.9 & $1.3 \times 10^{5}$  & 78.7 \\
        \hline
    \end{tabular}
\end{table*}

\subsection{Orientation with Respect to Local Filaments in the Cosmic Web} 
\label{sec_orientation}
One feature that can be clearly seen in Figure \ref{fig_pos_star_300} is that the angle between the plane formed by the satellites and the stellar disk of the main galaxy encompasses a wide range of values. The angle can be close to 90$^{\circ}$ as in the case of the Milky Way or close to zero. This is likely due to the fact that the formation and evolution of the main galaxy and the formation and accretion of the satellites both involve interaction with their environment, but at different times and scales. Hence, the orientations of the PoS and galactic disk are not directly correlated with each other.

Indeed, it has been suggested \citep{Laigle2014} that in-fall along a filamentary structure could affect the orbit orientations of satellites in a PoS system. Moreover, numerical simulations \citep{Arag07,Hahn07,Paz08,Zhang09,Shao:2019} seem to indicate an environmental and temporal dependence on the orientation of the dark-matter halo spin within filaments. There is also observational evidence [e.g. \cite{Navarro04,Trujillo06,Jones10}] for an environmental effect on the galaxy spin vector orientation. One possible explanation for this is that tidal forces on galaxies within filaments can lead to a tendency for the angular momentum vector of galaxies to be preferentially torqued perpendicular to the axis of the filament \citep{Jones10}.

\begin{figure*}
\centering
\includegraphics[trim=0cm 0cm 0cm 0cm, clip, width=0.32\textwidth]{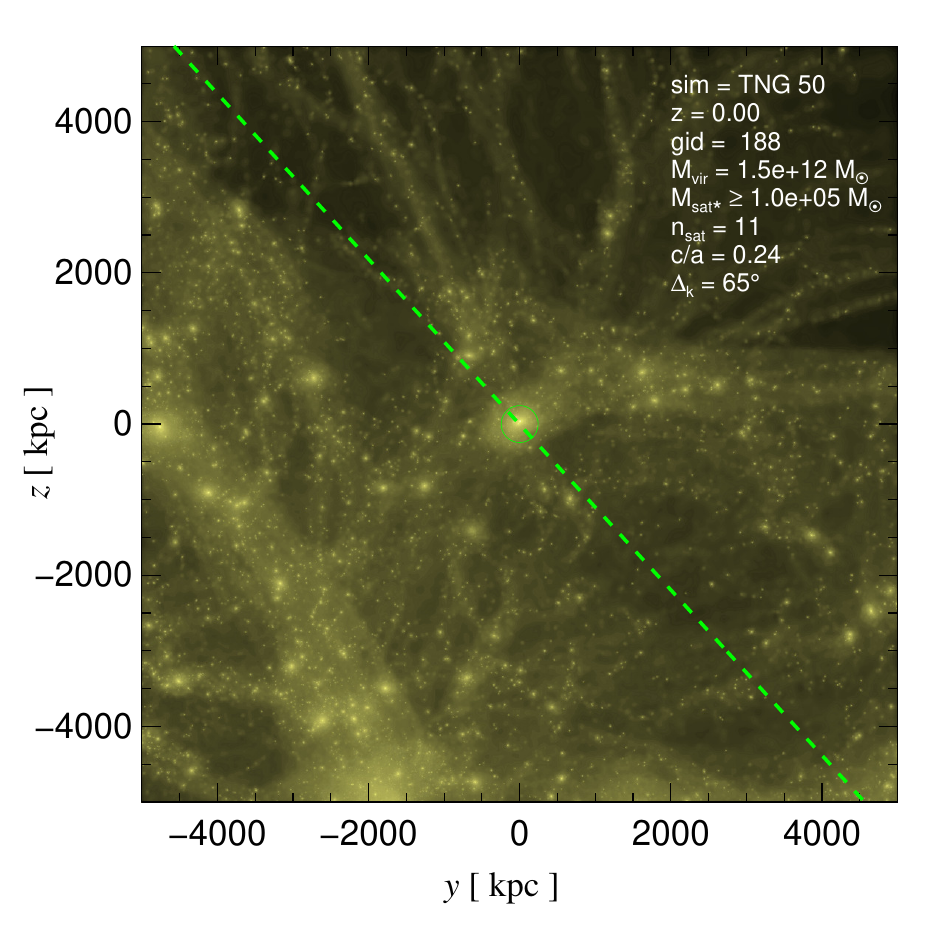} 
\includegraphics[trim=0cm 0cm 0cm 0cm, clip, width=0.32\textwidth]{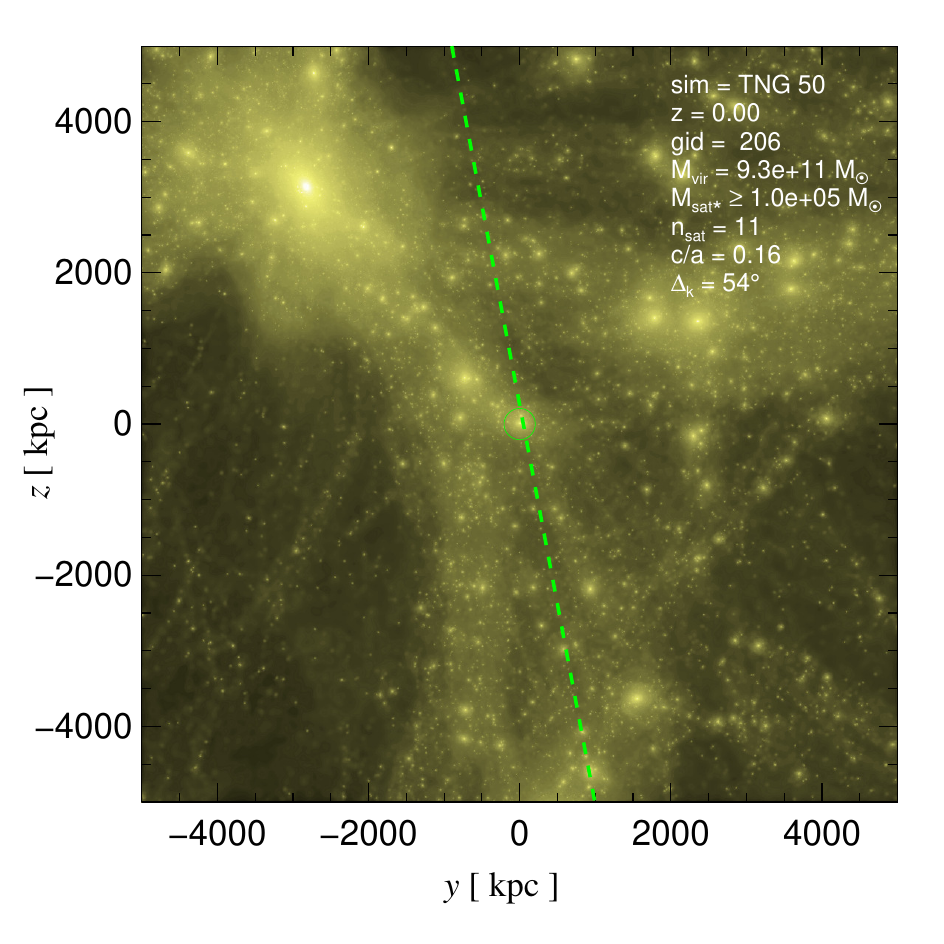} 
\includegraphics[trim=0cm 0cm 0cm 0cm, clip, width=0.32\textwidth]{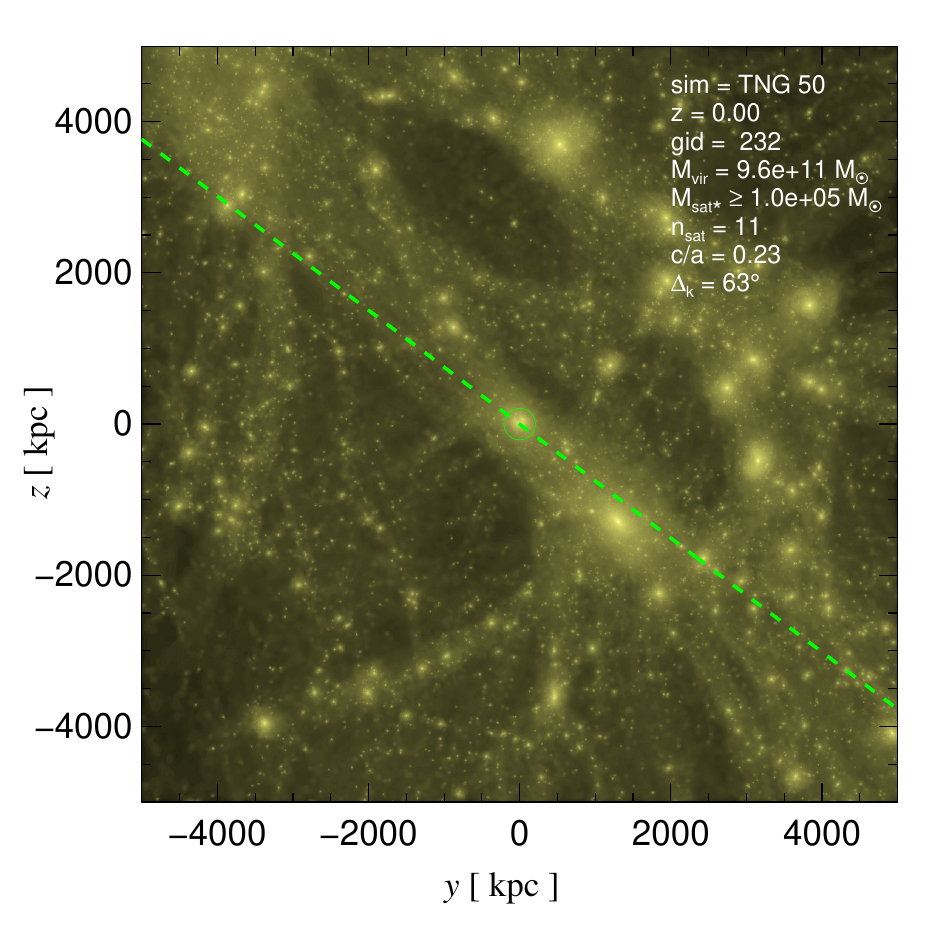} 

\includegraphics[trim=0cm 0cm 0cm 0cm, clip, width=0.32\textwidth]{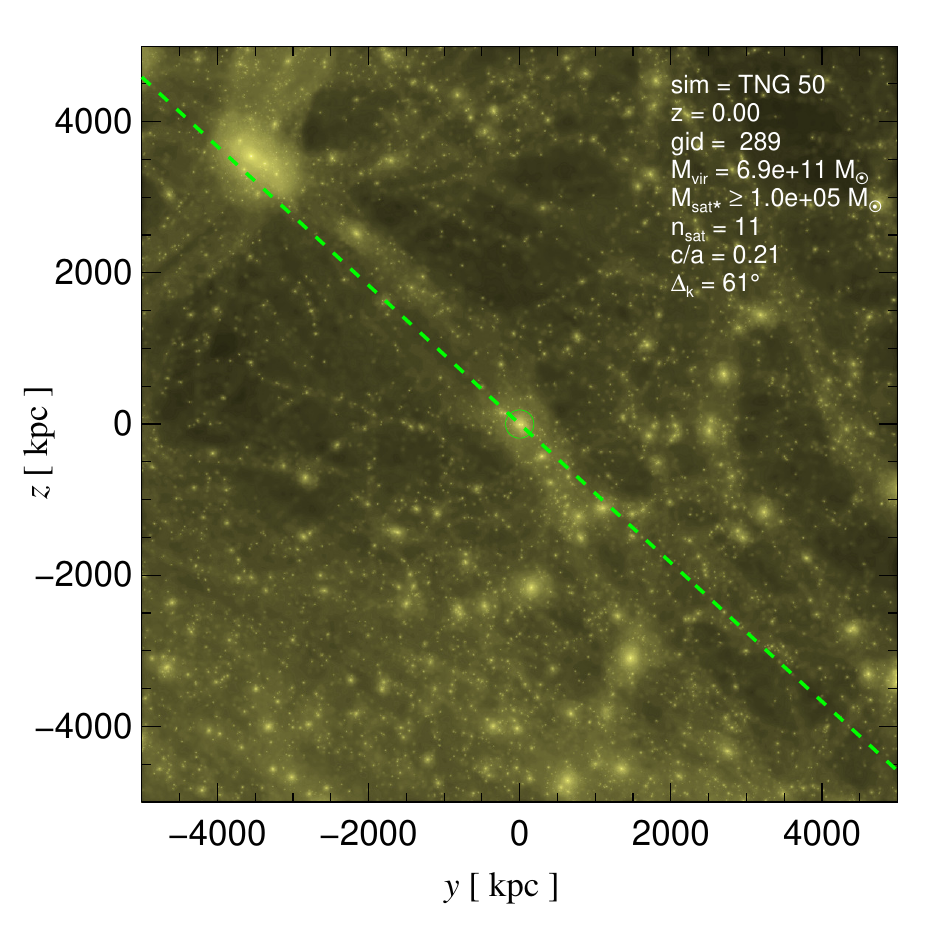} 
\includegraphics[trim=0cm 0cm 0cm 0cm, clip, width=0.32\textwidth]{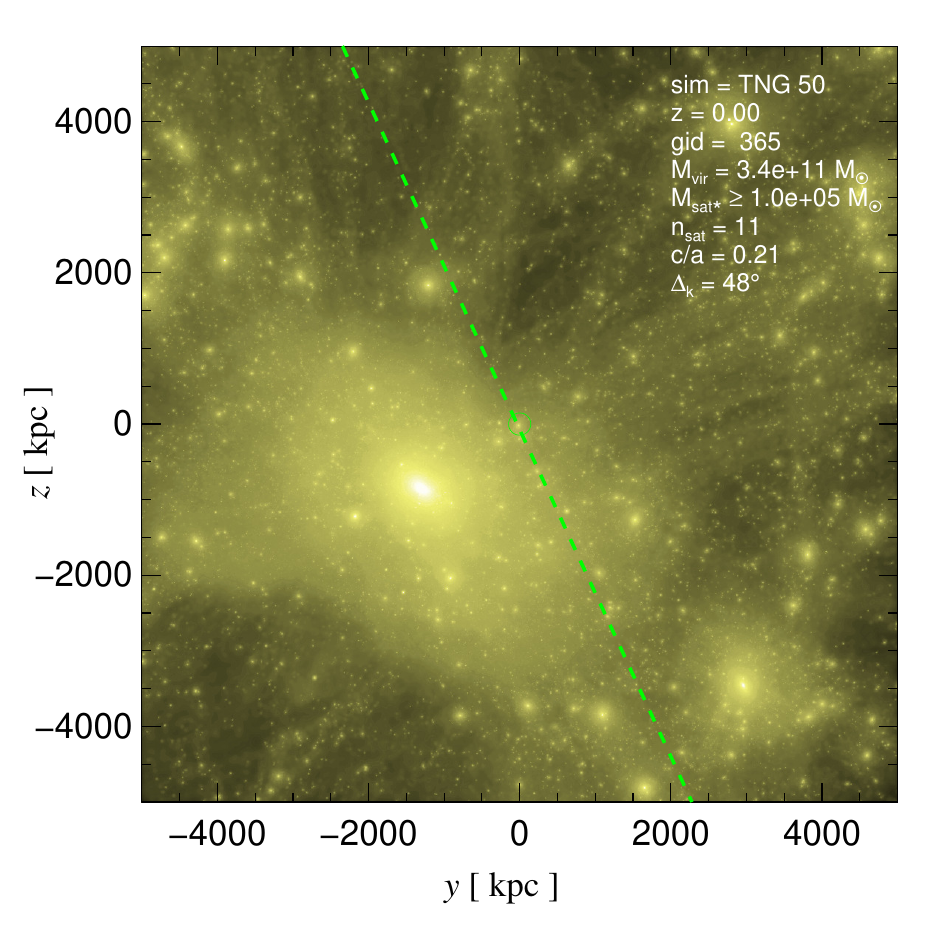} 
\includegraphics[trim=0cm 0cm 0cm 0cm, clip, width=0.32\textwidth]{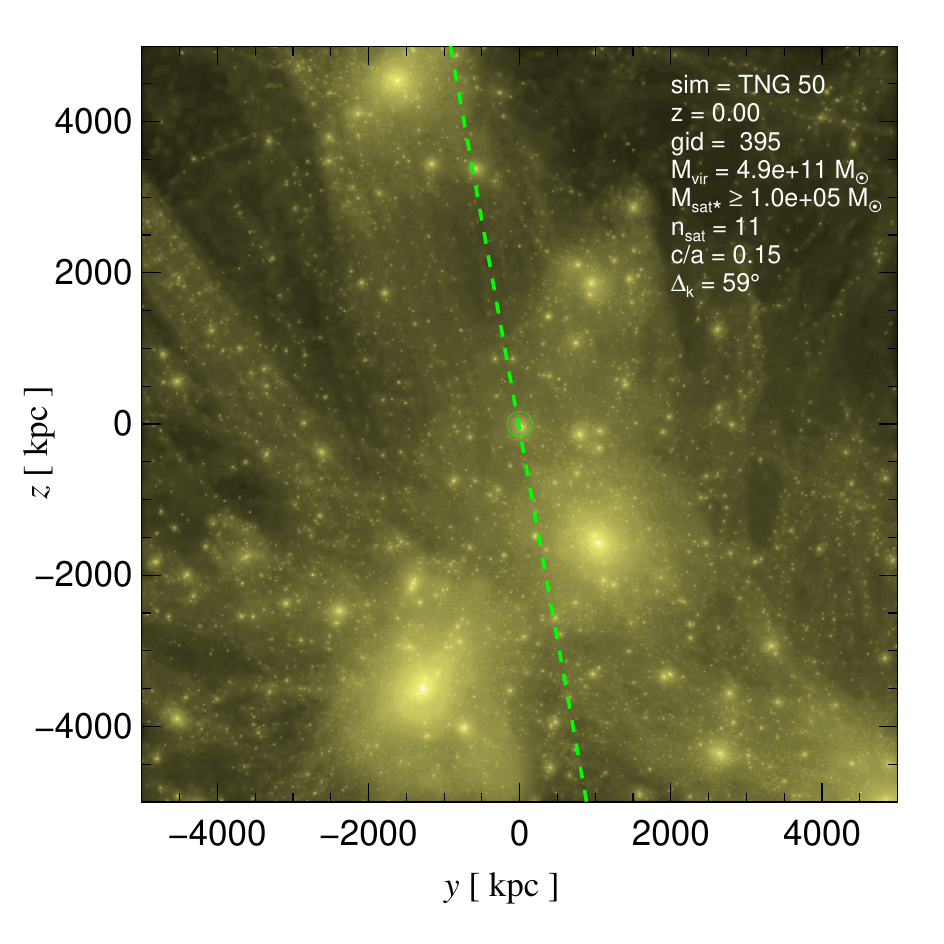} 

\includegraphics[trim=0cm 0cm 0cm 0cm, clip, width=0.32\textwidth]{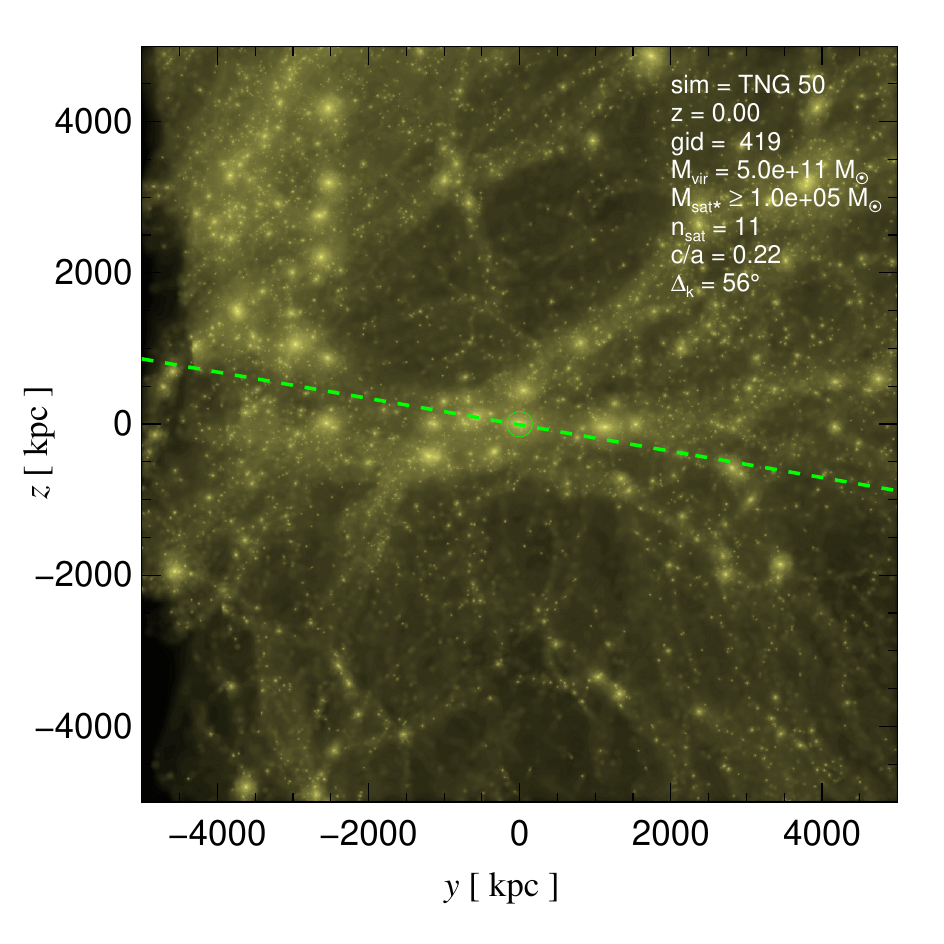} 
\includegraphics[trim=0cm 0cm 0cm 0cm, clip, width=0.32\textwidth]{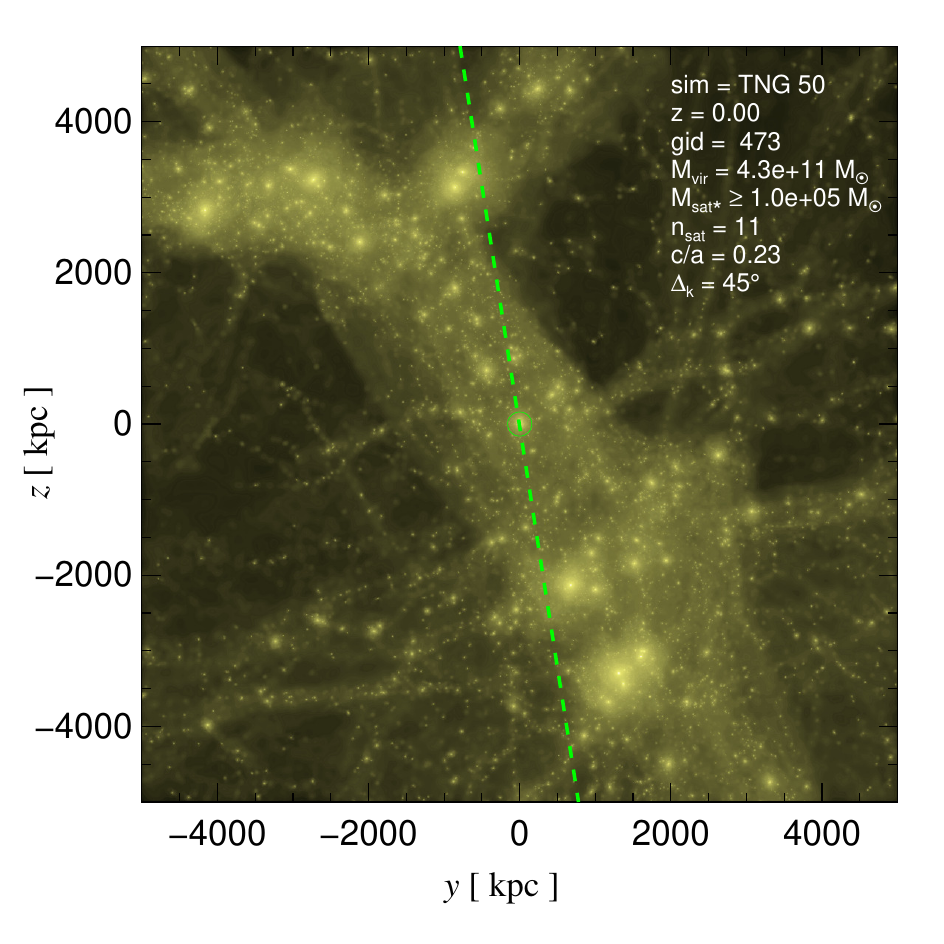} 
\includegraphics[trim=0cm 0cm 0cm 0cm, clip, width=0.32\textwidth]{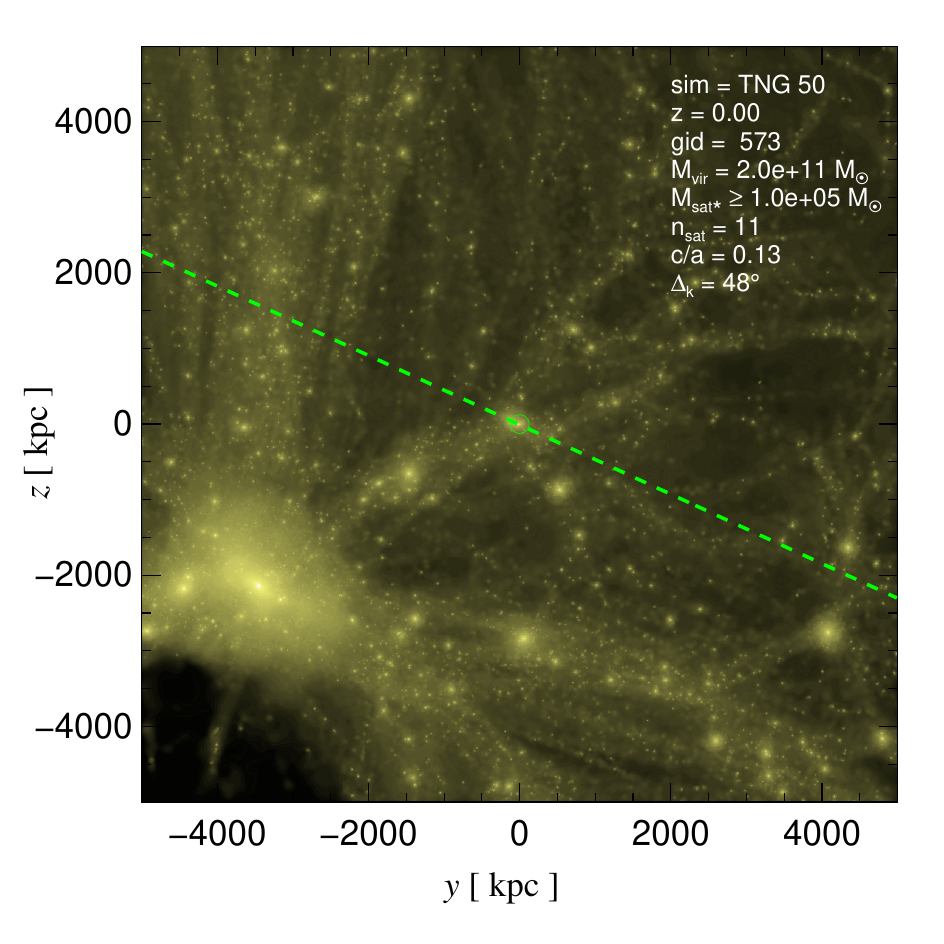} 

\caption{Edge-on views on a 5 Mpc scale of the nine MW-like PoS systems in the TNG50 simulation. The dark matter component is shown as 2D column density; brightness indicates density. Small green circles mark the virial radius of the host galaxy. Each image is oriented so the central galaxy is at the origin with its stellar disk in the $xy$ plane, and the fitted PoS is edge-on (dashed line).}
\label{fig_pos_dm_5000}
\end{figure*}

\begin{figure*}
\centering
\includegraphics[trim=0cm 0cm 0cm 0cm, clip, width=0.32\textwidth]{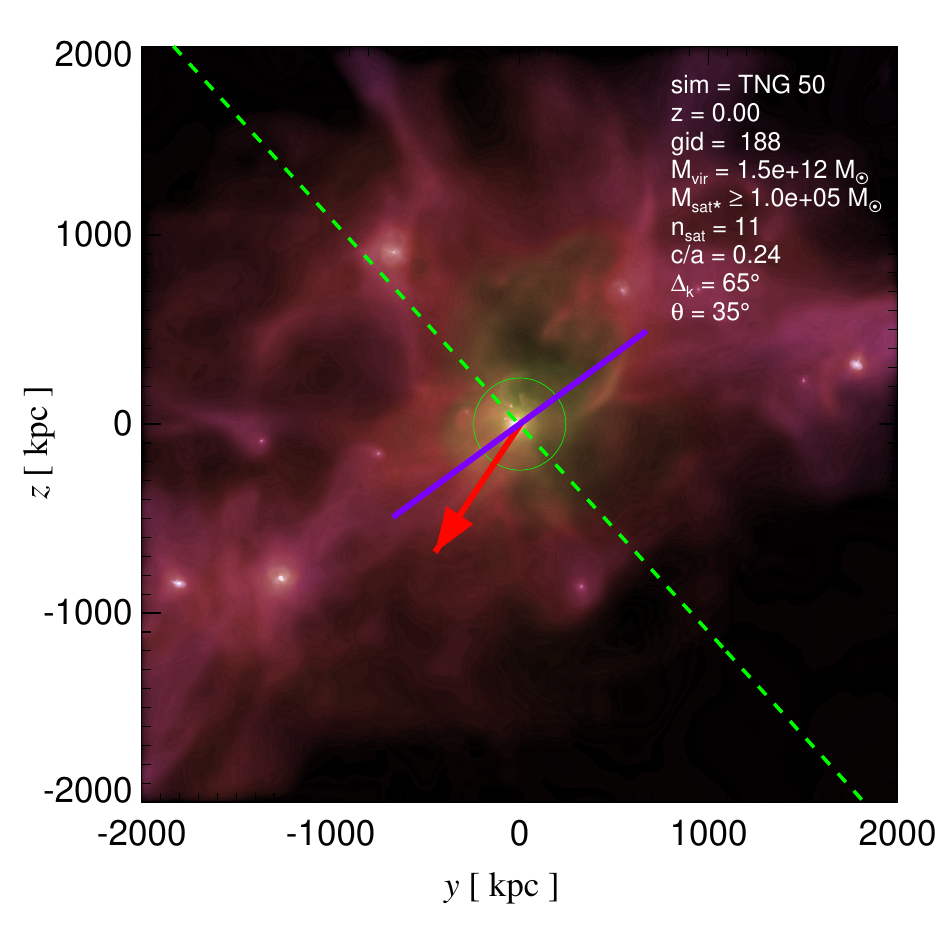} 
\includegraphics[trim=0cm 0cm 0cm 0cm, clip, width=0.32\textwidth]{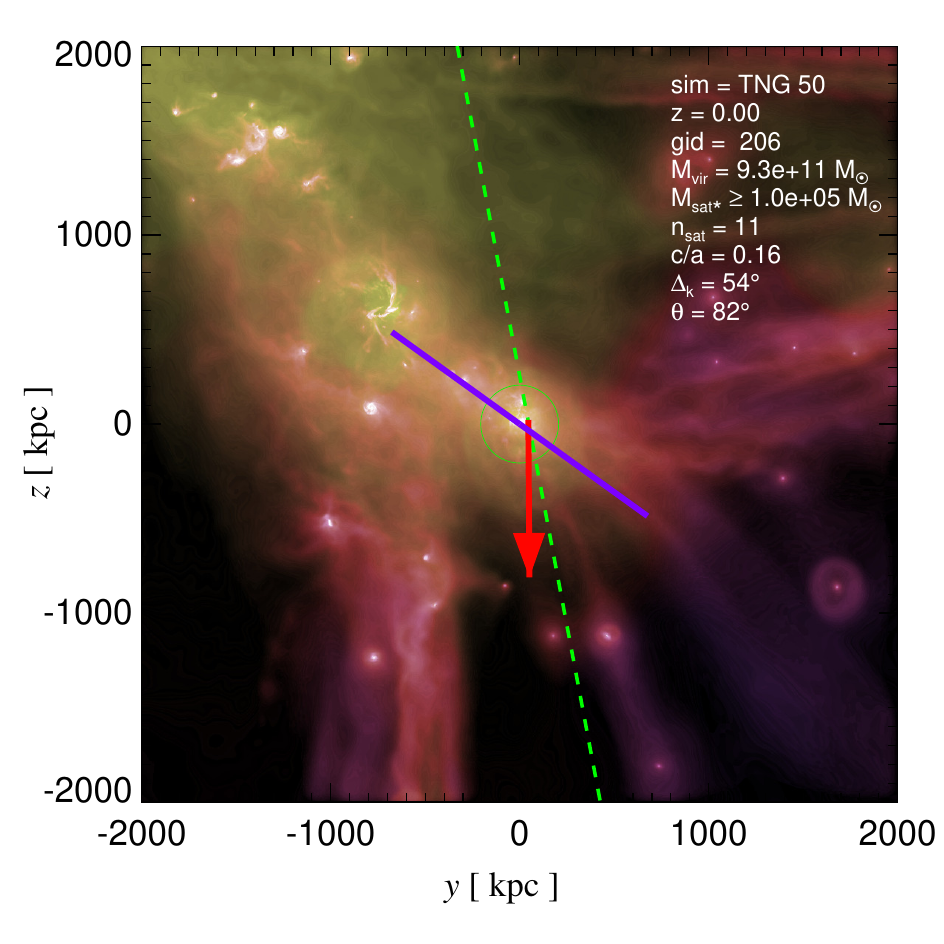} 
\includegraphics[trim=0cm 0cm 0cm 0cm, clip, width=0.32\textwidth]{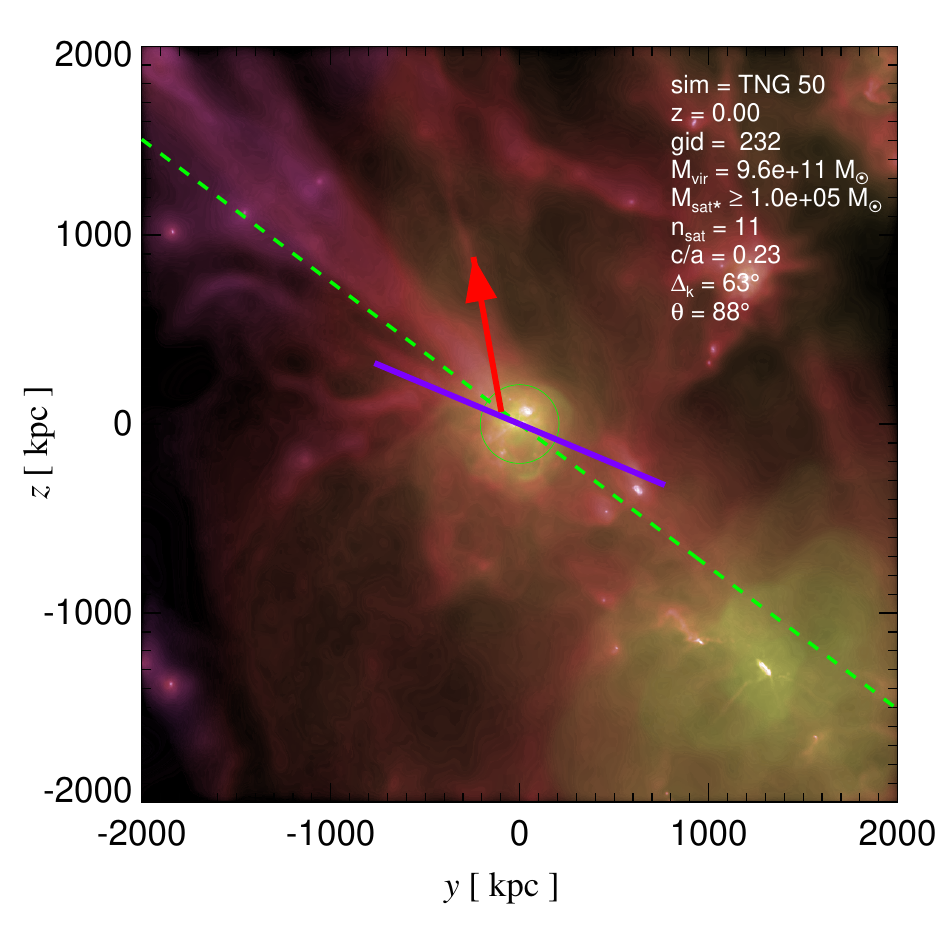} 

\includegraphics[trim=0cm 0cm 0cm 0cm, clip, width=0.32\textwidth]{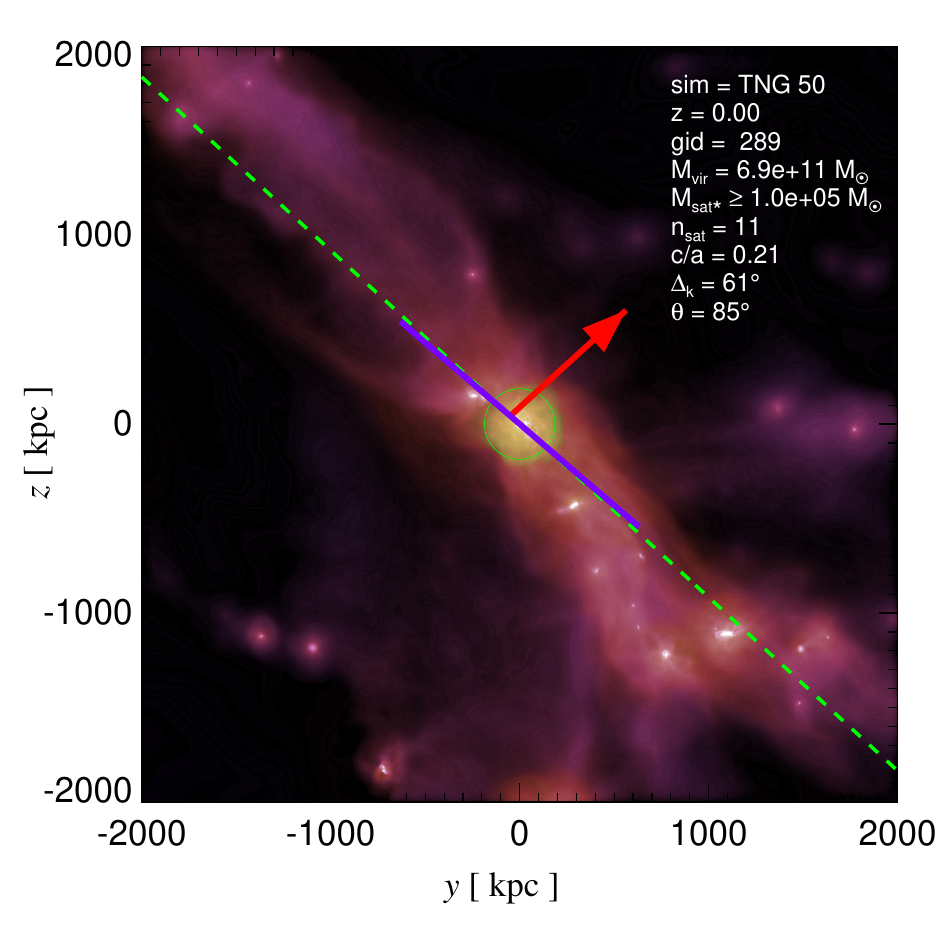} 
\includegraphics[trim=0cm 0cm 0cm 0cm, clip, width=0.32\textwidth]{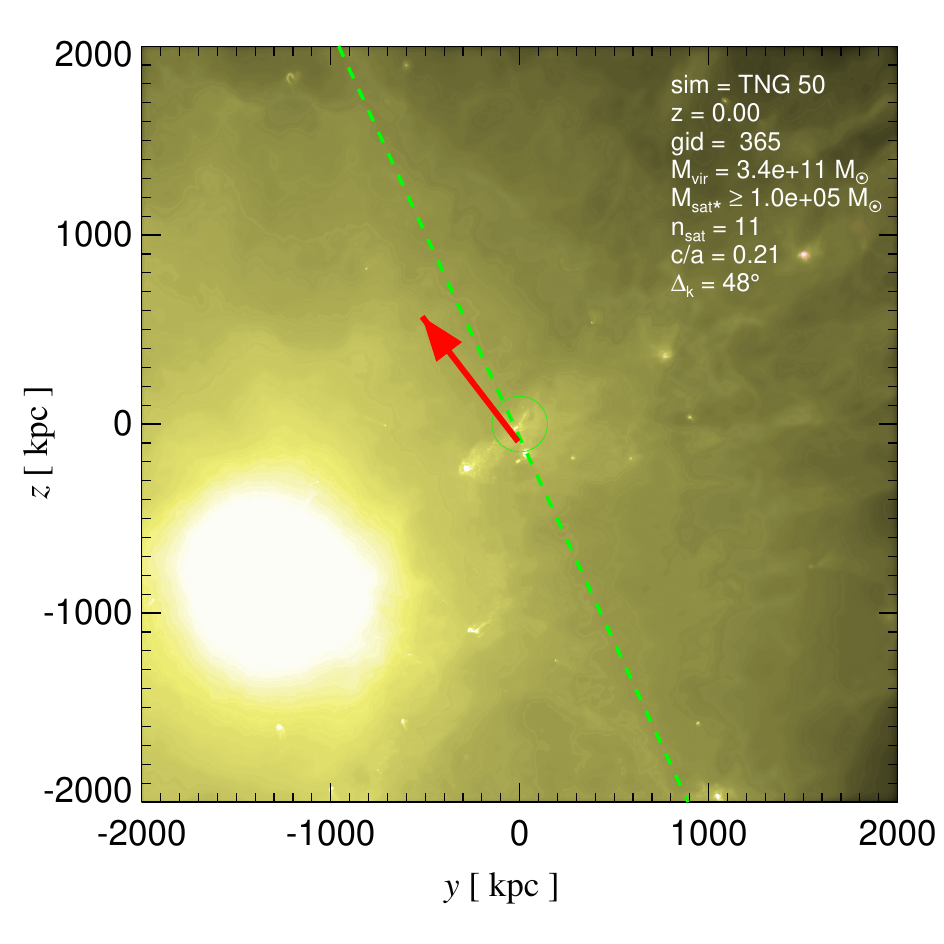} 
\includegraphics[trim=0cm 0cm 0cm 0cm, clip, width=0.32\textwidth]{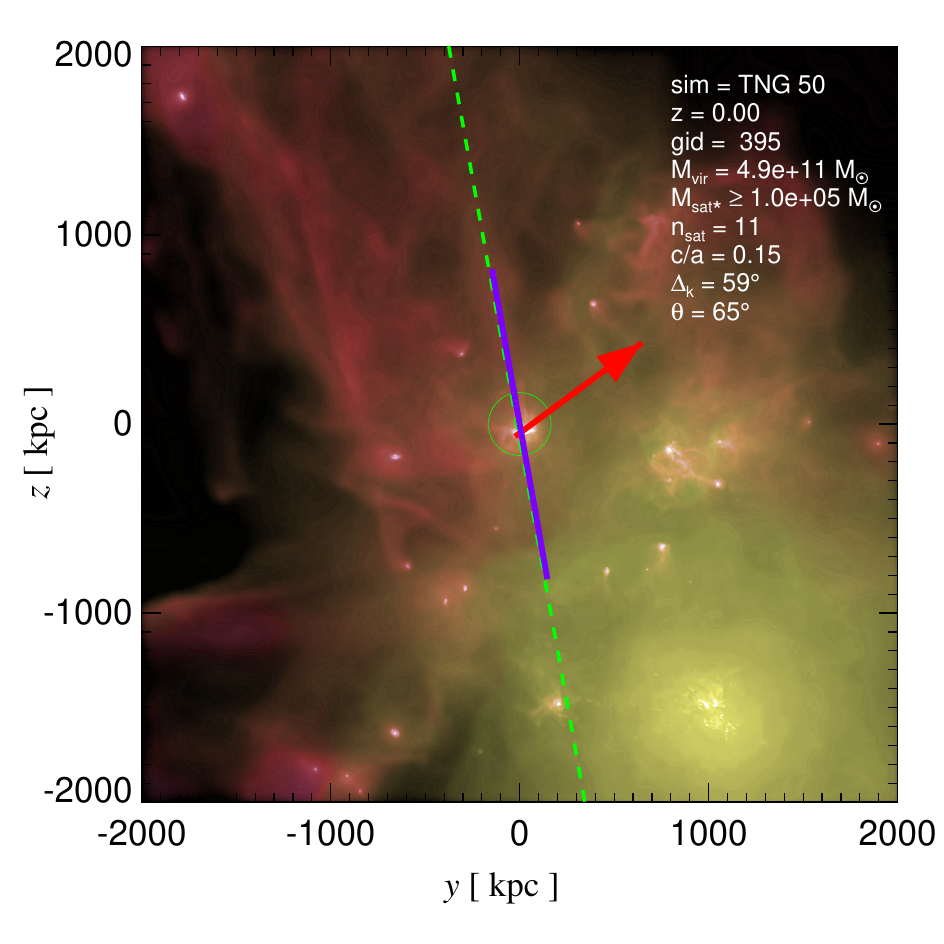} 

\includegraphics[trim=0cm 0cm 0cm 0cm, clip, width=0.32\textwidth]{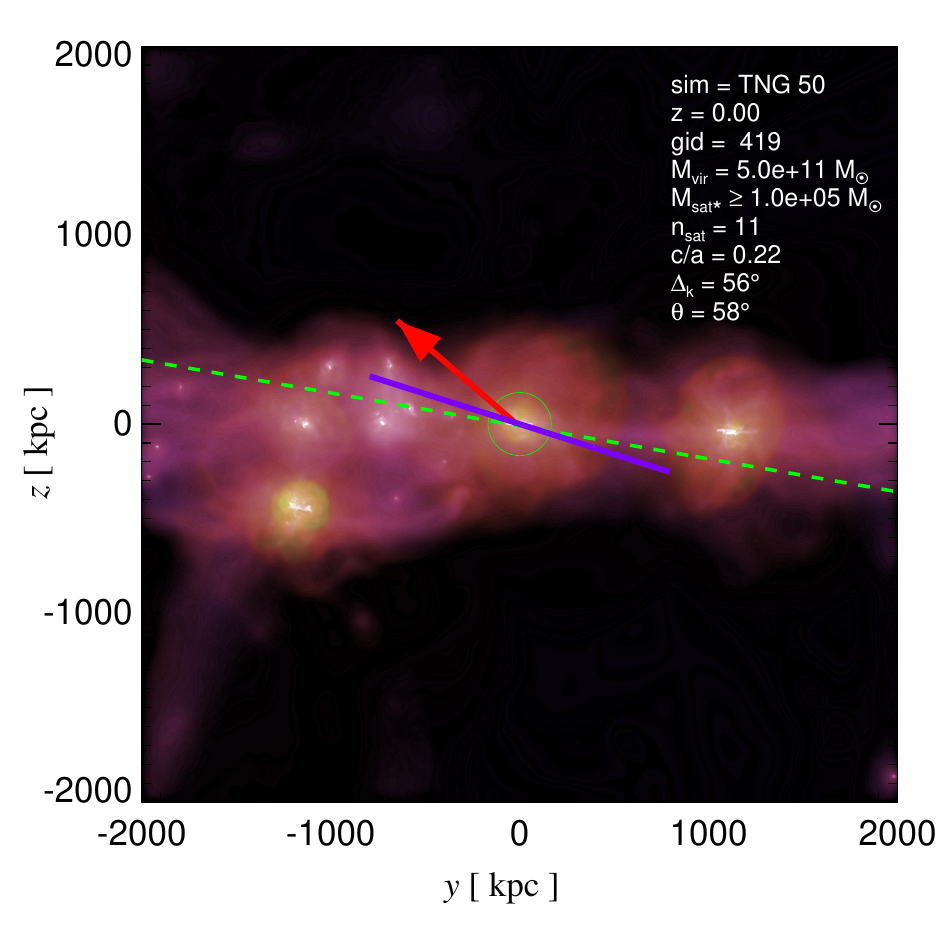} 
\includegraphics[trim=0cm 0cm 0cm 0cm, clip, width=0.32\textwidth]{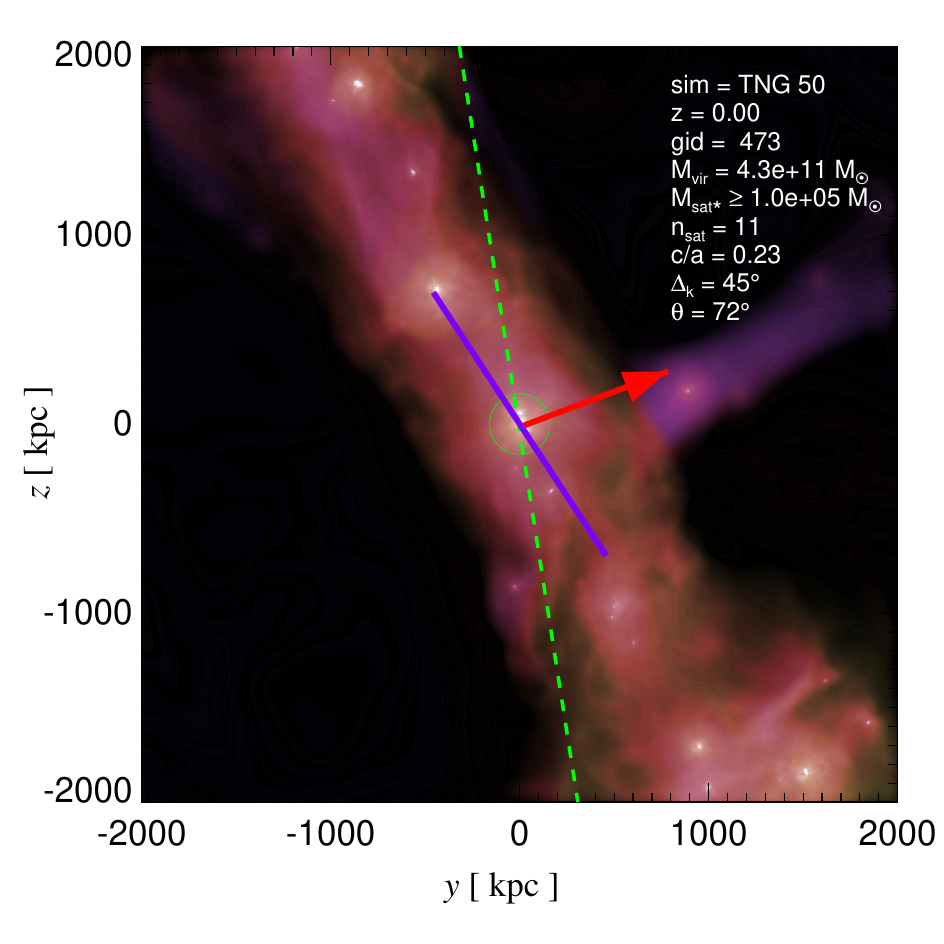} 
\includegraphics[trim=0cm 0cm 0cm 0cm, clip, width=0.32\textwidth]{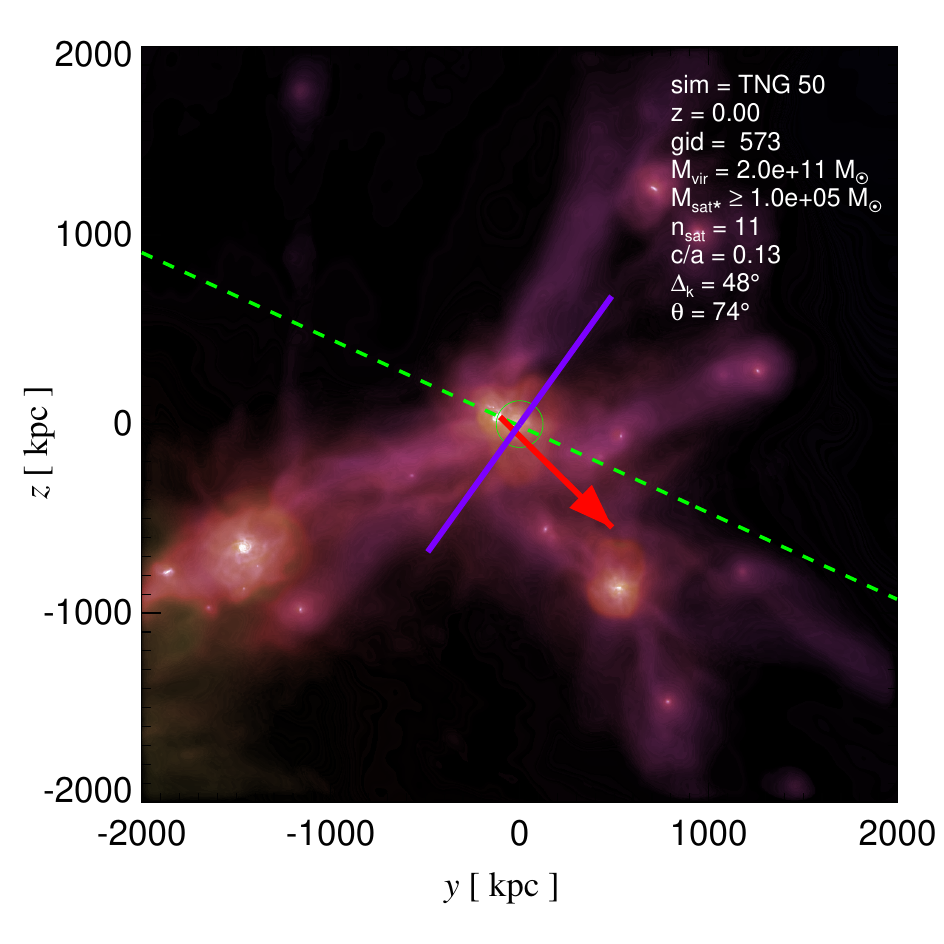} 

\caption{Edge-on views on a 2 Mpc scale of the nine MW-like PoS systems in the TNG50 simulation. The gas component is shown as 2D column density; color indicates gas temperature (blue→red/yellow for lower→higher temperature). Green circles mark the virial radius of each system. Each panel is oriented so the central galaxy is at the origin with its stellar disk in the $xy$ plane; the fitted PoS is edge-on (dashed line). The red arrow shows the direction of the total angular momentum of each PoS system. {The violet line shows the direction of the deduced spine of the filament that the central galaxy resides in.}}
\label{fig_pos_gas_2000}
\end{figure*}

Next, we focus on the formation and evolution of the plane of satellites on larger scales.
In Figure \ref{fig_pos_dm_5000}, we examine dark matter in the main and satellite galaxies and the surrounding structure for the systems listed in Table \ref{tab_pos_par} at $z=0$.  These are plotted in a cubic space of 5 megaparsecs along each dimension in the TNG50 simulation. In Figure \ref{fig_pos_gas_2000}, baryonic matter (gas) is rendered on a 2 megaparsec scale with the same setup as in Figure \ref{fig_pos_dm_5000} except that we also add the direction of the total angular momentum vector of the 11 satellites, and { the direction of the spine of the filament in which  the central galaxy resides. Group 365 is in a rich galaxy cluster.  It does not reside in well defined filaments.} 

We focus on PoS systems formed in the simulations that have environments somewhat similar to the Milky Way.  Hence, we do not categorize a PoS that is in a rich galaxy cluster (e.g. Group 365) or has a satellite with stellar mass well above 40\% of the host galaxy (i.e. Group 365 and 573) as a MW-like system for the purpose of the discussion on the impact of the large-scale environment. For the same purpose, { we adopt $V_{PoS}$ as more indicative of bulk motion of the satellites and the large-scale environment than the Gini coefficient.  This is because the Gini coefficient is mainly a local measure of the satellite radial distribution in the PoS and the interaction between satellites and the main galaxy, particularly whether there are massive Magellanic-Cloud (MC)-like satellites in the system.}

{We now examine the environments of the PoS systems at redshift $z=0$. 
As shown in Figure~\ref{fig_pos_dm_5000} and Figure~\ref{fig_pos_gas_2000}, in most cases, the host systems appear embedded within filamentary structures. }
{Using the filament spine direction estimated from the local dark matter density field with the algorithm outlined in Section~\ref{filament_spine_tensor}, we examine the geometric relation between the local filament orientation and the angular momentum direction of each PoS system, as illustrated in Figure~\ref{fig_pos_gas_2000}. Since the filament spine represents an unoriented axis, the alignment angle is defined within the range $0^\circ$--$90^\circ$. The resulting alignment angles and filament-confidence parameters as defined in Equation~(\ref{c_para}) for the eight MW-like systems are summarized in Table~\ref{angle_tab}.}

\begin{table}
\centering
\caption{{Relative orientation between the PoS angular momentum and the local filament spine direction for the 8 MW-like systems. The confidence parameter $C$ quantifies the coherence of the local filament orientation estimate, as defined in Section~\ref{filament_spine_tensor}.}}
\label{angle_tab}
{
\begin{tabular}{ccc}
\hline
Group ID & $\theta_{\rm PoSA-Spi}$ [$^\circ$] & $C$ \\
\hline
188 & 35 & 0.903 \\
206 & 82 & 0.602 \\
232 & 88 & 0.896 \\
289 & 85 & 0.935 \\
395 & 65 & 0.938 \\
419 & 58 & 0.832 \\
473 & 72 & 0.813 \\
573 & 74 & 0.862 \\
\hline
\end{tabular}
}
\end{table}

{Now, we consider each PoS system mainly using Figure~\ref{fig_pos_gas_2000} and Table~\ref{angle_tab}, we note that Group 188 resides near the center of a relatively extended filament. The satellite plane has formed nearly perpendicular to the filament, while the direction of the angular momentum is roughly along the filament. The angel ($\theta_{\rm PoSA-Spi}$) between the PoS angular momentum direction and the filament spine direction is only $35^{\circ}$. We speculate, therefore,  that the satellites could have accreted to the main galaxy along a direction perpendicular to the filament spine, while retaining the PoS angular momentum.  Hence, the PoS system rotates around the spine of the filament.  
Also, the most massive satellite has about one quarter stellar mass of the main galaxy, so its interaction with the main galaxy may have significant impacts on the formation of a PoS in this system.} The PoS velocity, $V_{PoS}=1711$ km s$^{-1}$, is much higher than that of the MW. This is indicative of the dynamical nature of this system.

{ Group 206 is a few Mpc away from a massive rich cluster and has $V_{PoS}=1782$ km s$^{-1}$.  This is the largest  among the nine galaxies listed in Table \ref{tab_pos_par}. Although $\theta_{\rm PoSA-Spi}=85^{\circ}$ indicates that the satellite motion within the PoS is close to the direction of the local filament, the filament itself is less well defined (the confidence parameter C=0.602) than in the other systems. Together with the observation that this system appears to be impacted by  outflow from the nearby massive rich cluster, we therefore suspect that the outflow probably also aided in the formation of the satellites themselves. That is, the outflow would have caused a compression of the local intergalactic medium along a front and the formation of satellites along that front.}


{ We find that 7 out of 9 MW-like PoS systems are located within a filament.  This is close to double the number expected from the fraction of galaxies observed to reside in filaments \citep{Tempel:2014}.  For Groups 232, 289, 419, and 473, in particular, the main galaxies are all clearly located near to the spines of  well-defined filaments with $C=0.896, 0.935, 0.832, 0.813$ and their satellite planes all roughly line up with filaments. The angle between the angular momentum direction of the PoS and the filament spine varies from {about 60$^{\circ}$ to about 90$^{\circ}$ with $\theta_{\rm PoSA-Spi}=88^{\circ}, 85^{\circ}, 58^{\circ}, 72^{\circ}$}. All of these characteristics of the PoS systems suggest that their satellite planes may have formed by accretion of the satellites to the main galaxies along the filaments (at least very recently).} 

{ Also, the bulk velocity of these PoS systems is close to that of the MW PoS, $V_{PoS}= 668$ km s$^{-1}$. (i.e. $V_{PoS}=972$ km s$^{-1}, 630$ km s$^{-1}, 683$ km s$^{-1}, 795$ km s$^{-1}$, respectively).  This suggests that the PoS formed in the simulation may have similar dynamics to that of the MW. (An exception, however, is the most luminous satellite of  Group 473 which  has a mass of $\approx$ 40\%  that of the main galaxy.)   Thus,  accretion along filaments seems to be a plausible explanation for the formation of the MW PoS. If we also take into account the low orbital pole dispersion ($\Delta_k$ and $\Delta_{k7}$) of these 4 systems, and the fact that their angular momentum direction is nearly perpendicular to the local filament spine, we suspect that at least some of the satellites around the main galaxies are rotationally supported.}

Group 395 is also near ($\sim 2$ Mpc) a rather massive galaxy cluster visible in the lower right corner of the respective panel in Figure \ref{fig_pos_gas_2000}.  However, its $V_{PoS}=156$ km s$^{-1}$. This is the lowest velocity in the 9 PoS systems we have selected. Also, it is not in a major filament. Hence, we believe that the PoS of Group 395 likely formed via accretion in a relatively virialized environment along a minor filament without the presence of significant intergalactic medium flow. Other than $V_{PoS}$, its PoS parameters match the MW well. However, the  Milky Way is not near a massive galaxy cluster.

The formation of the PoS for Group 365 is probably {also} related to a shock front generated by the outflow from the massive galaxy cluster {at about 1.5 Mpc away}. On the other hand, the PoS of Group 573 is probably formed via a merging process of two or more filaments. The $V_{PoS}$ for both groups is higher than that of the MW.   These two are not considered to be a MW PoS analog because of the presence of a massive companion to the central galaxy. It is possible that the tidal force created by the massive galaxies in each group may have played an important role in the formation of the PoS.  However, we will leave that aspect for a future study.

Although a more detailed analysis is needed, from the nine examples we have discussed, we speculate on the possible formation of a PoS from various physical processes. These are:
\begin{enumerate}
    \item {\bf {Satellite accretion along the spine of a filament.}}  In this case, the satellite plane will be along the spine of the filament. At least some of the satellites in the plane will rotate around the main galaxy in the filament (Group 232, 289, 419, and 473).
    \item {\bf Satellites formation  and alignment caused by the outflow of intergalactic medium from the center of a nearby massive galaxy cluster.}  In this case, the orientation of the satellite plane and its direction of motion will depend upon the direction of the outflow (Group 206 and 365).
    \item {\bf {Satellite accretion perpendicular to the spine  of a filament.}}  In this case, the satellite plane will be perpendicular to the spine of the filament, and at least some of the satellites in the plane will rotate around the main galaxy and the spine (Group 188).
    \item {\bf Satellites accrete and line up in a relatively isolated and virialized environment.} In this case the orientation of the satellite plane and satellite motion will depend on the local environment (Group 395).
    \item {\bf Satellites accrete and align during the merging process {within} the filament}.  In this case, the orientation of the satellite plane and the satellite motion will depend upon individual merging processes (Group 573).
\end{enumerate}

For the MW, we consider it most likely that the Plane of Satellites has formed via accretion along filaments (Process 1). This is consistent with other recent studies. For example \citep{Dupuy:2022} have deduced that satellites in the Local Group did accrete along the axis of the spine  of the local filament. Note, that if we know the spatial distribution and motion of the PoS on a large scale (the alignment and angular momentum direction of the PoS), we might be able to determine the major axis and spin direction of the central galaxy and host dark matter halo. This is of interest because there are possible alignments between the direction of the angular momentum of the satellite system and the host halo as suggested in \cite{Mezini:2025}. Also, the satellites tend to distribute along the major axis of the host halo \citep{Wang:2020,Xia:2021}. Future studies may offer further confirmation of these alignments. 

We believe that the satellite rotation around the spine of the filament in Process 1 is due to the retention of the satelite angular momentum from the spin of the filament \citep{Wang:2021}. Also, contrary to the findings in \cite{Madhani:2025}, we did not find that satellite accretion perpendicular to filaments (Process 3) would create satellite planes thinner than satellite accretion along filaments (Process 1). Rather, we find that accretion along filaments can produce satellite planes as thin as that of the MW. 
 
\subsection{Presence of MC-like Satellites}
\label{sec_lmc_smc}
\begin{figure}
\centering
\includegraphics[width=\columnwidth]{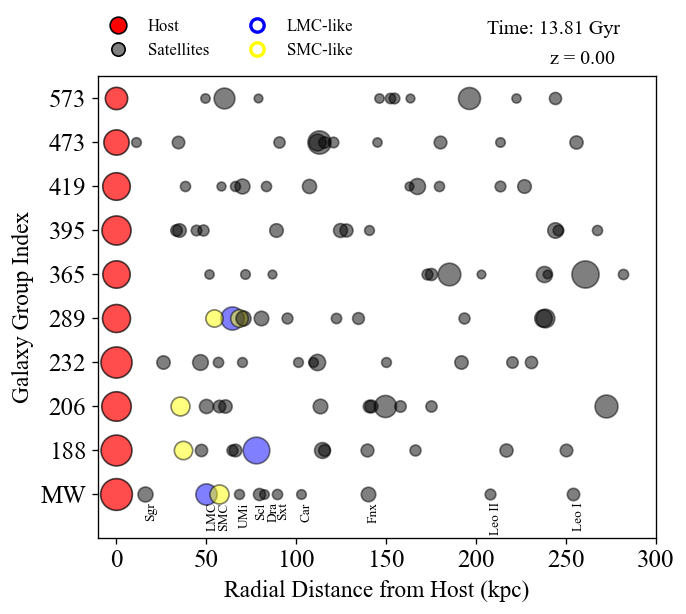}
\caption{Illustration of satellite masses and distances from the host for the most MW-like PoS systems. Circle size represents relative mass on a logarithmic scale. Red circles mark host galaxies; blue and yellow circles indicate LMC- and SMC-like satellites, respectively; gray circles are other satellites. The bottom row shows the Milky Way (MW) and its satellites.}
\label{fig_sat-r}
\end{figure}

Another possible contributing factor to PoS formation is the impact of relatively massive satellites such as the LMC and SMC. Figure \ref{fig_sat-r} illustrates the stellar masses and satellite distances for these 9 identified MW-like PoS systems.  The bottom row shows the MW and its PoS.  The presence of an SMC or LMC-like satellite is identified by the colors. From this figure one can see that the arrangement of distances and masses of the MW PoS is not particularly unusual.  The existence of a close LMC- and/or SMC-like system is apparent in groups 188, 206 and 289.

A striking feature of the MW PoS is the presence of the nearby LMC and SMC massive satellites.   Indeed, 1/3 of the massive satellites identified in the SAGA DR3 are Large-Magellanic Cloud (LMC)-like systems.  In the SAGA DR3 \citep{Mao:2024}, it was pointed out that the best predictor of the number of satellites in the system is the existence of a massive  satellite. However, the MW differs from many SAGA DR3 systems in that it has fewer satellites despite having an LMC-mass satellite. A possible explanation for this difference is the recent arrival of the LMC in the Milky~Way system.  Hence, we next discuss the dynamics of the simulated MW-like PoS systems that have an MC-like satellite.
 
The LMC and SMC masses inferred by observation are given in Table \ref{tab_obs_par}.  Since these are derived using intrinsic properties such as luminosity, metallicity or rotational curves, they are not affected by an ambiguity in the mass of the Milky-Way. 
For example,  in \cite{Stanimirovic04}  the dynamic mass of the SMC was derived using its HI rotation curve.
In \cite{vandermarel02,vandermarel14} the mass of the LMC was determined based upon kinematics.

Hence, we adopt the following constraints to identify systems similar to the LMC/SMC:
\begin{itemize}
    \item {\bf Isolation condition for the host system:} Satellites must have a stellar mass $\le 40\%$ of the host's stellar mass. If any satellite exceeds this, the PoS is not considered MW-like, but rather a galaxy merger system. This constraint is adapted from the discussion in SAGA DR3 \citep{Mao:2024}  to exclude systems with a satellite brighter than M$_K$+1 where M$_K$ is the host K-band magnitude.
\item {\bf Distance criterion:} Satellites must be located at $56\pm 28$ kpc from the host galaxy. [cf. \cite{Pillepich24}]. We adopt this constraint for the LMC and the SMC distance from  \cite{Pietrzynski19} and  \cite{Graczyk20}, respectively.
\item {\bf Stellar Mass Thresholds:} Adapted from the SAGA IV Survey \citep{Geha24} (we adopt a $\pm 0.5$ decade uncertainty)
\begin{itemize}
\item	LMC-like: Satellites with $\log{M_*} > 9$.
\item	SMC-like: Satellites with $\log{M_*} \sim  8-9$.
\end{itemize}
\end{itemize}

Based upon these selection criteria, we speculate the following with regard to the PoS systems identified in Tables \ref{sat-tab1}-\ref{sat-tab3}.
For the high-mass MW systems of Table \ref{sat-tab1}, satellites 1 and 2 of Group 188 are LMC- and SMC-like, respectively. This can be a MW Magellanic-Cloud system.
Satellite 3 of Group 206 is SMC-like. Group 232 does not contain an LMC or SMC-like system and, thus, is not considered a good MW-like PoS.

For the low-mass MW systems of Tables \ref{sat-tab2} and \ref{sat-tab3}, satellite 1 of Group 289 is LMC-like, and satellites 3 and 5 are SMC-like. { Note that satellites 1 and 3 are distinct satellites even though they appear in nearly the same radial distance in the edge-on view of Figure \ref{fig_pos_star_300}}.

Groups 365, 473 and 573 are not MW-like since they contain a satellite in excess of 40\% of the host mass, although Group 473 barely exceeds this limit.  Groups 395 and 419 don't have MC-like satellites.  

So in summary, if we exclude Groups 365, 473 and 573, then 1/2 of the systems have at least a Magellanic-Cloud-like satellite, and 1/3 of the systems have at least one LMC-like satellite. For the high-mass MW criterion, we also found that one in three has an LMC-like satellite.

This is more-or-less in agreement with SAGA DR3 who also found that a factor of 1/3 (34/101) of the SAGA systems have at least one confirmed satellite with an inferred stellar mass above $10^9$ M$_\odot$. Moreover, considering the MW-like systems that were excluded because a companion is too massive,  the presence of a massive satellite is clearly a significant contributing factor to the low aspect ratio and orbital-pole dispersion. However, this is not the only factor that determines the properties of a MW-like PoS and the satellite plane created this way can be short-lived \citep{Kanehisa23}.

Groups 188 and 289 contain both a LMC- and a SMC-like satellites. As one can see from Figure \ref{fig_radial_dist}, although their satellite radial profiles do not match that of the MW, their five innermost satellites are all near 80 kpc. Thus, having massive satellites at about 50 kpc just like MW does have a significant impact on the satellite radial profile at least in the inner region of the system in these two examples.

We also wish to point out that Group 289 with a virial mass of $6.9 \times 10^{11} M_{\odot}$ is very similar to the MW in all parameters we have discussed. It is the most Milky Way-like { PoS} system in our sample. If it indeed resembles our MW, then the MW is favored to have a mass in the low mass range as in \cite{Ou:2024, Ou25, Jiao23, Roche24}. The satellites of the MW are likely to have accreted along the local filament that the MW resides in, and at least some of the satellites are rotating around the MW with a rotational axis perpendicular to the filament.

\subsection{Evolution of the PoS}
\label{sec_time_evo}
\begin{figure*}
\centering
\includegraphics[width=0.32\textwidth]{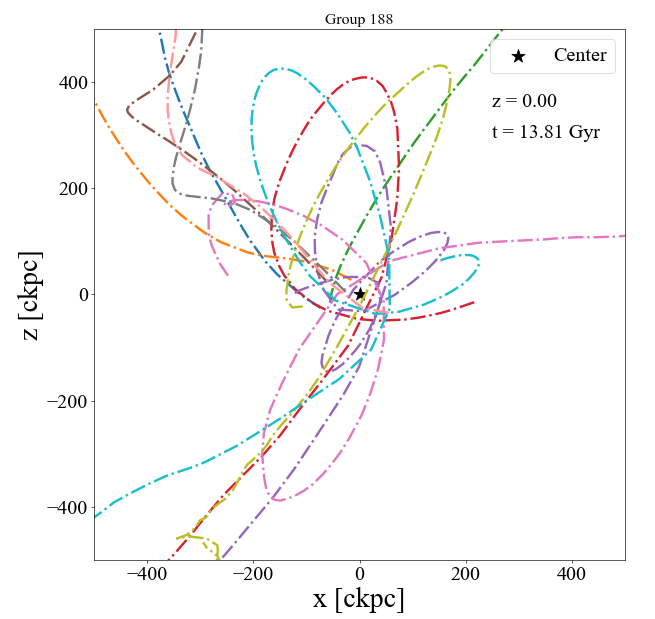}
\includegraphics[width=0.32\textwidth]{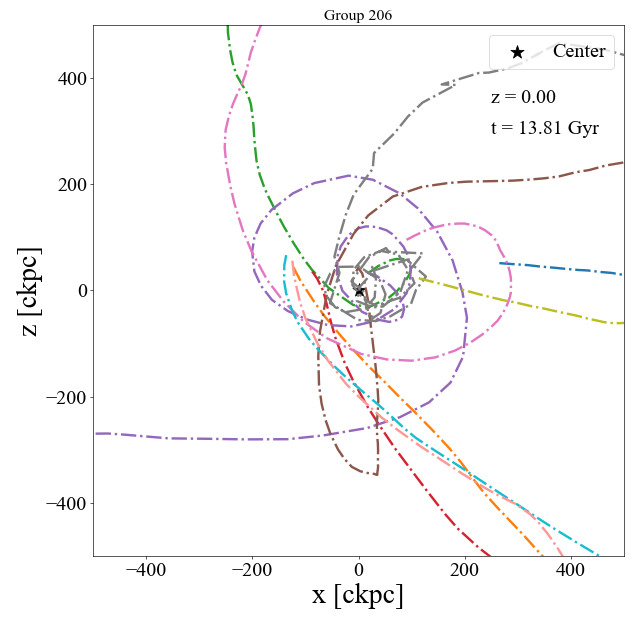}
\includegraphics[width=0.32\textwidth]{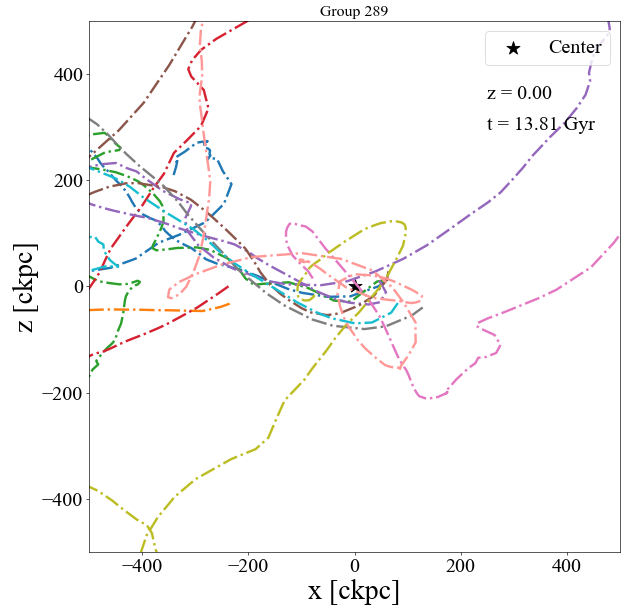}

\includegraphics[width=0.32\textwidth]{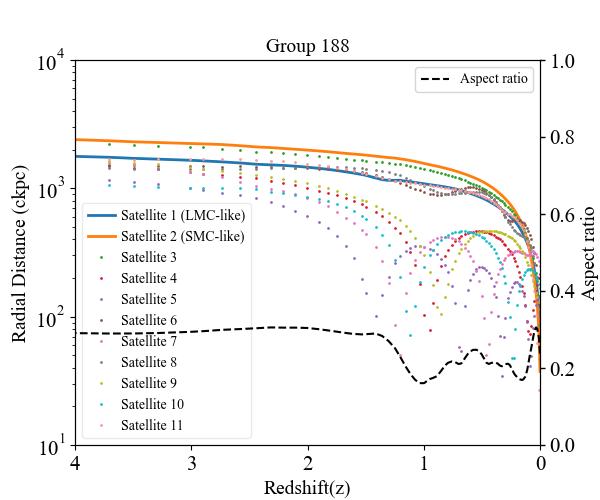}
\includegraphics[width=0.32\textwidth]{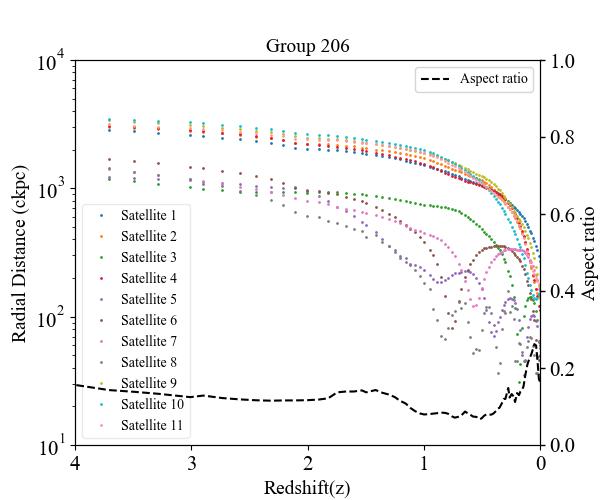}
\includegraphics[width=0.32\textwidth]{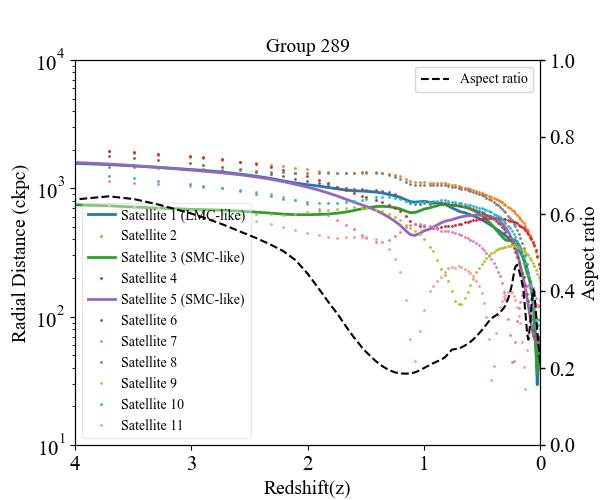}

\caption{Top row: PoS edge-on views of the trajectories of the 11 closest satellites for the three most Milky Way‚ like groups (188, 206, 289). Trajectories are shown from the time each satellite is within 500 kpc of the host, in comoving coordinates, with $x$ aligned with the PoS major axis at $z=0$ and $z$ with the minor axis at $x=0$. The star marks the central host galaxy. Bottom row: evolution of the radial distance from the host for the same satellites; the dashed line shows the aspect ratio $c/a$. Roughly half of the satellites arrived at high redshift and follow elliptical orbits, while the rest arrived more recently at low redshift, highlighting the transient nature of the PoS.}
\label{figure9}
\end{figure*}

We will discuss the dynamics and history of MW-like PoS systems in a separate paper.   Nevertheless, to provide a glimpse into the evolution of the PoS in these most MW-like systems, the upper panels of Figure \ref{figure9}  shows a projection of the PoS evolution, with the trajectories of the 11 satellites plotted relative to their host galaxy. These are shown for the three high-mass most MW-like PoS systems. The location of the central host galaxy is indicated by a star.  The lower panels of these figures show the evolution of the radial distances for the 11 PoS satellites. The black dashed lines show aspect ratio $c/a$ of the PoS from $z = 4$ to $z = 0$. The color coding corresponds to the satellite numbers listed in Tables \ref{sat-tab1}-\ref{sat-tab2}, as labeled.

We interpret these figures as follows: About half of the satellites arrived with in 1 Mpc of the host galaxy at high redshift and have assumed elliptic orbits as evidenced by the radial oscillation in the lower panels. However, the remaining satellites have only arrived recently at low redshift $z < 0.2$. This highlights the transient nature of a  PoS with {  a low}  $c/a$. The appearance of the 11 satellites in proximity to the host galaxy is, thus, a recent phenomenon occurring at low redshift ($z < 0.2$).

We also note that in each case for which a heavy LMC-like galaxy is part of the PoS, it arrives as one of the late in-falling groups bringing other satellite galaxies with it. This is apparent in the dark blue lines drawn for systems 188 and 289 in Figure \ref{figure9}.   
Hence, these time series suggest that the observed MW-like PoS is assembled relatively recently, i.e. at low redshift. This is consistent with the interpretation that the present-day thin PoS is a transient configuration rather than a long-lived spatial structure, while some degree of orbital coherence may persist over longer timescales \citep{Shao:2019,Sawala:2022a}. On the other hand, about half of the PoS members were captured by the host into elliptical orbits several Gyr in the past.  Nevertheless, in each case, the appearance of 11 galaxies in a PoS with a narrow aspect ratio in close proximity to the host galaxy is a temporary alignment often (but not always) associated with a heavy in-falling MC-like satellite.

\section{Summary and Conclusions}
\label{sec_conclusions}
{We have made a new independent study of the existence of MW-like simulated PoS systems.  We used the IllustrisTNG simulations based upon $\Lambda$CDM cosmology to study the spatial alignment and properties of simulated PoS galaxies around Milky  Way-mass hosts. We utulized two different mass ranges to identify MW-mass systems.}  

We found that the spatial distribution of satellite galaxies, dark matter subhalos, and dark matter in the halo can be significantly different. Satellite galaxies do not strictly trace the dark matter distribution in the halo. If only a small number of satellites ({11 in the case of the MW}) are chosen to be analyzed, there is a small but non-negligible chance that satellites can be distributed in a thin plane.

We have identified MW-like PoS systems by requiring its $c/a \le 0.25, \Delta_k \le 66^o$, and $\Delta_{k7} \le 35^o$. {Based upon these criteria, we have identified a grouping of outliers with properties very similar to the PoS of the MW.}  For the two mass ranges that we have selected for the MW-like galaxies, 
 the probability of finding a MW-like PoS is 1.6 percent (3 out of {186}) in the high-mass range and {0.4} percent (6 out of {1392}) in the low-mass range in the TNG50 simulations.  Hence, a MW-like PoS can form in simulations within $\Lambda$CDM cosmology at the $\sim 1$ percent level even if we require the satellites in the fitted plane to have kinematic properties {(low aspect ratio and low orbital pole dispersion)} like those of the MW.

Similar to the findings from {the SAGA DR3 observations}, we find that {the MW-like PoS systems  are the} most radially concentrated satellite systems compared to the galaxies with comparable mass in the TNG50 simulations. However, we find that {a low Gini coefficient, although indicative of anisotropy, is not a good} indicator of a MW-like PoS. Even a high Gini coefficient can sometimes occur in conjunction with a small $c/a$ aspect ratio.

If we consider the higher mass range, we find that the {simulated MW-like PoS systems are} the most radially concentrated among all the systems in the high mass range. The reason for this may be due to the tidal disruption close to the stellar disk. However, if we consider the lower mass range, the radial concentration of the {MW-like PoS is not as much of an outlier in that several systems were identified with a similar concentration. Nevertheless, systems with the combination of a low aspect ratio and a low orbital pole dispersion remains as a unique class of outliers.}  

By studying the galaxies with a MW-like PoS and their large-scale environments, we find that seven of the nine best candidate systems reside within well-defined filaments. We also find that the orientations of the PoS and galactic disk of the main galaxy are not directly correlated with each other. This is due to the fact that the formation and evolution of the
main galaxy and the formation and accretion of the satellites both involved interaction with their environments, but at different times and scales.


Consistent with the SAGA DR3 conclusions, we find that a significant fraction (about 1/3) of the MW-like PoS systems contain a massive LMC-like (or SMC-like) satellite. Having MC-like satellites could also increase the number of satellites in the inner region of the system so that the system is more radially concentrated. In turn, this makes the occurrence of a fitted thin plane more likely.


 In summary, 1) A PoS like that of the Milky Way is indeed a rare occurrence to be found in only $\sim$1\% of simulated MW-mass systems;   2) Although rare, a PoS with a low aspect ratio and low orbital dispersion is a natural occurrence in $\Lambda CDM$ cosmology. 3) Our analysis suggests that filamentary environments are common among the most MW-like PoS systems identified in this work, while recent accretion events, including those involving Magellanic-Cloud-like satellites, may also contribute to their formation. What remains unresolved is a detailed understanding of  the time evolution of the formation of the PoS and how it is influenced by cosmic structure. These will be the subject of a subsequent work.

\section*{Acknowledgements}

Work at the University of Notre Dame is supported by the U.S. Department of Energy under Nuclear Theory Grant DE-FG02-95-ER40934. Work at Dalton State College is partially supported by the U.S. Department of Education Title III HSI grant number P031C210024. X.Z. is grateful for the support from the visiting program of the Center for Computational Astrophysics (CCA) at the Flatiron Institute. This work was performed in part at Aspen Center for Physics, which is supported by National Science Foundation grant PHY-2210452. Research of L.A.P. was supported in part by grant NSF PHY-2309135 to the Kavli Institute for Theoretical Physics (KITP). L.A.P. and G.T. are also grateful for the hospitality of Perimeter Institute where part of this work was carried out. Research at Perimeter Institute is supported in part by the Government of Canada through the Department of Innovation, Science and Economic Development and by the Province of Ontario through the Ministry of Colleges and Universities. This work was also supported by a grant from the Simons Foundation (1034867, Dittrich).
We also acknowledge support from the University of Notre Dame Center for Research Computing for providing computational resources and services that have contributed to the research results reported in this paper.

\section*{Data Availability}
The IllustrisTNG simulation data used in this work are publicly available at \url{https://www.tng-project.org/data/} \citep{Nelson:2019}.



\bibliographystyle{mnras}
\bibliography{dos} 



\clearpage
\appendix
\twocolumn[
\section{Resolution properties of the satellite sample}
\label{sec_appendix}
]
{   Tables~\ref{tab_resolution_188}--\ref{tab_resolution_573} show the numerical resolution at $z=0$ for each of the 9 most MW-like PoS systems as labelled.  The second and third  columns give the mass in stars and number of star particles in the host and each satellite galaxy.  Columns 4 and 5 give the corresponding mass in dark matter and the total number of resolution elements (star particles, DM particles, gas cells, and black hole particles). }
\begin{table}[ht]
{
    \centering
    \caption{Resolution properties of the satellites in G188.}
    \label{tab_resolution_188}
    \begin{tabular}{ccccc}
        \hline
        Sat. Num. & $M_\star\,(M_\odot)$ & $N_\star$ & $M_{\rm DM}\,(M_\odot)$ & $N_{\rm total}$ \\
        \hline
        Host & $4.6\times10^{10}$ & 800442 & $1.4\times10^{12}$ & 4054725 \\
        1 & $1.1\times10^{10}$ & 176324 & $7.2\times10^{10}$ & 524135 \\
        2 & $2.5\times10^{8}$ & 4528 & $4.0\times10^{9}$ & 14500 \\
        3 & $5.8\times10^{7}$ & 989 & $9.2\times10^{9}$ & 23943 \\
        4 & $8.7\times10^{6}$ & 156 & $5.6\times10^{8}$ & 1381 \\
        5 & $6.9\times10^{6}$ & 129 & $2.3\times10^{8}$ & 630 \\
        6 & $6.0\times10^{6}$ & 99 & $8.2\times10^{8}$ & 1907 \\
        7 & $5.6\times10^{6}$ & 95 & $2.9\times10^{8}$ & 732 \\
        8 & $4.5\times10^{6}$ & 77 & $1.2\times10^{9}$ & 2677 \\
        9 & $2.8\times10^{6}$ & 51 & $1.3\times10^{9}$ & 2951 \\
        10 & $1.5\times10^{6}$ & 28 & $3.8\times10^{8}$ & 864 \\
        11 & $1.5\times10^{6}$ & 23 & $2.2\times10^{8}$ & 509 \\
        \hline
    \end{tabular}
    }
\end{table}

\begin{table}
{
    \centering
    \caption{Resolution properties of the satellites in G206.}
    \label{tab_resolution_206}
    \begin{tabular}{ccccc}
        \hline
        Sat. Num. & $M_\star\,(M_\odot)$ & $N_\star$ & $M_{\rm DM}\,(M_\odot)$ & $N_{\rm total}$ \\
        \hline
        Host & $2.9\times10^{10}$ & 482328 & $6.8\times10^{11}$ & 2984079 \\
        1 & $2.3\times10^{9}$ & 37948 & $1.1\times10^{11}$ & 414972 \\
        2 & $1.6\times10^{9}$ & 27348 & $5.1\times10^{10}$ & 200371 \\
        3 & $3.4\times10^{8}$ & 6096 & $1.1\times10^{9}$ & 9505 \\
        4 & $2.5\times10^{7}$ & 444 & $2.3\times10^{9}$ & 5823 \\
        5 & $1.9\times10^{7}$ & 327 & $7.3\times10^{8}$ & 1933 \\
        6 & $1.0\times10^{7}$ & 178 & $1.5\times10^{9}$ & 3532 \\
        7 & $6.6\times10^{6}$ & 112 & $2.2\times10^{9}$ & 4936 \\
        8 & $5.3\times10^{6}$ & 97 & $3.5\times10^{7}$ & 174 \\
        9 & $5.2\times10^{6}$ & 96 & $3.5\times10^{9}$ & 7895 \\
        10 & $2.2\times10^{6}$ & 42 & $4.7\times10^{9}$ & 10340 \\
        11 & $1.8\times10^{6}$ & 32 & $1.1\times10^{9}$ & 2467 \\
        \hline
    \end{tabular}
    }
\end{table}

\begin{table}
{
    \centering
    \caption{Resolution properties of the satellites in G232.}
    \label{tab_resolution_232}
    \begin{tabular}{ccccc}
        \hline
        Sat. Num. & $M_\star\,(M_\odot)$ & $N_\star$ & $M_{\rm DM}\,(M_\odot)$ & $N_{\rm total}$ \\
        \hline
        Host & $4.9\times10^{10}$ & 821981 & $8.7\times10^{11}$ & 3466271 \\
        1 & $7.5\times10^{7}$ & 1316 & $3.3\times10^{9}$ & 8581 \\
        2 & $5.2\times10^{7}$ & 979 & $9.8\times10^{8}$ & 3147 \\
        3 & $1.1\times10^{7}$ & 186 & $5.7\times10^{7}$ & 311 \\
        4 & $1.0\times10^{7}$ & 175 & $1.2\times10^{9}$ & 2733 \\
        5 & $4.5\times10^{6}$ & 82 & $6.4\times10^{9}$ & 14882 \\
        6 & $2.2\times10^{6}$ & 37 & $1.7\times10^{8}$ & 407 \\
        7 & $7.5\times10^{5}$ & 14 & $9.1\times10^{7}$ & 214 \\
        8 & $4.6\times10^{5}$ & 8 & $2.7\times10^{8}$ & 604 \\
        9 & $4.5\times10^{5}$ & 9 & $1.5\times10^{9}$ & 3240 \\
        10 & $4.2\times10^{5}$ & 7 & $8.0\times10^{7}$ & 183 \\
        11 & $4.0\times10^{5}$ & 7 & $3.1\times10^{8}$ & 687 \\
        \hline
    \end{tabular}
    }
\end{table}

\begin{table}
{
    \centering
    \caption{Resolution properties of the satellites in G289.}
    \label{tab_resolution_289}
    \begin{tabular}{ccccc}
        \hline
        Sat. Num. & $M_\star\,(M_\odot)$ & $N_\star$ & $M_{\rm DM}\,(M_\odot)$ & $N_{\rm total}$ \\
        \hline
        Host & $1.7\times10^{10}$ & 281753 & $5.6\times10^{11}$ & 2235035 \\
        1 & $2.6\times10^{9}$ & 39631 & $1.8\times10^{10}$ & 205664 \\
        2 & $2.9\times10^{8}$ & 4958 & $2.1\times10^{10}$ & 96743 \\
        3 & $1.6\times10^{8}$ & 2466 & $4.3\times10^{8}$ & 6478 \\
        4 & $1.6\times10^{8}$ & 2855 & $5.2\times10^{9}$ & 21384 \\
        5 & $1.4\times10^{8}$ & 2636 & $1.6\times10^{9}$ & 6394 \\
        6 & $3.6\times10^{7}$ & 622 & $1.3\times10^{9}$ & 5480 \\
        7 & $2.5\times10^{7}$ & 450 & $1.8\times10^{9}$ & 4466 \\
        8 & $3.0\times10^{6}$ & 55 & $1.1\times10^{9}$ & 2410 \\
        9 & $1.4\times10^{6}$ & 26 & $3.2\times10^{9}$ & 7000 \\
        10 & $1.3\times10^{6}$ & 25 & $1.3\times10^{9}$ & 2839 \\
        11 & $7.7\times10^{5}$ & 12 & $1.9\times10^{8}$ & 433 \\
        \hline
    \end{tabular}
    }
\end{table}

\begin{table}
{
    \centering
    \caption{Resolution properties of the satellites in G365.}
    \label{tab_resolution_365}
    \begin{tabular}{ccccc}
        \hline
        Sat. Num. & $M_\star\,(M_\odot)$ & $N_\star$ & $M_{\rm DM}\,(M_\odot)$ & $N_{\rm total}$ \\
        \hline
        Host & $1.4\times10^{10}$ & 232499 & $3.1\times10^{11}$ & 1047733 \\
        1 & $1.3\times10^{10}$ & 223303 & $9.2\times10^{10}$ & 426036 \\
        2 & $2.0\times10^{9}$ & 36706 & $8.1\times10^{10}$ & 304372 \\
        3 & $6.7\times10^{7}$ & 1219 & $1.5\times10^{9}$ & 4624 \\
        4 & $3.7\times10^{6}$ & 74 & $7.3\times10^{8}$ & 1680 \\
        5 & $1.3\times10^{6}$ & 25 & $2.2\times10^{7}$ & 73 \\
        6 & $7.0\times10^{5}$ & 14 & $4.0\times10^{9}$ & 8939 \\
        7 & $3.6\times10^{5}$ & 7 & $1.5\times10^{8}$ & 344 \\
        8 & $2.5\times10^{5}$ & 4 & $1.5\times10^{8}$ & 328 \\
        9 & $1.4\times10^{5}$ & 3 & $8.9\times10^{7}$ & 200 \\
        10 & $1.2\times10^{5}$ & 3 & $3.7\times10^{8}$ & 825 \\
        11 & $1.1\times10^{5}$ & 3 & $3.4\times10^{8}$ & 753 \\
        \hline
    \end{tabular}
    }
\end{table}

\begin{table}
{
    \centering
    \caption{Resolution properties of the satellites in G395.}
    \label{tab_resolution_395}
    \begin{tabular}{ccccc}
        \hline
        Sat. Num. & $M_\star\,(M_\odot)$ & $N_\star$ & $M_{\rm DM}\,(M_\odot)$ & $N_{\rm total}$ \\
        \hline
        Host & $2.5\times10^{10}$ & 432379 & $4.0\times10^{11}$ & 1905529 \\
        1 & $4.5\times10^{7}$ & 839 & $6.9\times10^{9}$ & 20066 \\
        2 & $1.5\times10^{7}$ & 267 & $3.6\times10^{9}$ & 8132 \\
        3 & $1.2\times10^{7}$ & 218 & $1.9\times10^{8}$ & 637 \\
        4 & $1.2\times10^{7}$ & 199 & $4.5\times10^{8}$ & 1193 \\
        5 & $7.9\times10^{6}$ & 141 & $1.7\times10^{9}$ & 3797 \\
        6 & $2.3\times10^{6}$ & 39 & $1.5\times10^{8}$ & 363 \\
        7 & $1.5\times10^{6}$ & 31 & $2.5\times10^{7}$ & 85 \\
        8 & $1.0\times10^{6}$ & 18 & $2.0\times10^{9}$ & 4394 \\
        9 & $9.8\times10^{5}$ & 16 & $2.2\times10^{8}$ & 504 \\
        10 & $6.3\times10^{5}$ & 13 & $3.2\times10^{8}$ & 717 \\
        11 & $4.5\times10^{5}$ & 8 & $2.3\times10^{8}$ & 521 \\
        \hline
    \end{tabular}
    }
\end{table}

\begin{table}
{
    \centering
    \caption{Resolution properties of the satellites in G419.}
    \label{tab_resolution_419}
    \begin{tabular}{ccccc}
        \hline
        Sat. Num. & $M_\star\,(M_\odot)$ & $N_\star$ & $M_{\rm DM}\,(M_\odot)$ & $N_{\rm total}$ \\
        \hline
        Host & $1.5\times10^{10}$ & 247704 & $4.5\times10^{11}$ & 1797398 \\
        1 & $6.3\times10^{7}$ & 1215 & $3.9\times10^{8}$ & 2077 \\
        2 & $3.2\times10^{7}$ & 570 & $3.2\times10^{8}$ & 1283 \\
        3 & $1.9\times10^{7}$ & 341 & $5.8\times10^{9}$ & 14762 \\
        4 & $1.3\times10^{7}$ & 233 & $5.8\times10^{9}$ & 13328 \\
        5 & $1.1\times10^{6}$ & 20 & $3.1\times10^{8}$ & 703 \\
        6 & $6.7\times10^{5}$ & 13 & $3.7\times10^{8}$ & 839 \\
        7 & $6.4\times10^{5}$ & 11 & $1.8\times10^{8}$ & 405 \\
        8 & $6.3\times10^{5}$ & 12 & $1.0\times10^{9}$ & 2318 \\
        9 & $5.5\times10^{5}$ & 10 & $1.3\times10^{8}$ & 294 \\
        10 & $1.7\times10^{5}$ & 3 & $1.1\times10^{8}$ & 244 \\
        11 & $1.5\times10^{5}$ & 3 & $3.2\times10^{8}$ & 711 \\
        \hline
    \end{tabular}
    }
\end{table}

\begin{table}
{
    \centering
    \caption{Resolution properties of the satellites in G473.}
    \label{tab_resolution_473}
    \begin{tabular}{ccccc}
        \hline
        Sat. Num. & $M_\star\,(M_\odot)$ & $N_\star$ & $M_{\rm DM}\,(M_\odot)$ & $N_{\rm total}$ \\
        \hline
        Host & $6.1\times10^{9}$ & 102682 & $2.8\times10^{11}$ & 1019616 \\
        1 & $2.6\times10^{9}$ & 44066 & $6.7\times10^{10}$ & 399285 \\
        2 & $1.1\times10^{8}$ & 1824 & $1.5\times10^{8}$ & 3034 \\
        3 & $9.3\times10^{6}$ & 165 & $5.2\times10^{9}$ & 12166 \\
        4 & $6.4\times10^{6}$ & 111 & $2.6\times10^{9}$ & 5755 \\
        5 & $5.3\times10^{6}$ & 96 & $7.6\times10^{7}$ & 264 \\
        6 & $2.2\times10^{6}$ & 30 & $4.5\times10^{5}$ & 1278 \\
        7 & $1.8\times10^{6}$ & 32 & $2.2\times10^{8}$ & 522 \\
        8 & $10.0\times10^{5}$ & 19 & $7.7\times10^{8}$ & 1712 \\
        9 & $4.1\times10^{5}$ & 7 & $8.2\times10^{7}$ & 188 \\
        10 & $3.4\times10^{5}$ & 5 & $1.3\times10^{9}$ & 2822 \\
        11 & $2.5\times10^{5}$ & 5 & $2.4\times10^{8}$ & 543 \\
        \hline
    \end{tabular}
    }
\end{table}

\begin{table}
{
    \centering
    \caption{Resolution properties of the satellites in G573.}
    \label{tab_resolution_573}
    \begin{tabular}{ccccc}
        \hline
        Sat. Num. & $M_\star\,(M_\odot)$ & $N_\star$ & $M_{\rm DM}\,(M_\odot)$ & $N_{\rm total}$ \\
        \hline
        Host & $1.8\times10^{9}$ & 29536 & $1.5\times10^{11}$ & 505625 \\
        1 & $1.5\times10^{9}$ & 25465 & $8.6\times10^{10}$ & 355921 \\
        2 & $8.8\times10^{8}$ & 14344 & $1.8\times10^{10}$ & 90920 \\
        3 & $4.0\times10^{6}$ & 74 & $4.3\times10^{9}$ & 11478 \\
        4 & $1.1\times10^{6}$ & 22 & $2.9\times10^{8}$ & 663 \\
        5 & $7.9\times10^{5}$ & 15 & $5.5\times10^{8}$ & 1222 \\
        6 & $2.8\times10^{5}$ & 5 & $3.2\times10^{8}$ & 719 \\
        7 & $2.7\times10^{5}$ & 5 & $1.1\times10^{9}$ & 2467 \\
        8 & $2.6\times10^{5}$ & 3 & $0$ & 1652 \\
        9 & $2.2\times10^{5}$ & 4 & $4.9\times10^{8}$ & 1093 \\
        10 & $1.4\times10^{5}$ & 3 & $4.4\times10^{8}$ & 976 \\
        11 & $1.3\times10^{5}$ & 3 & $1.9\times10^{8}$ & 431 \\
        \hline
    \end{tabular}
    }
\end{table}



\bsp	
\label{lastpage}
\end{document}